\documentclass[a4paper,11pt]{article}
\pdfoutput=1 

\usepackage{jheppub} 
 \usepackage{array} 
\usepackage[T1]{fontenc}
\usepackage[utf8]{inputenc}
\usepackage{amsmath,amssymb,amstext,amsfonts}

\usepackage[dvipsnames]{xcolor}
\usepackage{graphicx}
\usepackage{subfig}
\usepackage{hyperref}
\usepackage{float}
\usepackage{mathtools}
\usepackage{cleveref}
\usepackage{slashed}
\usepackage{multirow}
\usepackage{booktabs}
\usepackage{placeins}
\usepackage{caption}

\usepackage{tikz}
\usepackage{tikz-feynman}
\usepackage{contour}
\usetikzlibrary{decorations.pathmorphing,arrows.meta}

\newcommand{\eps}{\epsilon}

\definecolor{cnblue}{RGB}{7,82,154}

\newcommand{\ie}{i.e.~}

\newcommand{\nn}{\nonumber}

\allowdisplaybreaks[4]

\title{Analytic two-loop corrections to top-quark pair production at hadron colliders}

\affiliation[a]{CERN, Theoretical Physics Department, 1211 Geneva 23, Switzerland}

\affiliation[b]{Max Planck Institute for Gravitational Physics (Albert Einstein Institute), Am Mühlenberg 1, 14476 Potsdam, Germany}

\affiliation[c]{Institut f\"ur Physik, Humboldt-Universit\"at zu Berlin, 12489 Berlin, Germany}

\affiliation[d]{Physik Department, James-Franck-Stra{\ss}e 1, Technische Universit\"at M\"unchen, 85748 Garching, Germany}

\author[a]{Federico Buccioni,}
\author[b]{Christoph Nega,}
\author[c,d]{Benjamin Sauer,}
\author[d]{Lorenzo Tancredi,}
\author[d]{Fabian J. Wagner}

\emailAdd{federico.buccioni@cern.ch}
\emailAdd{christoph.nega@aei.mpg.de}
\emailAdd{benjamin.sauer@tum.de}
\emailAdd{lorenzo.tancredi@tum.de}
\emailAdd{fabianjohannes.wagner@tum.de}

\preprint{\begin{minipage}[t]{8cm}\begin{flushright}
CERN-TH-2026-261 \\ TUM-HEP-1617/26 \\ HU-EP-26/31-RTG
\end{flushright}\end{minipage}}

\abstract{
We present the complete two-loop QCD amplitudes for the production of a top-quark pair in the quark--antiquark annihilation and gluon-fusion channels in analytic form up to order $\mathcal{O}(\varepsilon^2)$. 
Our analysis maintains full dependence on the top-quark mass and on the number of colours.
The amplitudes are expressed in terms of 417 master integrals, which we compute by means of
differential equations in $\varepsilon$-factorised form. Their calculation involves 
several elliptic curves
and a hyperelliptic curve of genus two, 
which makes top-quark pair production an ideal arena for 
recent methods to construct canonical bases beyond polylogarithms.
To the best of our knowledge, this is the first scattering 
amplitude in which iterated integrals related to a genus-two surface 
survive in the finite remainder in four space-time dimensions.
We provide high-precision series expansions at the threshold
and combine them with the numerical solution of the differential equations
in a code that evaluates the UV-renormalised infrared-finite
remainders at arbitrary points of the physical phase space.
}

\begin{document}

\maketitle
\raggedbottom

%
\section{Introduction}
\label{sec:intro}

The top quark is the heaviest known elementary particle. 
Its large mass and the correspondingly large coupling to the Higgs boson give it a special role in the mechanism of electroweak symmetry breaking and make it a natural portal to physics beyond the Standard Model. 
At the Large Hadron Collider (LHC), pairs of top quarks are produced in large numbers, predominantly through their couplings to gluons such that
top-quark pair production has become one of the most precisely measured processes 
at hadron colliders. 
In particular, inclusive and differential $t\bar{t}$ cross sections are measured with an accuracy of a few per cent~\cite{CMS:2019esx, ATLAS:2023gsl, CMS:2023qyl, ATLAS:2023slx}. 
These observables can be used to determine fundamental parameters of the Standard Model, 
such as the top-quark mass and the strong coupling constant, 
and to constrain the gluon density of the proton at large momentum fractions~\cite{CMS:2018fks, CMS:2019esx, Czakon:2016olj}. 
The spin correlations of the top-quark pair, which are transmitted to its decay products, are measured with 
high precision and have been used to test the Standard Model, including observables inspired by quantum information~\cite{ATLAS:2023fsd, CMS:2024pts, CMS:2024zkc}.
Finally, top-quark pairs are an important background to many searches for new physics.

Exploiting these measurements requires theoretical predictions of matching precision. 
The next-to-leading order (NLO) QCD corrections to on-shell $t\bar{t}$ production have been known 
for more than three decades~\cite{Nason:1987xz, Nason:1989zy, Beenakker:1988bq, Beenakker:1990maa, Czakon:2008ii}. Next-to-next-to-leading order (NNLO) QCD predictions became available first for the total cross section~\cite{Barnreuther:2012wtj, Czakon:2012zr, Czakon:2012pz, Czakon:2013goa}, 
and subsequently for fully differential observables~\cite{Czakon:2014xsa, Czakon:2015owf, Catani:2019iny, Catani:2019hip,Catani:2020tko},
including as well electroweak corrections~\cite{Czakon:2017wor}, and matching to a parton shower
with the same perturbative accuracy~\cite{Mazzitelli:2020jio,Mazzitelli:2021mmm}.
A central ingredient of all these calculations are the two-loop virtual amplitudes for the partonic channels $q\bar{q}\to t\bar{t}$ and $gg\to t\bar{t}$. 
For the landmark calculation of the total cross section of~\cite{Czakon:2013goa}, 
the two-loop
amplitudes were computed numerically~\cite{Czakon:2008zk, Barnreuther:2013qvf}, by integrating the differential equations for the master integrals in the bulk of the phase space and matching these results to  high-energy and  threshold expansions~\cite{Czakon:2007ej, Czakon:2007wk}. 
The infrared-finite remainders 
are available in the form of an interpolation grid covering the physical phase space, 
supplemented by expansions around the production threshold and in the high-energy limit. 
This approach was later extended to include the full dependence on the spins and colours of the external partons~\cite{Chen:2017jvi}. These advances enabled the study of NNLO QCD corrections to resonant top-pair production with spin-correlated decays and realistic selection criteria applied to the top-quark decay products~\cite{Behring:2019iiv,Czakon:2020qbd,Buonocore:2025fqs}.

The motivation for resorting to numerical methods lies in the mathematical 
complexity of these amplitudes.
In fact, it has been known for a long time that 
the gluon-fusion channel involves large numbers of 
Feynman integrals which cannot be expressed in terms of multiple polylogarithms (MPLs)~\cite{Kummer, Goncharov:1995, Goncharov:1998kja, Remiddi:1999ew, Vollinga:2004sn}.
In particular, some of the integrals with 
closed top-quark loops were soon found to involve elliptic curves~\cite{Czakon:2008ii,vonManteuffel:2017hms,Adams:2018bsn,Adams:2018kez} and,
more recently, it was also observed that one of the non-planar graphs with the largest number of propagators involves a hyper-elliptic curve of genus two~\cite{Marzucca:2023gto}. 
Despite this, many intermediate analytic results were obtained for subsets of the amplitudes that only involve MPLs, 
such as the contributions with closed light-quark loops and the 
leading-colour contributions 
without closed heavy-quark loops~\cite{Bonciani:2008az, Bonciani:2009nb, Bonciani:2010mn, Bonciani:2013ywa,vonManteuffel:2013uoa}. 
Moreover, analytic results for the helicity amplitudes, including 
the elliptic contributions, are available at leading colour~\cite{Badger:2021owl}. 
From this point of view, the quark-annihilation channel is substantially simpler. 
Its master integrals evaluate entirely to MPLs~\cite{Bonciani:2008az, Bonciani:2009nb, DiVita:2019lpl, Becchetti:2019tjy}, which recently made it possible to recompute this partonic channel analytically~\cite{Mandal:2022vju}, and produce 
a numerical grid analogous to that of~\cite{Chen:2017jvi}. 

In recent years, there has been considerable progress in understanding 
Feynman integrals beyond MPLs. 
In particular, the concept of canonical differential equations~\cite{Henn:2013pwa}, 
which has revolutionised the calculation of polylogarithmic Feynman integrals, 
has been extended to integrals associated with elliptic curves, Calabi-Yau varieties and curves of higher genus~\cite{Adams:2018yfj, Dlapa:2022wdu, Pogel:2022ken, Frellesvig:2023iwr, Gorges:2023zgv, Frellesvig:2024rea, Duhr:2024uid, Duhr:2025lbz, Chen:2025hzq, Chaubey:2025adn, e-collaboration:2025frv, Bree:2025tug, Forner:2026vby, Chaubey:2026egr}, 
and efficient methods have been developed to evaluate the resulting iterated integrals numerically~\cite{Baur:2026zlw, Czakon:2026tog}. 
These ideas have already been applied, among the other cases, 
 to realistic two-loop $2\to2$ QCD amplitudes involving elliptic 
curves~\cite{Becchetti:2025rrz, Coro:2025vgn}, and they 
offer the opportunity to reconsider the two-loop amplitudes for $t\bar{t}$ 
production in a fully analytic way. 
From a formal point of view, this process is particularly interesting: 
its master integrals 
involve 13 different elliptic curves as well as a genus-two hyperelliptic curve, 
several of which appear 
together in the same system of differential equations. Importantly, 
this constitutes to the best of our knowledge the first scattering amplitude 
that requires the evaluation of 
iterated integrals related to a genus-two surface in its four-dimensional
finite remainder.
It is therefore an ideal laboratory to stress-test our current 
understanding of Feynman integrals beyond polylogarithms in a realistic, 
phenomenologically relevant setting.

Motivated by the considerations above,
in this paper 
we compute the complete two-loop QCD corrections to $q\bar{q}\to t\bar{t}$ and $gg\to t\bar{t}$ analytically, keeping the full dependence on the number of colours and including all contributions from closed loops of light and top quarks. 
To deal with the two-loop Feynman integrals, we construct a canonical basis with the method of~\cite{Gorges:2023zgv, Duhr:2025lbz, Forner:2026vby}, 
which allows us to treat  both the homogeneous and inhomogeneous sectors of the differential equations seamlessly, including the many non-trivial couplings between 
the different non-polylogarithmic sectors.\footnote{We stress that alternative approaches to the construction of canonical bases beyond 
polylogarithms~\cite{Adams:2018yfj, Dlapa:2022wdu, Pogel:2022ken, e-collaboration:2025frv, Bree:2025tug} and to deal with elliptic problems avoiding $\varepsilon$-factorisation~\cite{Chaubey:2025adn, Chaubey:2026egr} exist.} 
In this way, defining the dimensional regulator $\varepsilon = (4-d)/2$, we obtain differential equations in $\varepsilon$-factorised form for all master integrals required to express all configurations. 
One of the important advantages of $\varepsilon$-factorised differential equations is that
their solution can be effortlessly expressed in terms of linearly independent iterated integrals and that results can be obtained, at least in principle, to arbitrary orders in $\varepsilon$. 
Moreover, having equations in this form, it is especially simple to compute series expansion solutions close to arbitrary singular or regular points and, as it was recently argued, to obtain fast and reliable numerical 
evaluation of the corresponding integrals by numerical transport~\cite{Czakon:2026tog}. 
We employ these ideas to further extend the calculation of 
the two-loop amplitudes through $\mathcal{O}(\varepsilon^2)$, two orders beyond what is required for NNLO predictions. These higher-order terms will be important, among other things, to study the infrared structure of the three-loop amplitudes for $t\bar{t}$ production.

Together with the analytic results, we provide a numerical code that evaluates the ultraviolet (UV)-renormalised infrared-finite remainders of the two-loop amplitudes for both channels at arbitrary phase-space points. The code works dynamically in double and quadruple precision with internal switches to guarantee a reliable result across the whole phase space. Since the higher orders in $\varepsilon$ do not enter physical predictions at NNLO, they are not included in the standard code. Instead, we provide benchmark values for them in this paper, together with a separate code that evaluates the master integrals to the required order in $\varepsilon$.

All our results are published in electronic format together with the submission of this paper.
The analytic expressions for the amplitudes and master integrals, together with their boundary values at threshold, are provided in the accompanying Zenodo repository~\cite{Zenodo_ttbar}, while the codes for the numerical evaluation of polarised amplitudes up to two-loops \texttt{hqpAmp},
and of the master integrals, \texttt{hqpInt}, are available in separate git repositories~\cite{hqpamp}.

The rest of the paper is organised as follows. In \cref{sec:setup}, we introduce our notation and the kinematics of the process. In \cref{sec:amplitude}, we describe the tensor and colour decomposition of the amplitudes, the projectors onto the form factors, and the spin-polarised amplitudes. \Cref{sec:integrals} is devoted to the master integrals: 
we discuss the integral families, the construction of the canonical basis, with special emphasis on the genus-two sector, and the solution of the differential equations together with their boundary conditions. In \cref{sec:ren}, we discuss the ultraviolet (UV) renormalisation and the infrared (IR) structure of the amplitudes, and in \cref{sec:numerics}, their numerical evaluation and our results. We conclude in \cref{sec:conc}. Our conventions, the definition of the integral families, and further details on the renormalisation and the infrared structure are collected in the appendices.

%
\section{Notation and kinematics}
\label{sec:setup}

We consider the production of a top-antitop ($t\bar{t}$) pair at hadron colliders
\begin{equation}
    A(p_1) + B(p_2) \to t(p_3) + \bar{t}(p_4)\,,
\end{equation}
where $A$ and $B$ are massless partons, i.e.~$p_1^2=p_2^2=0$, 
and 
$(A,B) = (g,g)$, $(A,B) = (q,\bar{q})$ or $(A,B) = (\bar{q},q)$.
The momenta of the top quarks fulfil the on-shell conditions $p_3^2 = p_4^2 = m_t^2$.
We further introduce the following usual Mandelstam invariants
\begin{equation}
\label{eq:mandelstaminv}
    s = (p_1 + p_2)^2, \quad t = (p_1 - p_3)^2, \quad u = (p_1 - p_4)^2,
\end{equation}
which are related by momentum conservation via
\begin{equation}
    s  + t + u = 2m_t^2\,.
\end{equation}

\begin{figure}[ht!]
    \centering
    \includegraphics[scale=1.2]{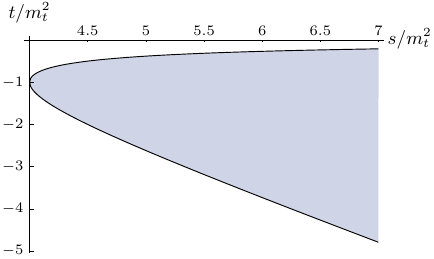}
    \caption{The shaded area represents the physical region in the $(s,t)$-plane. Both $s$ and $t$ are rescaled by $m_t^2$. The boundaries of the phase space are given by $\Delta=0$, see~\cref{eq:gramdet}.}
    \label{fig:physregionst}
\end{figure}

Without loss of generality, we study the kinematics of the production of two heavy quarks  
in the centre-of-mass frame of the scattering partons.
In this frame, defining $\hat{n}_1$ as the normal vector along the beam axis,
one has
\begin{equation}
\label{eq:kinexample}
    p_{1,2} = \frac{\sqrt{s}}{2} \left(1,\pm \hat{n}_1\right),\quad\quad
    p_{3,4} = \frac{\sqrt{s}}{2} \left(1,\pm \beta \,\hat{n}_t\right),\quad \text{with}
    \quad \beta = \sqrt{1-\frac{4m_t^2}{s}}\,,
\end{equation}
where the normal vector $\hat{n}_t$ indicates instead the direction of the top quark.
The $t$ and $u$ invariants can thus be expressed as
\begin{equation}
\label{eq:mandelstamt}
    t = m_t^2 -  \frac{s}{2}\left(1 - \beta \,\cos\theta\right),\quad\quad
    u = m_t^2 -  \frac{s}{2}\left(1 + \beta \, \cos\theta\right)\,,
\end{equation}
where $\cos\theta\equiv\hat{n}_1\cdot\hat{n}_t$, with $\theta$ denoting the polar angle of the top-quark direction relative to the beam axis.
Given that the $t\bar{t}$ production threshold is given by $s\ge 4m_t^2$, and taking into account that $-1 \le \cos\theta \le 1$, \cref{eq:mandelstamt} defines the physical scattering region
\begin{equation}
    \frac{\left( 1 - \beta\right)}{2}\frac{s}{m_t^2} - 1 \le \frac{-t}{m_t^2} \le \frac{\left( 1 + \beta\right)}{2}\frac{s}{m_t^2} - 1\,.
\end{equation}
The graphical visualisation of the physical region in the $(s,t)$-plane is presented in~\cref{fig:physregionst}.
Moreover, we define the Gram determinant $\Delta$ as
\begin{equation}
    \Delta = \det(G_{ij}) =\frac{s}{4}\left(u\, t -m_t^4\right), \quad \text{with} \quad 
    G_{ij}=p_i\cdot p_j,\;\; i,j \in \{1,2,3\} \, . \label{eq:gramdet}
\end{equation}   
In terms of the velocity $\beta$ and the scattering angle $\theta$, one finds 
$$\Delta = s^3\beta^2\sin^2\theta/16,$$ 
so that $\Delta$ vanishes at the boundaries of the physical phase space, 
namely in the forward and backward regions,
see also~\cref{fig:physregionst}.
Finally, for later convenience,
we introduce an auxiliary vector $a^\mu$ defined through the fully antisymmetric Levi-Civita symbol $\varepsilon^{\mu\nu\rho\sigma}$ as
\begin{equation}
\label{eq:avector}
    a^\mu = \frac{i \epsilon^{\mu \alpha_1 \alpha_2 \alpha_3}}{\sqrt{\Delta}}\, p_{1,\alpha_1}p_{2,\alpha_2}p_{3,\alpha_3} \, ,
\end{equation}
which fulfils the relations
\begin{equation}
    p_{1,\mu} a^\mu = p_{2,\mu} a^\mu = p_{3,\mu} a^\mu = 0 \quad\quad\text{and}\quad\quad
    a_{\mu}a^\mu = 1\,.
\end{equation}
The vector $a^\mu$ is by construction four-dimensional, and orthogonal to the three momenta $p_i$ with $i=1,2,3$ and can therefore be used to span the full four-dimensional space in which the external states are confined. Importantly, 
the norm of the vector $a^\mu$ can be consistently continued to $d$ space-time dimensions
by a standard continuation of the metric tensor $g^{\mu \nu}$
\begin{equation}
    a_{\mu}a^\mu = d-3\,.
\end{equation}

Let us investigate the scaling of $a^\mu$ with respect to $\Delta$. Without loss of generality, we choose the frame of \cref{eq:kinexample} with $\hat{n}_1$ along the $z$-axis and define the $(y,z)$-plane as the one spanned by $t\bar{t}$-system, i.e.
\begin{equation}
    p_{1,2} = \frac{\sqrt{s}}{2}\left(1,0,0,\pm 1\right)\,,\qquad
    p_{3,4} = \left(\frac{\sqrt{s}}{2},\,0,\,\pm p_{t,\perp},\,\pm p_{t,z}\right)\,,
    \label{eq:framespin}
\end{equation}
where $p_{t,\perp}$ is the transverse momentum of the top quark.\footnote{One can always find a rotation around the beam axis so that $p_{t,y}=p_{t,\perp}>0$.}
One can easily show that
\begin{equation}
    \epsilon^{\mu \alpha_1 \alpha_2 \alpha_3} p_{1,\alpha_1}p_{2,\alpha_2}p_{3,\alpha_3} =
    \left(0,\frac{s}{2} p_{t,\perp},0,0\right)\,.
\end{equation}
We can then impose the on-shell conditions and use \cref{eq:mandelstaminv},
together with $\varepsilon^{0123}=-1$ (see \cref{app:spinorconventions}),
to get
\begin{equation}
    p_{t,\perp} = \frac{2}{s}\sqrt{\Delta}\quad \text{and} \quad
    a^\mu = \left(0,i,0,0\right)\,.
\end{equation}
This choice fixes the sign of the components of $a^\mu$, which 
will be useful in the following to construct explicit helicity amplitudes.
More generally, $a^\mu$ is a unit-norm complex vector transverse to the plane spanned by the collision axis and the $t\bar{t}$-system.


%
\section{Tensor and colour decomposition}
\label{sec:amplitude}

Scattering amplitudes in QCD can be  decomposed into independent Lorentz and Dirac tensor structures as well as colour structures. In this section we summarise how these decompositions work
for the production of two heavy quarks.

Let us start with the Lorentz and Dirac structures. We work in the 't~Hooft--Veltman (tHV) scheme~\cite{tHooft:1972tcz}, with four-dimensional external states. Following~\cite{Peraro:2019cjj,Peraro:2020sfm}, we only retain the tensor structures that are relevant in $d=4$, \ie the ``physical'' tensors, and discard those that vanish for strictly four-dimensional external states. The form factors are then the coefficients of these physical tensors, and are extracted with projectors, constructed with $d$-dimensional Dirac and Lorentz algebra. Our results are fully equivalent to the ones previously obtained in~\cite{Chen:2017jvi}.

We will employ the parity odd vector $a^\mu$
to span the full four-dimensional space.
In practice, since the process we are considering is parity invariant,
the vector $a^\mu$ can only appear in pairs in the tensor structures that contribute to the amplitude. For this reason, it is convenient to introduce the projector onto the space transverse to $p_1,p_2,p_3$,
\begin{equation}
\label{eq:transverseproj}
    T_A^{\mu\nu} \equiv a^\mu a^\nu = -\frac{1}{\Delta}\,\epsilon^{\mu p_1p_2p_3}\epsilon^{\nu p_1p_2p_3}
    = g^{\mu\nu} - \sum_{i,j=1}^{3} \left(G^{-1}\right)_{ij}\, p_i^\mu p_j^\nu\,,
\end{equation}
which is well defined in $d$ dimensions, and satisfies
\begin{equation}
    p_{i,\mu}\,T_A^{\mu\nu} = 0 \quad (i=1,2,3)\,, \qquad T_{A\,\mu}^{\ \ \mu} = d-3\,, \qquad T_A^{\mu\alpha}T_{A\,\alpha}^{\ \ \nu} = T_A^{\mu\nu}\,.
\end{equation}
Since $\det G=\Delta$, see \cref{eq:gramdet}, the entries of $G^{-1}$, and hence $T_A^{\mu\nu}$, are proportional to $1/\Delta$, which is the origin of the $1/\Delta$ factors in the tensors introduced below. Explicitly,
\begin{align}
    T_A^{\mu\nu} = g^{\mu\nu} &+ \frac{(s+t-m_t^2)^2}{4\Delta}\,p_1^\mu p_1^\nu + \frac{(m_t^2-t)^2}{4\Delta}\,p_2^\mu p_2^\nu + \frac{s^2}{4\Delta}\,p_3^\mu p_3^\nu \nn\\
    &+ \frac{m_t^4+m_t^2 s-2m_t^2 t+s t+t^2}{4\Delta}\left(p_1^\mu p_2^\nu+p_2^\mu p_1^\nu\right) \nn\\
    &- \frac{s\,(s+t-m_t^2)}{4\Delta}\left(p_1^\mu p_3^\nu+p_3^\mu p_1^\nu\right)
    - \frac{s\,(m_t^2-t)}{4\Delta}\left(p_2^\mu p_3^\nu+p_3^\mu p_2^\nu\right)\,.
 \end{align}
As discussed below, we will use this relation differently in the two partonic channels. 

\subsection{The gluon scattering channel}
\label{sec:ggchannel}
The gluon-gluon partonic channel is defined as
\begin{equation}
    g(p_1) + g(p_2) \to t(p_3) + \bar{t}(p_4)\,.
\end{equation}
We start by decomposing the corresponding scattering amplitude into independent Lorentz and Dirac tensor structures $T_i$ as
\begin{equation}
\label{eq:ggdecomp}
    \boldsymbol{\mathcal{A}}^{(gg)} = (4 \pi \alpha_s) \, \sum_{i=1}^{16} \mathbf{F}_i\, T_i\,,
\end{equation}
where we have pulled out an overall leading order coupling $g_s^2= 4 \pi \alpha_s$, and $\alpha_s$ is the strong coupling constant.
The form factors $\mathbf{F}_i$ admit a perturbative expansion in the strong coupling as
\begin{equation}
\label{eq:loopexpansion}
    \mathbf{F}_i = \sum_{\ell\ge 0}\left(\frac{\alpha_s}{4\pi}\right)^{\!\ell} \mathbf{F}_i^{(\ell)}\,.
\end{equation}

We first describe the colour structure of the scattering amplitude, and then move to a detailed discussion of the tensors $T_i$.
Each form factor is a vector in colour space, 
\begin{equation}
    \mathbf{F}^{(\ell)}_i = \sum_{k=1}^{3} F_i^{(\ell),[k]}\, \big|C_k^{(gg)}\big\rangle\,, \qquad
\end{equation}
and we present our results in the basis
\begin{equation}
\label{eq:ggcolour}
    \big|C_1^{(gg)}\big\rangle = \left(T^aT^b\right)_{i_3i_4}\,,\qquad
    \big|C_2^{(gg)}\big\rangle = \left(T^bT^a\right)_{i_3i_4}\,,\qquad
    \big|C_3^{(gg)}\big\rangle = \delta^{ab}\,\delta_{i_3i_4}\,.
\end{equation}
Here, $a,b$ are the colour indices in the adjoint representation of $SU(N_c)$,
while $i_{3,4}$ are in the fundamental one, and the $SU(N_c)$ generators $T^a_{ij}$ satisfy the normalisation
$\mathrm{Tr}(T^a T^b) = T_F \,\delta^{ab}$ with $T_F = 1/2$.
An alternative basis resolves the amplitude into irreducible
$s$-channel colour representations
\begin{equation}
\label{eq:octetsinglet}
    \big|\bar{C}_1^{(gg)}\big\rangle = i f^{abc}\,T^{c}_{i_3i_4}\,,\qquad
    \big|\bar{C}_2^{(gg)}\big\rangle = d^{abc}\,T^{c}_{i_3i_4}\,,\qquad
    \big|\bar{C}_3^{(gg)}\big\rangle = \delta^{ab}\,\delta_{i_3i_4}\,,
\end{equation}
where the first two tensors describe adjoint exchange and are respectively
antisymmetric and symmetric under interchange of the two gluon
colour indices, while the third describes singlet exchange.
For $N_c=3$, the adjoint representation is the octet.
 Denoting the form factors in this basis by barred quantities, one has
\begin{gather}
    \bar{F}_i^{(\ell),[1]} =
    \frac{1}{2}\left(F_i^{(\ell),[1]} - F_i^{(\ell),[2]}\right),\quad\quad
    \bar{F}_i^{(\ell),[2]} =
    \frac{1}{2}\left(F_i^{(\ell),[1]} + F_i^{(\ell),[2]}\right),
    \label{eq:octettotrace}\\
    \bar{F}_i^{(\ell),[3]} =
    \frac{1}{2N_c}\left(F_i^{(\ell),[1]} + F_i^{(\ell),[2]}\right)
    + F_i^{(\ell),[3]}.
    \notag
\end{gather}

We now turn to the discussion of the spin structure of the amplitude.
In the tHV scheme there is a total of 16 tensors, matching the number of different helicity amplitudes. 
First, we impose the following orthogonality conditions on the gluon polarisation vectors
\begin{equation}
\label{eq:orthogonality}
\epsilon(p_{1}) \cdot p_{1}=
\epsilon(p_{1}) \cdot p_{2}=0\,,
\quad\quad
\epsilon(p_{2}) \cdot p_{2}=
\epsilon(p_{2}) \cdot p_{1}=0\,.
\end{equation}
Second, we exploit the fact that for a $2 \to 2$ process, 
the tensors can be neatly organised into two subsets, the first parity-even and the second parity odd.
The parity-odd tensors fully decouple from the parity even ones and do not contribute to the $gg\to t\bar{t}$ process, so we ignore them from now on, while the parity even ones read
\begin{align}
T_{1} &=
\frac{s^{2}}{4m_t\Delta}\,
\bar{u}(p_{3})v(p_{4})\,
\epsilon^{*}_{\mu}(p_{1})\epsilon^{*}_{\nu}(p_{2})\,
p_{3}^{\mu}p_{3}^{\nu},
&\;\;\;
T_{5} &=
\frac{m_t^{2}s}{4\Delta}\,
\bar{u}(p_{3})\slashed{a}v(p_{4})\,
\epsilon^{*}_{\mu}(p_{1})\epsilon^{*}_{\nu}(p_{2})\,
p_{3}^{\mu}a^{\nu},
\nn\\
T_{2} &=
\frac{s^{2}}{4m_t^{2}\Delta}\,
\bar{u}(p_{3})\slashed{p}_1v(p_{4})\,
\epsilon^{*}_{\mu}(p_{1})\epsilon^{*}_{\nu}(p_{2})\,
p_{3}^{\mu}p_{3}^{\nu},
&\;\;\;
T_{6} &=
\frac{m_t s}{4\Delta}\,
\bar{u}(p_{3})\slashed{p}_1\slashed{a}v(p_{4})\,
\epsilon^{*}_{\mu}(p_{1})\epsilon^{*}_{\nu}(p_{2})\,
p_{3}^{\mu}a^{\nu},
\nn\\
T_{3} &=
\frac{1}{m_t}\,
\bar{u}(p_{3})v(p_{4})\,
\epsilon^{*}_{\mu}(p_{1})\epsilon^{*}_{\nu}(p_{2})\,
a^{\mu}a^{\nu},
&\;\;\;
T_{7} &=
\frac{m_t^{2}s}{4\Delta}\,
\bar{u}(p_{3})\slashed{a}v(p_{4})\,
\epsilon^{*}_{\mu}(p_{1})\epsilon^{*}_{\nu}(p_{2})\,
a^{\mu}p_{3}^{\nu},
\nn\\
T_{4} &=
\frac{1}{m_t^{2}}\,
\bar{u}(p_{3})\slashed{p}_1v(p_{4})\,
\epsilon^{*}_{\mu}(p_{1})\epsilon^{*}_{\nu}(p_{2})\,
a^{\mu}a^{\nu},
&\;\;\;
T_{8} &=
\frac{m_t s}{4\Delta}\,
\bar{u}(p_{3})\slashed{p}_1\slashed{a}v(p_{4})\,
\epsilon^{*}_{\mu}(p_{1})\epsilon^{*}_{\nu}(p_{2})\,
a^{\mu}p_{3}^{\nu}\,.
\label{eq:Teven}
\end{align}
We notice that in $T_{1,\ldots,8}$, the vector $a^\mu$ appears only in pairs, and every product $a^\mu a^\nu$ should be understood as the transverse projector $T_A^{\mu\nu}$ of \cref{eq:transverseproj}.

Adopting the conventions for massive and massless spinors spelt out in \cref{app:spinorconventions}, we can specialise the tensors above 
to the independent spin configurations of the top quarks and the gluons and, in this way, 
define spin-polarised amplitudes.
All polarised amplitudes are defined in the all-outgoing notation
$0\to g\,g\,t\,\bar{t}$. This choice explains the appearance of the conjugate vectors $\epsilon^*$ in the tensors $T_i$ of \cref{eq:Teven} also for initial-state gluons.
Consistently with the choice in~\cref{eq:orthogonality}, we choose the polarisations of the gluons as 
\begin{align}
    \epsilon^{*\mu}_{+}(p_1) &= \frac{1}{\sqrt{2}}\frac{\langle 2 \gamma^\mu 1]}{\langle 2 1\rangle}\,, &
    \epsilon^{*\mu}_{-}(p_1) &= -\frac{1}{\sqrt{2}}\frac{[2\gamma^\mu 1\rangle}{[2 1]}\,,
    \nn\\
    \epsilon^{*\nu}_{+}(p_2) &= \frac{1}{\sqrt{2}}\frac{\langle 1 \gamma^\nu 2]}{\langle 1 2\rangle}\,, &
    \epsilon^{*\nu}_{-}(p_2) &= -\frac{1}{\sqrt{2}}\frac{[1\gamma^\nu 2\rangle}{[1 2]}\,.
\end{align}

Let us start considering the case of two equal-helicity gluons, which we choose to be
positive in the outgoing convention. We then get
\begin{equation}
    \epsilon^{*\mu}_{+}(p_1)  \epsilon^{*\nu}_{+}(p_2) = -\frac{1}{2\langle12\rangle^2}\mathrm{Tr}\left(P_L \slashed{p}_2 \gamma^\mu\slashed{p}_1 \gamma^\nu\right)\,.
\end{equation}
After some algebra, one can easily show that for generic top-quark polarisations the amplitude can be organised as
\begin{align}
    \boldsymbol{\mathcal{A}}_{++}^{(gg)} = \Phi_{++}&\left[\left(\frac{\mathbf{F}_3-\mathbf{F}_1}{2}\right)\frac{\bar{u}(p_3)v(p_4)}{m_t}
    + \left(\frac{\mathbf{F}_7-\mathbf{F}_5}{2}\right)\frac{m_t^2}{2\sqrt{\Delta}}\,\bar{u}(p_3)\slashed{a}v(p_4)
    \right.\nn\\
    &+\left.\left(\frac{\mathbf{F}_4-\mathbf{F}_2}{2}\right)\frac{1}{m_t^2}\,\bar{u}(p_3)\slashed{p}_1 v(p_4)
    +\left(\frac{\mathbf{F}_8-\mathbf{F}_6}{2}\right)\frac{m_t}{2\sqrt{\Delta}}\,\bar{u}(p_3)\slashed{p}_1 \slashed{a} v(p_4) \right]
    \label{eq:Appgg}
\end{align}
where the overall phase reads
\begin{equation}
\Phi_{++} = \frac{s}{\langle12\rangle^2}\,.
\end{equation}
For gluons with opposite helicities, one finds instead
\begin{align}
    \boldsymbol{\mathcal{A}}_{+-}^{(gg)} = \Phi_{+-}&\left[\left(\frac{\mathbf{F}_1+\mathbf{F}_3}{2}\right)\frac{\bar{u}(p_3)v(p_4)}{m_t}
    - \left(\frac{\mathbf{F}_5+\mathbf{F}_7}{2}\right)\frac{m_t^2}{2\sqrt{\Delta}}\,\bar{u}(p_3)\slashed{a}v(p_4)
    \right.\nn\\
    &+\left.\left(\frac{\mathbf{F}_2+\mathbf{F}_4}{2}\right)\frac{1}{m_t^2}\,\bar{u}(p_3)\slashed{p}_1 v(p_4)
    -\left(\frac{\mathbf{F}_6+\mathbf{F}_8}{2}\right)\frac{m_t}{2\sqrt{\Delta}}\,\bar{u}(p_3)\slashed{p}_1 \slashed{a} v(p_4) \right],
    \label{eq:Apmgg}
\end{align}
with the phase
\begin{equation}
\Phi_{+-} = -\frac{4\Delta}{s\,\langle1|\slashed{p}_3|2]^2} = -\frac{\langle2|\slashed{p}_3|1]}{\langle1|\slashed{p}_3|2]}\,,
\end{equation}
which, like $\Phi_{++}$, has unit modulus, since $\langle1|\slashed{p}_3|2]\langle2|\slashed{p}_3|1]=4\Delta/s$. The amplitudes $\boldsymbol{\mathcal{A}}_{--}^{(gg)}$ and $\boldsymbol{\mathcal{A}}_{-+}^{(gg)}$ can be obtained from $\boldsymbol{\mathcal{A}}_{++}^{(gg)}$ and $\boldsymbol{\mathcal{A}}_{+-}^{(gg)}$ via a parity transformation, which involves $a^\mu \to - a^\mu$. 
Note that the presence of the parity odd vector $a^\mu$ in~\cref{eq:Appgg,eq:Apmgg}
reflects the fact that, after fixing the helicities of the incoming gluons, 
the resulting amplitudes do not have to be parity invariant.
The explicit expressions for the complete set of polarised amplitudes are provided in \cref{app:helcoeff}.

As expected, the $gg$ amplitude is determined by the eight form factors $\mathbf{F}_{1,\ldots,8}$ of the parity-even tensors, which are however not all independent. In particular, using Bose symmetry and charge 
conjugation invariance of the amplitude one can prove that
\begin{equation}
\label{eq:FFlindep}
    \mathbf{F}_7 = \mathbf{F}_5 + 2\mathbf{F}_6\,, \qquad \mathbf{F}_8 = -\mathbf{F}_6\,,
\end{equation}
for every colour component, at all loop orders.
As a consequence, only six independent combinations of form factors enter the polarised amplitudes. Inserting \cref{eq:FFlindep} in \cref{eq:Appgg,eq:Apmgg}, $\boldsymbol{\mathcal{A}}_{++}^{(gg)}$ depends on $\mathbf{F}_3-\mathbf{F}_1$, $\mathbf{F}_4-\mathbf{F}_2$ and $\mathbf{F}_6$, and $\boldsymbol{\mathcal{A}}_{+-}^{(gg)}$ on $\mathbf{F}_1+\mathbf{F}_3$, $\mathbf{F}_2+\mathbf{F}_4$ and $\mathbf{F}_5+\mathbf{F}_6$, while the term with $\bar{u}(p_3)\slashed{p}_1\slashed{a}v(p_4)$ drops out of $\boldsymbol{\mathcal{A}}_{+-}^{(gg)}$. At tree level, one further has
\begin{equation}
    \mathbf{F}_3^{(0)} = 0\,, \qquad \mathbf{F}_4^{(0)} = \mathbf{F}_2^{(0)}\,.
\end{equation}

The scalar form factors, or the combinations thereof defining the polarised amplitudes, 
can be extracted from the Feynman diagram
decomposition of the corresponding amplitudes by suitably defined projector operators.
To define them, we start by 
recalling that the form of the polarisation sums for the external gluons, consistent with the gauge choice of \cref{eq:orthogonality}, reads
\begin{align}
    &\sum_{\rm pol}\epsilon_\mu(p_1)\epsilon^*_\nu(p_1) = -g_{\mu\nu} + \frac{p_{1,\mu}p_{2,\nu}+p_{2,\mu}p_{1,\nu}}{p_1\cdot p_2}\,, \\
    &\sum_{\rm pol}\epsilon_\mu(p_2)\epsilon^*_\nu(p_2) = -g_{\mu\nu} + \frac{p_{1,\mu}p_{2,\nu}+p_{2,\mu}p_{1,\nu}}{p_1\cdot p_2}\,.
\end{align}
Similarly, the sum over spins of the top quarks with 
\begin{align}
&\sum u(p_3)\bar{u}(p_3)=\slashed{p}_3+m_t\,, 
\qquad \sum v(p_4)\bar{v}(p_4)=\slashed{p}_4-m_t\,.
\label{eq:polmassiveq}
\end{align} 

We can then employ the polarisation sums to define a scalar product
operation among the tensors and their duals as
$$\langle T_j, T_i \rangle = \sum_{\rm pol,\,spins} T_i\, T_j^\dagger $$
and with this we can compute their ``Gram matrix''
\begin{equation}
\label{eq:gramtensors}
    \mathcal{G}_{ij} = \sum_{\rm pol,\,spins} T_i\, T_j^\dagger\,, \qquad i,j = 1,\ldots,8\,.
\end{equation}
which is real and symmetric and depends on $s$, $t$, $m_t^2$ and $d$. 
The form factors are then given by
\begin{equation}
\label{eq:projectors}
    \mathbf{F}_i = \sum_{j=1}^{8}(\mathcal{G}^{-1})_{ij} \sum_{\rm pol,\,spins} T_j^\dagger\,\boldsymbol{\mathcal{A}} \ \equiv \sum_{\rm pol,\,spins}\mathcal{P}_i^\dagger\,\boldsymbol{\mathcal{A}}\,,
\end{equation}
which defines the projectors $\mathcal{P}_i$. Since they act only on the Lorentz and spin indices, the colour structure of $\boldsymbol{\mathcal{A}}$ is inherited by the form factors $\mathbf{F}_i$. We note that $\mathcal{G}^{-1}$ is block diagonal, with $2\times2$ blocks acting on the pairs $\{1,2\}$, $\{3,4\}$, $\{5,6\}$ and $\{7,8\}$. 
With $\Delta$ as in \cref{eq:gramdet} and $\kappa\equiv(m_t^2-t)(m_t^2-u)$, the non-vanishing blocks are
\begin{align}
    \mathcal{G}^{(-1)}_{\{1,2\}} &= (d-3)\,\mathcal{G}^{(-1)}_{\{3,4\}} = \frac{m_t^2}{8\Delta}
    \begin{pmatrix} \kappa & -m_t^2\,(t-u)\\[0.2em] -m_t^2\,(t-u) & m_t^2\,(s-4m_t^2)\end{pmatrix},\nn\\[0.4em]
    \mathcal{G}^{(-1)}_{\{5,6\}} &= \mathcal{G}^{(-1)}_{\{7,8\}} = \frac{1}{2(d-3)\,m_t^2}
    \begin{pmatrix} \kappa/m_t^2 & -s\\[0.2em] -s & s\end{pmatrix}.
    \label{eq:Mgg}
\end{align}

\subsection{The quark scattering channel}
\label{sec:qqchannel}

We consider now the quark-antiquark annihilation channel
\begin{equation}
    q(p_1) + \bar{q}(p_2) \to t(p_3) + \bar{t}(p_4)\,.
\end{equation}
At variance with the gluon channel, the amplitude is now decomposed in terms of two colour-basis
elements and four form factors $\mathbf{G}_i$ multiplying the Dirac structures $\mathcal{T}_i$
\begin{align}
\label{eq:qqbtensordec}
    \boldsymbol{\mathcal{A}}^{q\bar{q}} = (4 \pi \alpha_s)\sum_{i=1}^4 \mathbf{G}_i \mathcal{T}_i\,.
\end{align}
The form factors $\mathbf{G}_i$ admit a perturbative expansion in $\alpha_s$ equivalent to \cref{eq:loopexpansion}.
As for the colour structure, we have
\begin{equation}
\label{eq:qqbtrace}
    \mathbf{G}^{(\ell)}_i = \sum_{k=1}^{2} G_i^{(\ell),[k]}\,\big|C_k^{(q\bar q)}\big\rangle\,,
\end{equation}
and we present our results in the basis
\begin{equation}
\label{eq:qqcolour}
    \big|C_1^{(q\bar q)}\big\rangle = \delta_{i_2 i_4}\,\delta_{i_3 i_1}\,,\qquad
    \big|C_2^{(q\bar q)}\big\rangle = \delta_{i_2 i_1}\,\delta_{i_3 i_4}\,,
\end{equation}
where $i_{1,2}$ are colour indices for the initial-state quarks.
Similarly to the $gg$ channel, one can define a colour basis in terms of irreducible representations of $SU(N_c)$ in the $s$-channel, explicitly,
\begin{equation}
\label{eq:octetsingqqb}
    \big|\bar{C}_1^{(q\bar q)}\big\rangle = T^{a}_{i_2 i_1}\,T^{a}_{i_3 i_4}\,,\qquad
    \big|\bar{C}_2^{(q\bar q)}\big\rangle = \delta_{i_2 i_1}\,\delta_{i_3 i_4}\,,
\end{equation}
where, for $N_c = 3$, the first one is the octet exchange, 
while the second one is the singlet exchange. 
The corresponding form factors in this basis, denoted with barred quantities, are related to the ones in
\cref{eq:qqbtrace} via
\begin{equation}
    \bar{G}_i^{(\ell),[1]} = 2 G_i^{(\ell),[1]}\,,\quad\quad
    \bar{G}_i^{(\ell),[2]} = \frac{G_i^{(\ell),[1]}}{N_c} + G_i^{(\ell),[2]}\,.
\end{equation}

The four Dirac structures of \cref{eq:qqbtensordec} are chosen as
\begin{align}
    \mathcal{T}_1 &= \frac{m_t\,s}{4\Delta}\left[\bar{v}(p_2)\slashed{p}_3u(p_1)\right]\left[\bar{u}(p_3)v(p_4)\right]\,, &
    \mathcal{T}_3 &= \frac{m_t^2\,s}{4\Delta}\left[\bar{v}(p_2)\gamma^\mu u(p_1)\right]\left[\bar{u}(p_3)\gamma_\mu v(p_4)\right]\,, \nn\\
    \mathcal{T}_2 &= \frac{s}{4\Delta}\left[\bar{v}(p_2)\slashed{p}_3u(p_1)\right]\left[\bar{u}(p_3)\slashed{p}_1v(p_4)\right]\,, &
    \mathcal{T}_4 &= \frac{m_t\,s}{4\Delta}\left[\bar{v}(p_2)\gamma^\mu u(p_1)\right]\left[\bar{u}(p_3)\gamma_\mu\slashed{p}_1v(p_4)\right]\,,
    \label{eq:qqtensors}
\end{align}
where $u(p_1)$ and $\bar{v}(p_2)$ are the spinors of the incoming massless quark and antiquark and $\bar{u}(p_3)$ and $v(p_4)$ those of the outgoing top and antitop. 
We choose all tensors to be dimensionless, and in analogy with the $gg$ channel, the normalisation factors guarantee that the 
form factors contain no spurious singularities at tree-level for $\Delta=0$. 
As for $gg$, we neglected all parity-odd structures, which can be obtained by inserting the vector $a^\mu$ of \cref{eq:avector} as in $\bar{v}(p_2)\slashed{a}u(p_1)$. 
We notice also that, at tree level, only $\mathcal{T}_3$ contributes. 

Also in this case, not all four form factors contribute to the physical scattering amplitude. Indeed, invoking once again charge conjugation invariance it is possible to show that 
\begin{equation}
    \label{eq:vanishingffqqb}
\mathbf{G}_4 = 0\,.
\end{equation}
Furthermore, at tree-level the only non vanishing form factor is $\mathbf{G}^{(0)}_3$\,.

As done in the $gg$ case, one can easily obtain spin-polarised amplitudes for definite helicities of the massless quarks and spin states of the top quarks. We start by fixing the polarisations of the massless quarks using the standard spinor-helicity formalism. Since the vector current 
$$J_{\lambda_1 \lambda_2}^\mu = \bar{v}_{\lambda_2}(p_2)\gamma^\mu u_{\lambda_1}(p_1)$$ conserves chirality,
only two helicity assignments of the massless quarks contribute, i.e.
$$ J_{-+}^\mu 
= \langle2|\gamma^\mu|1]\,, \qquad 
J_{+-}^\mu
= [2|\gamma^\mu|1\rangle=(J_{-+}^\mu)^*\,.$$ 
With these conventions, the amplitude for generic top-quark spinors follows from \cref{eq:qqtensors},
\begin{equation}
    \boldsymbol{\mathcal{A}}_{\lambda\bar\lambda}^{q\bar{q}} = \frac{s}{4\Delta}\Big\{ (J_{\lambda\bar\lambda}\!\cdot p_3)\left[m_t \mathbf{G}_1\,\bar{u}(p_3)v(p_4) + \mathbf{G}_2\,\bar{u}(p_3)\slashed{p}_1v(p_4)\right]
    + m_t^2 \mathbf{G}_3 \,\bar{u}(p_3)\,\slashed{J}_{\lambda\bar\lambda} 
    v(p_4)\Big\}\,,
    \label{eq:Aqq}
\end{equation}
with $J_{\lambda\bar\lambda}\!\cdot p_3=\langle2|\slashed{p}_3|1]$ or $[2|\slashed{p}_3|1\rangle$. 
Decomposing the current for $(\lambda,\bar\lambda)=(\mp,\pm)$ on the vectors transverse to $p_1,p_2$ as
$$J_{\lambda\bar\lambda}^\mu=-\frac{(J_{\lambda\bar\lambda}\!\cdot p_3)\,p_{3\perp}^\mu}{|p_{3\perp}|^2}+(J_{\lambda\bar\lambda}\!\cdot a)\,a^\mu ,\qquad 
J_{\lambda\bar\lambda}\!\cdot a=\mp J_{\lambda\bar\lambda}\!\cdot p_3\,\frac{s}{(2\sqrt{\Delta})}\,,$$ 
where we used 
$|p_{3\perp}|= 2\sqrt{\Delta}/s $,
 and using the Dirac equation to remove $\slashed{p}_2$ and $\slashed{p}_3$, one arrives at the same structures found also in the gluon channel,
\begin{align}
    \boldsymbol{\mathcal{A}}_{\mp\pm}^{q\bar{q}} = \Phi_{\mp\pm}\,\frac{\sqrt{s}}{2\sqrt{\Delta}}\Bigg\{
    &\left[m_t^2\mathbf{G}_1-\frac{m_t^4s^2}{4\Delta}\,\mathbf{G}_3\right]\frac{\bar{u}(p_3)v(p_4)}{m_t}
\mp s\,\mathbf{G}_3\frac{m_t^2}{2\sqrt{\Delta}}\,\bar{u}(p_3)\slashed{a}v(p_4)
    \nn\\
    &+\left[m_t^2\mathbf{G}_2-\frac{m_t^4\,s\,(u-t)}{4\Delta}\,\mathbf{G}_3\right]\frac{1}{m_t^2}\bar{u}(p_3)\slashed{p}_1v(p_4)\Bigg\}\,,
    \label{eq:Aqqred}
\end{align}
with the phases defined as
\begin{equation}
    \Phi_{-+}=\frac{\langle2|\slashed{p}_3|1]}{|\langle2|\slashed{p}_3|1]|}\,,\qquad
    \Phi_{+-}=\frac{[2|\slashed{p}_3|1\rangle}{|[2|\slashed{p}_3|1\rangle|}=\Phi_{-+}^*\,.
\end{equation}

As for the gluon case, parity relates the two initial-helicity configurations when the final-state spins and momenta are also transformed. In the expressions above, with momenta and top-quark spinors held fixed, switching between the two initial-helicity configurations amounts to reversing the sign of the term proportional 
to $\slashed a$ and complex conjugating the overall phase, while leaving the form factors unchanged.
As for the gluons,
the complete set of polarised amplitudes 
is provided in \cref{app:helcoeff}.

We construct the projector operators to extract the form factors in a similar way as for the gluon case.
In particular, we use the same conventions for the polarisation sums of the massive quarks, see~\cref{eq:polmassiveq} and
in addition
\begin{equation}
\sum u(p_1)\bar{u}(p_1)=\slashed{p}_1\,, \qquad \sum v(p_2)\bar{v}(p_2)=\slashed{p}_2 
\label{eq:polmasslessq}
\end{equation}
 for the massless quarks.

With this, we can write
\begin{equation}
    \mathcal{G}_{ij}=\sum_{\rm spins}\mathcal{T}_i\,\mathcal{T}_j^\dagger\,, \qquad \mathbf{G}_i=\sum_{j=1}^{4}(\mathcal{G}^{(-1)})_{ij}\sum_{\rm spins}\mathcal{T}_j^\dagger\,\boldsymbol{\mathcal{A}}\,,
\end{equation}
with the tensors $\mathcal{T}_i$ of \cref{eq:qqtensors}. The matrix $\mathcal{G}^{(-1)}$ is real and symmetric, with $\mathcal{G}^{(-1)}_{13}=\mathcal{G}^{(-1)}_{24}=0$ and the other independent entries given by
\begin{align}
    \mathcal{G}^{(-1)}_{11} &= \frac{(d-2)\,\kappa}{4(d-3)\,m_t^2\,s}\,, &
    \mathcal{G}^{(-1)}_{12} &= -\frac{(d-2)(t-u)}{4(d-3)\,s}\,, \nn\\
    \mathcal{G}^{(-1)}_{14} &= -\frac{\Delta}{(d-3)\,m_t^2s^2}\,, &
    \mathcal{G}^{(-1)}_{22} &= \frac{(d-2)(s-4m_t^2)}{4(d-3)\,s}-\frac{4\Delta}{(d-3)\,s^3}\,, \nn\\
    \mathcal{G}^{(-1)}_{23} &= -\frac{(t-u)\,\Delta}{(d-3)\,m_t^2s^3}\,, &
    \mathcal{G}^{(-1)}_{33} &= \frac{\kappa\,\Delta}{(d-3)\,m_t^4s^3}\,,  \nn\\
    \mathcal{G}^{(-1)}_{34} &= 
    \mathcal{G}^{(-1)}_{44} = \frac{\Delta}{(d-3)\,m_t^2s^2}\,. &&&&
    \label{eq:Mqq}
\end{align}


%
\section{Master integrals and canonical basis}
\label{sec:integrals}

\subsection{Amplitude computation and integral families}
\label{sec:ampcomp}
We compute the complete two-loop QCD corrections to the partonic processes defined in~\cref{sec:setup}, without restrictions on the colour or flavour structure and retaining full dependence on the top-quark mass. In particular, we keep the full dependence on the number of colours $N_c$, accounting for the entire tower of sub-leading colour contributions. Moreover, we include all the contributions of the $n_l$ massless quark flavours, which can appear in closed loops, and of the closed top-quark loops, with $n_h=1$ heavy flavours.

We generate all contributing Feynman diagrams with \texttt{qgraf}~\cite{Nogueira:1991ex}. 
The number of diagrams is reported in \cref{tab:ndiagrams}. 
We perform the colour and Dirac algebra, the insertion of the Feynman rules, and the application of the projectors of  in $d$ dimensions with \texttt{FORM}~\cite{Vermaseren:2000nd, Kuipers:2012rf, Ruijl:2017dtg, Davies:2026cci}. 
Finally, we match each diagram by a suitable shift of the loop momenta onto the integral families defined in \cref{app:families} using \texttt{Reduze~2}~\cite{Studerus:2009ye, vonManteuffel:2012np}.
\begin{table}[ht!]
\centering
\begin{tabular}{lccc}
\toprule
Process & tree level & one loop & two loops \\
\midrule
$gg\to t\bar{t}$ & 3 & 34 & 789 \\
$q\bar{q}\to t\bar{t}$ & 1 & 11 & 218 \\
\bottomrule
\end{tabular}
\caption{Number of Feynman diagrams generated with \texttt{qgraf} for the two partonic channels, at tree level, one loop and two loops.}
\label{tab:ndiagrams}
\end{table}

After applying the projectors, all form factors are expressed as linear combinations of scalar Feynman integrals of the form
\begin{equation}
    \int \prod_{i=1}^2\left(\frac{\mathrm{d}^dk_i}{i\pi^{d/2}} e^{\varepsilon\gamma_E}\right)\,
    \frac{1}{D_1^{a_1}\cdots D_9^{a_9}}\,, \qquad D_i = q_i^2 - m_i^2\,,
\end{equation}
with $a_i\in\mathbb{Z}$.
We group them into $14$ two-loop integral families with nine propagators each: five planar families \texttt{pA}--\texttt{pE} and nine non-planar ones \texttt{nA}--\texttt{nI}. Their propagators are collected in \cref{app:families} and we indicate the loop momenta by $k_1,k_2$. Families that are related by the exchange $p_1\leftrightarrow p_2$, \ie, by $t\leftrightarrow u$, are treated as crossings of the same family, which we denote by the suffix \texttt{x12}, following the notation introduced in \texttt{Reduze~2}~\cite{Studerus:2009ye, vonManteuffel:2012np}.
Before the reduction, we use \texttt{Reduze~2} to exploit the shift symmetries of the integral families, which relate integrals within a family and across different families. In this way, the scalar integrals are reduced to a minimal subset. We then reduce this subset to a basis of master integrals by means of integration-by-parts identities (IBPs)~\cite{Tkachov:1981wb, Chetyrkin:1981qh}, which we solve with the Laporta algorithm~\cite{Laporta:2000dsw}. The final IBP reduction is performed with \texttt{Kira~3}~\cite{Maierhofer:2017gsa, Klappert:2020nbg,Lange:2025fba}, which relies on \texttt{FireFly}~\cite{Klappert:2019emp, Klappert:2020aqs} for the reconstruction of the rational functions from finite-field samples.

The reduction leads to a basis of $417$ master integrals, distributed among the families as in \cref{tab:mastercounts}. We stress that this basis includes all the master integrals of the crossed families with $p_1\leftrightarrow p_2$ (\texttt{x12}), that are needed to cover the full physical amplitude: no further integral has to be computed to obtain the amplitudes for both channels 
in the whole physical region. 

\subsection{Canonical basis}
\label{sec:canonical}
We compute the master integrals by means of differential equations~\cite{Kotikov:1990kg, Remiddi:1997ny, Gehrmann:1999as}. In particular, we exploit recent developments in the theory of Feynman integrals 
beyond polylogarithms~\cite{Gorges:2023zgv,Duhr:2024uid,Duhr:2025lbz,Forner:2026vby} to 
find a set of master integrals such that their differential equations are in canonical form~\cite{Henn:2013pwa} 
\begin{equation}
\label{eq:canonicalDE}
  \mathrm{d}\vec{J}=\varepsilon\,{A}\,\vec{J}\,.
\end{equation}
Importantly, when integrals are chosen in this way, 
the dependence on the dimensional regulator $\varepsilon$ is factorised from the connection matrix 
${A}$ and its entries are linearly independent differential 
forms with at most simple poles.
For the sectors that evaluate to multiple polylogarithms (MPLs), 
these are $\mathrm{d}\!\log$-forms only. 
We recall here that for many of the planar integrals addressed in this calculation,
a polylogarithmic canonical form was already known in the literature, see \cite{Adams:2018kez,Becchetti:2019tjy,Mandal:2022vju}.

As a starting point, we perform for most sectors a full integrand analysis in a suitable
parametrisation.
For this, we often choose the Baikov representation~\cite{Baikov:1996iu} and, when convenient,
we parametrise the loops iteratively following the so-called loop-by-loop approach~\cite{Frellesvig:2017aai,Frellesvig:2024ymq}.
We use this analysis to first identify all integrals whose integrands are 
in $\mathrm{d}\!\log$-form and which are
multiplied by one single leading singularity. Following~\cite{Henn:2013pwa,Henn:2020lye}, 
we normalise these candidates by their leading singularity, obtaining so-called unit leading singularity integrals, which are expected to fulfil canonical differential equations.
Already in the polylogarithmic sectors, this normalisation step forces us to 
introduce ten algebraic roots of the kinematic invariants, related to the various leading singularities:
\begin{equation}
\begin{array}{@{}r@{}l@{\quad}r@{}l@{\quad}r@{}l@{\quad}r@{}l@{}}
  r_1 & {}=\sqrt{s}\,,
  & r_2 & {}=\sqrt{s-4m_t^2}\,,
  & r_3 & {}=\sqrt{s+4m_t^2}\,,
  & r_4 & {}=\sqrt{m_t^2-t}\,,
  \\[0.6ex]
  r_5 & {}=\sqrt{s+t-m_t^2}\,,
  & r_6 & {}=\sqrt{-3m_t^2-t}\,,
  & r_7 & {}=\sqrt{s+t-5m_t^2}\,,
  & &
  \\[0.6ex]
  r_8 & \multicolumn{7}{@{}l@{}}{
    {}=\sqrt{-m_t^6+6m_t^4s-m_t^2s^2
             +m_t^4t-2m_t^2st+s^2t}\,,
  }
  \\[0.6ex]
r_9 & \multicolumn{7}{@{}l@{}}{
    {}=\sqrt{m_t^6+m_t^4s+3m_t^2s^2-s^3
             -m_t^4t+2m_t^2st-s^2t}\,,
    \hspace{1.7cm} r_{10}=\sqrt{m_t^2}\,.
  }
\end{array}
\end{equation}
In sectors where this strategy does not provide a complete set of independent canonical integrals, we complete the basis with derivatives of canonical integrals and determine the rotation to a canonical basis from the differential equations by hand. 
Using derivatives of canonical integrals guarantees that 
the required rotation is $\varepsilon$-independent, see also~\cite{Duhr:2025lbz,Forner:2026zhk,Forner:2026vby}.
In this way, we obtain a canonical basis for all polylogarithmic sectors.

The sectors that do not admit integrands in $\mathrm{d}\!\log$-forms are associated with non-trivial geometries, 
and their treatment requires a generalisation
of the method described above. 
We employ the approach presented in~\cite{Gorges:2023zgv,Duhr:2024uid,Duhr:2025lbz,Forner:2026vby},
which starts with the analysis of the maximal cuts of the various graphs to identify the respective irreducible geometries~\cite{Primo:2016ebd,Primo:2017ipr}. In doing so, 
we find 13 elliptic curves and one curve of genus two.
We denote the elliptic curves by $E_1,\ldots,E_8$ and $E_1^{\rm x12}$, $E_2^{\rm x12}$, $E_3^{\rm x12}$, $E_7^{\rm x12}$ and $E_8^{\rm x12}$, where the last five are associated with the corresponding sectors of the crossed families obtained by exchanging $p_1\leftrightarrow p_2$, i.e.\ $t\leftrightarrow u$.
The curve of genus 2 turns out to be associated with the non-planar sector 
\texttt{nB463} (see~\cref{tab:nonplanarfamilies1} for a definition of the corresponding integral family). 
Here, the number is the sector identifier, $\sum_i 2^{i-1}$ with the sum running over the propagators with positive powers.
For each of the curves $E_1,\ldots,E_8$ and the genus-two curve, \cref{tab:geometries} shows the graphs of all sectors whose homogeneous blocks contain it, together with the number of master integrals in the homogeneous block. The graphs of the sectors containing the crossed curves $E_i^{\rm x12}$ follow by swapping $p_1\leftrightarrow p_2$.
We stress that a curve is not confined to the sector in which it appears: $E_7$ and $E_8$, for instance, also enter the inhomogeneous blocks of several higher sectors, including non-planar ones, and $E_8$ and $E_8^{\rm x12}$ even enter the inhomogeneous block of the genus-two sector \texttt{nB463}.
%
\tikzset{
  gmass/.style={line width=1.0pt},
  gglue/.style={decorate,decoration={coil,aspect=0.55,segment length=3pt,amplitude=1.2pt},line width=0.4pt},
  lab/.style={fill=white,inner sep=0.4pt,font=\tiny}
}
\begin{table}[htbp]
\centering
\renewcommand{\arraystretch}{1.0}
\setlength{\tabcolsep}{4pt}
\small
\newcommand{\gscale}{0.24}
\newcommand{\sg}[4]{\begin{tabular}[c]{@{}c@{}}#1\\[-1pt]{\tiny #2}\\[-2pt]{\tiny #3\,(#4)}\end{tabular}}
\begin{tabular}{lc@{\hspace{14pt}}lc@{\hspace{14pt}}lc}
\toprule
Curve & Sectors & Curve & Sectors & Curve & Sectors \\
\midrule
\begin{tabular}[c]{@{}l@{}}$E_{1}$\\{\scriptsize elliptic}\end{tabular} &
\sg{%
\begin{tikzpicture}[scale=\gscale,baseline=(current bounding box.center)]
\draw[gmass,bend left=55] (0,0) to node[pos=0.5,lab] {3} (3,0);
\draw[gmass] (0,0) to node[pos=0.5,lab] {5} (3,0);
\draw[gmass,bend right=55] (0,0) to node[pos=0.5,lab] {8} (3,0);
\draw[gglue] (0,0) -- (-0.848528,0.848528);
\node[font=\tiny] at (-1.09602,1.09602) {$p_1$};
\draw[gmass,-{Stealth[length=4pt]}] (0,0) -- (-0.848528,-0.848528);
\node[font=\tiny] at (-1.09602,-1.09602) {$p_3$};
\draw[gglue] (3,0) -- (3.70711,0.707107);
\node[font=\tiny] at (3.95459,0.954594) {$p_2$};
\draw[gmass,-{Stealth[length=4pt]}] (3,0) -- (3.84853,-0.848528);
\node[font=\tiny] at (4.09602,-1.09602) {$p_4$};
\fill (0,0) circle (2.2pt);
\fill (3,0) circle (2.2pt);
\end{tikzpicture}}{\texttt{pB148}}{2}{0} &
\begin{tabular}[c]{@{}l@{}}$E_{2}$\\{\scriptsize elliptic}\end{tabular} &
\sg{%
\begin{tikzpicture}[scale=\gscale,baseline=(current bounding box.center)]
\draw[gglue] (0,1.6) to node[pos=0.5,lab] {1} (3,1.6);
\draw[gmass,bend left=35] (3,1.6) to node[pos=0.5,lab] {3} (3,-1.6);
\draw[gmass,bend right=35] (3,1.6) to node[pos=0.5,lab] {5} (3,-1.6);
\draw[gglue] (3,-1.6) to node[pos=0.5,lab] {6} (0,-1.6);
\draw[gmass] (0,1.6) to node[pos=0.5,lab] {8} (0,-1.6);
\draw[gmass,-{Stealth[length=4pt]}] (0,1.6) -- (-1.03923,2.2);
\node[font=\tiny] at (-1.34234,2.375) {$p_3$};
\draw[gglue] (3,1.6) -- (4.03923,2.2);
\node[font=\tiny] at (4.34234,2.375) {$p_1$};
\draw[gglue] (3,-1.6) -- (3.86603,-2.1);
\node[font=\tiny] at (4.16913,-2.275) {$p_2$};
\draw[gmass,-{Stealth[length=4pt]}] (0,-1.6) -- (-1.03923,-2.2);
\node[font=\tiny] at (-1.34234,-2.375) {$p_4$};
\fill (0,1.6) circle (2.2pt);
\fill (3,1.6) circle (2.2pt);
\fill (3,-1.6) circle (2.2pt);
\fill (0,-1.6) circle (2.2pt);
\end{tikzpicture}}{\texttt{pB181}}{3}{0} &
\begin{tabular}[c]{@{}l@{}}$E_{3}$\\{\scriptsize elliptic}\end{tabular} &
\sg{%
\begin{tikzpicture}[scale=\gscale,baseline=(current bounding box.center)]
\draw[gmass] (0,1.5) to node[pos=0.5,lab] {1} (0,-1.5);
\draw[gmass] (2.4,-1.5) to node[pos=0.5,lab] {2} (2.4,1.5);
\draw[gmass] (0,1.5) to node[pos=0.5,lab] {3} (2.4,1.5);
\draw[gglue] (0,-1.5) to node[pos=0.5,lab] {6} (2.4,-1.5);
\draw[gglue] (2.4,1.5) to node[pos=0.5,lab] {7} (4.6,0);
\draw[gmass] (0,-1.5) to node[pos=0.3,lab] {9} (4.6,0);
\draw[gglue] (0,1.5) -- (-0.848528,2.34853);
\node[font=\tiny] at (-1.09602,2.59602) {$p_1$};
\draw[gglue] (0,-1.5) -- (-0.707107,-2.20711);
\node[font=\tiny] at (-0.954594,-2.45459) {$p_2$};
\draw[gmass,-{Stealth[length=4pt]}] (2.4,-1.5) -- (3.24853,-2.34853);
\node[font=\tiny] at (3.49602,-2.59602) {$p_4$};
\draw[gmass,-{Stealth[length=4pt]}] (4.6,0) -- (5.8,0);
\node[font=\tiny] at (6.15,0) {$p_3$};
\fill (0,1.5) circle (2.2pt);
\fill (0,-1.5) circle (2.2pt);
\fill (2.4,-1.5) circle (2.2pt);
\fill (2.4,1.5) circle (2.2pt);
\fill (4.6,0) circle (2.2pt);
\end{tikzpicture}}{\texttt{nB359}}{5}{0} \\
\begin{tabular}[c]{@{}l@{}}$E_{4}$\\{\scriptsize elliptic}\end{tabular} &
\sg{%
\begin{tikzpicture}[scale=\gscale,baseline=(current bounding box.center)]
\draw[gmass] (0,0) to node[pos=0.5,lab] {2} (1.6,1.6);
\draw[gmass] (0,0) to node[pos=0.5,lab] {3} (3.2,0);
\draw[gmass] (1.6,1.6) to node[pos=0.5,lab] {5} (3.2,0);
\draw[gmass] (3.2,0) to node[pos=0.5,lab] {6} (1.6,-1.6);
\draw[gmass] (0,0) to node[pos=0.5,lab] {8} (1.6,-1.6);
\draw[gmass,-{Stealth[length=4pt]}] (0,0) -- (-1.2,1.46958e-16);
\node[font=\tiny] at (-1.55,1.8982e-16) {$p_3$};
\draw[gglue] (1.6,1.6) -- (1.6,2.8);
\node[font=\tiny] at (1.6,3.15) {$p_1$};
\draw[gmass,-{Stealth[length=4pt]}] (3.2,0) -- (4.4,0);
\node[font=\tiny] at (4.75,0) {$p_4$};
\draw[gglue] (1.6,-1.6) -- (1.6,-2.6);
\node[font=\tiny] at (1.6,-2.95) {$p_2$};
\fill (0,0) circle (2.2pt);
\fill (1.6,1.6) circle (2.2pt);
\fill (3.2,0) circle (2.2pt);
\fill (1.6,-1.6) circle (2.2pt);
\end{tikzpicture}}{\texttt{pE182}}{3}{0} &
\begin{tabular}[c]{@{}l@{}}$E_{5}$\\{\scriptsize elliptic}\end{tabular} &
\sg{%
\begin{tikzpicture}[scale=\gscale,baseline=(current bounding box.center)]
\draw[gmass] (0,1.5) to node[pos=0.5,lab] {1} (2.4,1.5);
\draw[gmass] (0,-1.5) to node[pos=0.5,lab] {2} (2.4,-1.5);
\draw[gmass] (0,1.5) to node[pos=0.25,lab] {3} (2.4,-1.5);
\draw[gmass] (2.4,1.5) to node[pos=0.25,lab] {4} (0,-1.5);
\draw[gglue] (2.4,-1.5) to node[pos=0.5,lab] {7} (4.6,0);
\draw[gglue] (2.4,1.5) to node[pos=0.5,lab] {8} (4.6,0);
\draw[gglue] (0,1.5) -- (-0.848528,2.34853);
\node[font=\tiny] at (-1.09602,2.59602) {$p_1$};
\draw[gglue] (0,-1.5) -- (-0.707107,-2.20711);
\node[font=\tiny] at (-0.954594,-2.45459) {$p_2$};
\draw[gmass,-{Stealth[length=4pt]}] (4.6,0) -- (5.63923,-0.6);
\node[font=\tiny] at (5.94234,-0.775) {$p_3$};
\draw[gmass,-{Stealth[length=4pt]}] (4.6,0) -- (5.63923,0.6);
\node[font=\tiny] at (5.94234,0.775) {$p_4$};
\fill (0,1.5) circle (2.2pt);
\fill (2.4,1.5) circle (2.2pt);
\fill (0,-1.5) circle (2.2pt);
\fill (2.4,-1.5) circle (2.2pt);
\fill (4.6,0) circle (2.2pt);
\end{tikzpicture}}{\texttt{nB207}}{2}{0} &
\begin{tabular}[c]{@{}l@{}}$E_{6}$\\{\scriptsize elliptic}\end{tabular} &
\sg{%
\begin{tikzpicture}[scale=\gscale,baseline=(current bounding box.center)]
\draw[gmass] (0,1.5) to node[pos=0.5,lab] {1} (2.4,1.5);
\draw[gmass] (0,-1.5) to node[pos=0.5,lab] {2} (2.4,-1.5);
\draw[gmass] (0,1.5) to node[pos=0.25,lab] {3} (2.4,-1.5);
\draw[gmass] (2.4,1.5) to node[pos=0.25,lab] {4} (0,-1.5);
\draw[gglue] (2.4,-1.5) to node[pos=0.5,lab] {7} (4.6,0);
\draw[gmass] (2.4,1.5) to node[pos=0.5,lab] {9} (4.6,0);
\draw[gglue] (0,1.5) -- (-0.848528,2.34853);
\node[font=\tiny] at (-1.09602,2.59602) {$p_1$};
\draw[gmass,-{Stealth[length=4pt]}] (2.4,1.5) -- (3,2.53923);
\node[font=\tiny] at (3.175,2.84234) {$p_4$};
\draw[gglue] (0,-1.5) -- (-0.707107,-2.20711);
\node[font=\tiny] at (-0.954594,-2.45459) {$p_2$};
\draw[gmass,-{Stealth[length=4pt]}] (4.6,0) -- (5.8,0);
\node[font=\tiny] at (6.15,0) {$p_3$};
\fill (0,1.5) circle (2.2pt);
\fill (2.4,1.5) circle (2.2pt);
\fill (0,-1.5) circle (2.2pt);
\fill (2.4,-1.5) circle (2.2pt);
\fill (4.6,0) circle (2.2pt);
\end{tikzpicture}}{\texttt{nB335}}{4}{2} \\
\midrule
\begin{tabular}[c]{@{}l@{}}$E_{7}$\\{\scriptsize elliptic}\end{tabular} &
\multicolumn{5}{l}{%
\sg{%
\begin{tikzpicture}[scale=\gscale,baseline=(current bounding box.center)]
\draw[gglue] (0,1.6) to node[pos=0.5,lab] {1} (3,1.6);
\draw[gmass] (3,1.6) to node[pos=0.5,lab] {5} (3,-1.6);
\draw[gmass] (0,-1.6) to node[pos=0.5,lab] {7} (3,-1.6);
\draw[gmass] (0,1.6) to node[pos=0.5,lab] {8} (0,-1.6);
\draw[gmass] (3,1.6) to node[pos=0.35,lab] {3} (0,-1.6);
\draw[gmass,-{Stealth[length=4pt]}] (0,1.6) -- (-0.86603,2.1);
\node[font=\tiny] at (-1.14315,2.26) {$p_3$};
\draw[gglue] (3,1.6) -- (3.86603,2.1);
\node[font=\tiny] at (4.14315,2.26) {$p_1$};
\draw[gglue] (3,-1.6) -- (3.86603,-2.1);
\node[font=\tiny] at (4.14315,-2.26) {$p_2$};
\draw[gmass,-{Stealth[length=4pt]}] (0,-1.6) -- (-0.86603,-2.1);
\node[font=\tiny] at (-1.14315,-2.26) {$p_4$};
\fill (0,1.6) circle (2.2pt);
\fill (3,1.6) circle (2.2pt);
\fill (3,-1.6) circle (2.2pt);
\fill (0,-1.6) circle (2.2pt);
\end{tikzpicture}}{\texttt{pB213}}{4}{0}\hspace{5pt}\sg{%
\begin{tikzpicture}[scale=\gscale,baseline=(current bounding box.center)]
\draw[gmass] (0,1.6) to node[pos=0.5,lab] {2} (3,1.6);
\draw[gmass,bend left=35] (0,1.6) to node[pos=0.5,lab] {3} (0,-1.6);
\draw[gmass] (3,1.6) to node[pos=0.5,lab] {5} (3,-1.6);
\draw[gmass] (0,-1.6) to node[pos=0.5,lab] {7} (3,-1.6);
\draw[gmass,bend right=35] (0,1.6) to node[pos=0.5,lab] {8} (0,-1.6);
\draw[gmass,-{Stealth[length=4pt]}] (0,1.6) -- (-1.03923,2.2);
\node[font=\tiny] at (-1.34234,2.375) {$p_3$};
\draw[gglue] (3,1.6) -- (4.03923,2.2);
\node[font=\tiny] at (4.34234,2.375) {$p_1$};
\draw[gmass,-{Stealth[length=4pt]}] (0,-1.6) -- (-1.03923,-2.2);
\node[font=\tiny] at (-1.34234,-2.375) {$p_4$};
\draw[gglue] (3,-1.6) -- (3.86603,-2.1);
\node[font=\tiny] at (4.16913,-2.275) {$p_2$};
\fill (0,1.6) circle (2.2pt);
\fill (3,1.6) circle (2.2pt);
\fill (0,-1.6) circle (2.2pt);
\fill (3,-1.6) circle (2.2pt);
\end{tikzpicture}}{\texttt{pB214}}{3}{0}\hspace{5pt}\sg{%
\begin{tikzpicture}[scale=\gscale,baseline=(current bounding box.center)]
\draw[gmass] (0,0) to node[pos=0.5,lab] {1} (2,1.6);
\draw[gglue] (0,0) to node[pos=0.5,lab] {8} (2,-1.6);
\draw[gglue] (2,1.6) to node[pos=0.5,lab] {3} (2,-1.6);
\draw[gmass] (2,1.6) to node[pos=0.5,lab] {2} (4.6,1.6);
\draw[gmass] (2,-1.6) to node[pos=0.5,lab] {7} (4.6,-1.6);
\draw[gmass] (4.6,1.6) to node[pos=0.5,lab] {5} (4.6,-1.6);
\draw[gmass,-{Stealth[length=4pt]}] (0,0) -- (-1,0);
\node[font=\tiny] at (-1.32,0) {$p_3$};
\draw[gmass,-{Stealth[length=4pt]}] (2,-1.6) -- (1.5,-2.46603);
\node[font=\tiny] at (1.34,-2.74315) {$p_4$};
\draw[gglue] (4.6,1.6) -- (5.46603,2.1);
\node[font=\tiny] at (5.74315,2.26) {$p_1$};
\draw[gglue] (4.6,-1.6) -- (5.46603,-2.1);
\node[font=\tiny] at (5.74315,-2.26) {$p_2$};
\fill (0,0) circle (2.2pt);
\fill (2,1.6) circle (2.2pt);
\fill (2,-1.6) circle (2.2pt);
\fill (4.6,1.6) circle (2.2pt);
\fill (4.6,-1.6) circle (2.2pt);
\end{tikzpicture}}{\texttt{pC215}}{3}{0}\hspace{5pt}\sg{%
\begin{tikzpicture}[scale=\gscale,baseline=(current bounding box.center)]
\draw[gglue] (0,1.6) to node[pos=0.5,lab] {1} (2.4,1.6);
\draw[gmass] (2.4,1.6) to node[pos=0.5,lab] {2} (4.8,1.6);
\draw[gmass] (0,1.6) to node[pos=0.5,lab] {8} (0,-1.6);
\draw[gmass] (2.4,1.6) to node[pos=0.5,lab] {3} (2.4,-1.6);
\draw[gmass] (4.8,1.6) to node[pos=0.5,lab] {5} (4.8,-1.6);
\draw[gglue] (0,-1.6) to node[pos=0.5,lab] {6} (2.4,-1.6);
\draw[gmass] (2.4,-1.6) to node[pos=0.5,lab] {7} (4.8,-1.6);
\draw[gmass,-{Stealth[length=4pt]}] (0,1.6) -- (-0.86603,2.1);
\node[font=\tiny] at (-1.14315,2.26) {$p_3$};
\draw[gglue] (4.8,1.6) -- (5.66603,2.1);
\node[font=\tiny] at (5.94315,2.26) {$p_1$};
\draw[gglue] (4.8,-1.6) -- (5.66603,-2.1);
\node[font=\tiny] at (5.94315,-2.26) {$p_2$};
\draw[gmass,-{Stealth[length=4pt]}] (0,-1.6) -- (-0.86603,-2.1);
\node[font=\tiny] at (-1.14315,-2.26) {$p_4$};
\fill (0,1.6) circle (2.2pt);
\fill (2.4,1.6) circle (2.2pt);
\fill (4.8,1.6) circle (2.2pt);
\fill (0,-1.6) circle (2.2pt);
\fill (2.4,-1.6) circle (2.2pt);
\fill (4.8,-1.6) circle (2.2pt);
\end{tikzpicture}}{\texttt{pB247}}{5}{0}\hspace{5pt}\sg{%
\begin{tikzpicture}[scale=\gscale,baseline=(current bounding box.center)]
\draw[gglue] (0,1.5) to node[pos=0.5,lab] {1} (2.4,1.5);
\draw[gglue] (0,-1.5) to node[pos=0.5,lab] {2} (2.4,-1.5);
\draw[gmass] (0,1.5) to node[pos=0.25,lab] {3} (2.4,-1.5);
\draw[gmass] (2.4,1.5) to node[pos=0.25,lab] {4} (0,-1.5);
\draw[gmass] (2.4,-1.5) to node[pos=0.5,lab] {7} (4.6,-1.5);
\draw[gmass] (2.4,1.5) to node[pos=0.5,lab] {8} (4.6,1.5);
\draw[gmass] (4.6,-1.5) to node[pos=0.5,lab] {9} (4.6,1.5);
\draw[gmass,-{Stealth[length=4pt]}] (0,1.5) -- (-0.70711,2.20711);
\node[font=\tiny] at (-0.93338,2.43338) {$p_3$};
\draw[gmass,-{Stealth[length=4pt]}] (0,-1.5) -- (-0.70711,-2.20711);
\node[font=\tiny] at (-0.93338,-2.43338) {$p_4$};
\draw[gglue] (4.6,-1.5) -- (5.30711,-2.20711);
\node[font=\tiny] at (5.53338,-2.43338) {$p_2$};
\draw[gglue] (4.6,1.5) -- (5.30711,2.20711);
\node[font=\tiny] at (5.53338,2.43338) {$p_1$};
\fill (0,1.5) circle (2.2pt);
\fill (2.4,1.5) circle (2.2pt);
\fill (0,-1.5) circle (2.2pt);
\fill (2.4,-1.5) circle (2.2pt);
\fill (4.6,-1.5) circle (2.2pt);
\fill (4.6,1.5) circle (2.2pt);
\end{tikzpicture}}{\texttt{nC463}$^{\rm x12}$}{4}{0}} \\
\midrule
\begin{tabular}[c]{@{}l@{}}$E_{8}$\\{\scriptsize elliptic}\end{tabular} &
\multicolumn{5}{l}{%
\sg{%
\begin{tikzpicture}[scale=\gscale,baseline=(current bounding box.center)]
\draw[gmass,bend left=35] (0,1.6) to node[pos=0.5,lab] {2} (3,1.6);
\draw[gmass,bend right=35] (0,1.6) to node[pos=0.5,lab] {3} (3,1.6);
\draw[gglue] (3,1.6) to node[pos=0.5,lab] {4} (3,-1.6);
\draw[gmass] (3,-1.6) to node[pos=0.5,lab] {6} (0,-1.6);
\draw[gmass] (0,1.6) to node[pos=0.5,lab] {8} (0,-1.6);
\draw[gmass,-{Stealth[length=4pt]}] (0,1.6) -- (-1.03923,2.2);
\node[font=\tiny] at (-1.34234,2.375) {$p_3$};
\draw[gglue] (3,1.6) -- (4.03923,2.2);
\node[font=\tiny] at (4.34234,2.375) {$p_1$};
\draw[gmass,-{Stealth[length=4pt]}] (3,-1.6) -- (4.03923,-2.2);
\node[font=\tiny] at (4.34234,-2.375) {$p_4$};
\draw[gglue] (0,-1.6) -- (-0.866025,-2.1);
\node[font=\tiny] at (-1.16913,-2.275) {$p_2$};
\fill (0,1.6) circle (2.2pt);
\fill (3,1.6) circle (2.2pt);
\fill (3,-1.6) circle (2.2pt);
\fill (0,-1.6) circle (2.2pt);
\end{tikzpicture}}{\texttt{pE174}}{3}{0}\hspace{5pt}\sg{%
\begin{tikzpicture}[scale=\gscale,baseline=(current bounding box.center)]
\draw[gglue] (0,1.6) to node[pos=0.5,lab] {1} (3,1.6);
\draw[gmass] (3,1.6) to node[pos=0.5,lab] {5} (3,-1.6);
\draw[gglue] (0,-1.6) to node[pos=0.5,lab] {7} (3,-1.6);
\draw[gmass] (0,1.6) to node[pos=0.5,lab] {8} (0,-1.6);
\draw[gmass] (3,1.6) to node[pos=0.35,lab] {3} (0,-1.6);
\draw[gmass,-{Stealth[length=4pt]}] (0,1.6) -- (-0.86603,2.1);
\node[font=\tiny] at (-1.14315,2.26) {$p_3$};
\draw[gglue] (3,1.6) -- (3.86603,2.1);
\node[font=\tiny] at (4.14315,2.26) {$p_2$};
\draw[gmass,-{Stealth[length=4pt]}] (3,-1.6) -- (3.86603,-2.1);
\node[font=\tiny] at (4.14315,-2.26) {$p_4$};
\draw[gglue] (0,-1.6) -- (-0.86603,-2.1);
\node[font=\tiny] at (-1.14315,-2.26) {$p_1$};
\fill (0,1.6) circle (2.2pt);
\fill (3,1.6) circle (2.2pt);
\fill (3,-1.6) circle (2.2pt);
\fill (0,-1.6) circle (2.2pt);
\end{tikzpicture}}{\texttt{pE213}$^{\rm x12}$}{5}{0}\hspace{5pt}\sg{%
\begin{tikzpicture}[scale=\gscale,baseline=(current bounding box.center)]
\draw[gglue] (0,1.6) to node[pos=0.5,lab] {1} (2.6,1.6);
\draw[gmass] (0,1.6) to node[pos=0.5,lab] {8} (0,-1.6);
\draw[gmass] (0,-1.6) to node[pos=0.5,lab] {6} (2.6,-1.6);
\draw[gglue] (2.6,1.6) to node[pos=0.5,lab] {3} (2.6,-1.6);
\draw[gglue] (2.6,1.6) to node[pos=0.5,lab] {5} (4.6,0);
\draw[gmass] (2.6,-1.6) to node[pos=0.5,lab] {7} (4.6,0);
\draw[gmass,-{Stealth[length=4pt]}] (0,1.6) -- (-0.86603,2.1);
\node[font=\tiny] at (-1.14315,2.26) {$p_3$};
\draw[gglue] (2.6,1.6) -- (3.1,2.46603);
\node[font=\tiny] at (3.26,2.74315) {$p_2$};
\draw[gmass,-{Stealth[length=4pt]}] (4.6,0) -- (5.6,0);
\node[font=\tiny] at (5.92,0) {$p_4$};
\draw[gglue] (0,-1.6) -- (-0.86603,-2.1);
\node[font=\tiny] at (-1.14315,-2.26) {$p_1$};
\fill (0,1.6) circle (2.2pt);
\fill (2.6,1.6) circle (2.2pt);
\fill (2.6,-1.6) circle (2.2pt);
\fill (0,-1.6) circle (2.2pt);
\fill (4.6,0) circle (2.2pt);
\end{tikzpicture}}{\texttt{pD245}$^{\rm x12}$}{3}{0}\hspace{5pt}\sg{%
\begin{tikzpicture}[scale=\gscale,baseline=(current bounding box.center)]
\draw[gglue] (0,1.6) to node[pos=0.5,lab] {1} (2.4,1.6);
\draw[gmass] (2.4,1.6) to node[pos=0.5,lab] {2} (4.8,1.6);
\draw[gmass] (0,1.6) to node[pos=0.5,lab] {8} (0,-1.6);
\draw[gmass] (2.4,1.6) to node[pos=0.5,lab] {3} (2.4,-1.6);
\draw[gmass] (4.8,1.6) to node[pos=0.5,lab] {5} (4.8,-1.6);
\draw[gmass] (0,-1.6) to node[pos=0.5,lab] {6} (2.4,-1.6);
\draw[gglue] (2.4,-1.6) to node[pos=0.5,lab] {7} (4.8,-1.6);
\draw[gmass,-{Stealth[length=4pt]}] (0,1.6) -- (-0.86603,2.1);
\node[font=\tiny] at (-1.14315,2.26) {$p_3$};
\draw[gglue] (4.8,1.6) -- (5.66603,2.1);
\node[font=\tiny] at (5.94315,2.26) {$p_2$};
\draw[gmass,-{Stealth[length=4pt]}] (4.8,-1.6) -- (5.66603,-2.1);
\node[font=\tiny] at (5.94315,-2.26) {$p_4$};
\draw[gglue] (0,-1.6) -- (-0.86603,-2.1);
\node[font=\tiny] at (-1.14315,-2.26) {$p_1$};
\fill (0,1.6) circle (2.2pt);
\fill (2.4,1.6) circle (2.2pt);
\fill (4.8,1.6) circle (2.2pt);
\fill (0,-1.6) circle (2.2pt);
\fill (2.4,-1.6) circle (2.2pt);
\fill (4.8,-1.6) circle (2.2pt);
\end{tikzpicture}}{\texttt{pE247}$^{\rm x12}$}{4}{1}\hspace{5pt}\sg{%
\begin{tikzpicture}[scale=\gscale,baseline=(current bounding box.center)]
\draw[gglue] (0,1.5) to node[pos=0.5,lab] {1} (2.4,1.5);
\draw[gmass] (0,-1.5) to node[pos=0.5,lab] {2} (2.4,-1.5);
\draw[gmass] (0,1.5) to node[pos=0.25,lab] {3} (2.4,-1.5);
\draw[gmass] (2.4,1.5) to node[pos=0.25,lab] {4} (0,-1.5);
\draw[gglue] (2.4,-1.5) to node[pos=0.5,lab] {7} (4.6,-1.5);
\draw[gmass] (2.4,1.5) to node[pos=0.5,lab] {8} (4.6,1.5);
\draw[gmass] (4.6,-1.5) to node[pos=0.5,lab] {9} (4.6,1.5);
\draw[gmass,-{Stealth[length=4pt]}] (0,1.5) -- (-0.70711,2.20711);
\node[font=\tiny] at (-0.93338,2.43338) {$p_3$};
\draw[gglue] (0,-1.5) -- (-0.70711,-2.20711);
\node[font=\tiny] at (-0.93338,-2.43338) {$p_1$};
\draw[gmass,-{Stealth[length=4pt]}] (4.6,-1.5) -- (5.30711,-2.20711);
\node[font=\tiny] at (5.53338,-2.43338) {$p_4$};
\draw[gglue] (4.6,1.5) -- (5.30711,2.20711);
\node[font=\tiny] at (5.53338,2.43338) {$p_2$};
\fill (0,1.5) circle (2.2pt);
\fill (2.4,1.5) circle (2.2pt);
\fill (0,-1.5) circle (2.2pt);
\fill (2.4,-1.5) circle (2.2pt);
\fill (4.6,-1.5) circle (2.2pt);
\fill (4.6,1.5) circle (2.2pt);
\end{tikzpicture}}{\texttt{nI463}}{6}{1}} \\
\midrule
\begin{tabular}[c]{@{}l@{}}genus two\\{\scriptsize \phantom{x}}\end{tabular} &
\multicolumn{5}{l}{%
\sg{%
\begin{tikzpicture}[scale=\gscale,baseline=(current bounding box.center)]
\draw[gmass] (0,1.5) to node[pos=0.5,lab] {1} (2.4,1.5);
\draw[gmass] (0,-1.5) to node[pos=0.5,lab] {2} (2.4,-1.5);
\draw[gmass] (0,1.5) to node[pos=0.25,lab] {3} (2.4,-1.5);
\draw[gmass] (2.4,1.5) to node[pos=0.25,lab] {4} (0,-1.5);
\draw[gglue] (2.4,-1.5) to node[pos=0.5,lab] {7} (4.6,-1.5);
\draw[gglue] (2.4,1.5) to node[pos=0.5,lab] {8} (4.6,1.5);
\draw[gmass] (4.6,-1.5) to node[pos=0.5,lab] {9} (4.6,1.5);
\draw[gglue] (0,1.5) -- (-0.848528,2.34853);
\node[font=\tiny] at (-1.09602,2.59602) {$p_1$};
\draw[gglue] (0,-1.5) -- (-0.707107,-2.20711);
\node[font=\tiny] at (-0.954594,-2.45459) {$p_2$};
\draw[gmass,-{Stealth[length=4pt]}] (4.6,-1.5) -- (5.44853,-2.34853);
\node[font=\tiny] at (5.69602,-2.59602) {$p_3$};
\draw[gmass,-{Stealth[length=4pt]}] (4.6,1.5) -- (5.44853,2.34853);
\node[font=\tiny] at (5.69602,2.59602) {$p_4$};
\fill (0,1.5) circle (2.2pt);
\fill (2.4,1.5) circle (2.2pt);
\fill (0,-1.5) circle (2.2pt);
\fill (2.4,-1.5) circle (2.2pt);
\fill (4.6,-1.5) circle (2.2pt);
\fill (4.6,1.5) circle (2.2pt);
\end{tikzpicture}}{\texttt{nB463}}{5}{0}} \\
\bottomrule
\end{tabular}
 \caption{Non-polylogarithmic geometries of the canonical basis. For each curve all sectors are shown whose homogeneous diagonal block of the canonical differential equation contains it, with the number of master integrals $N_{\rm MI}$ and, in brackets, the number $N_{\log}$ of them that do not couple to the curve within the block (pure dlog). Thick (coiled) lines are massive (massless) propagators, labelled by the indices of \cref{app:families}; $p_1,p_2$ are the incoming gluons, $p_3,p_4$ the outgoing top quarks. ${\rm x12}$ marks the crossed family ($p_1\leftrightarrow p_2$); the crossed curves $E_1^{\rm x12}$, $E_2^{\rm x12}$, $E_3^{\rm x12}$, $E_7^{\rm x12}$ and $E_8^{\rm x12}$ appear in the crossed images of the sectors shown. The other elliptic curves are mapped to themselves under crossing.}
\label{tab:geometries}
\end{table}

For each geometry, the periods satisfy Picard--Fuchs differential equations.
For the elliptic curves, these are of second order, and we denote their holomorphic 
period by $\varpi_0^{(E)}$, where $E$ indicates the elliptic curve, 
and by $\varpi_0^{(E\rm x12)}$ for the corresponding crossed ones.
For the genus-two curve, the periods $\pi^{(1)}_j$ and $\pi^{(2)}_j$ of the two holomorphic differentials satisfy a coupled system of differential equations, which we discuss in \cref{sec:genustwo}.

To cast the integrals associated with these geometries in canonical form
we employ a natural extension of the integrand analysis performed in the polylogarithmic case.
In particular, the approach can be streamlined by choosing an initial basis aligned with the geometry underlying the maximal cut, normalising it by the inverse of the semi-simple part of the period matrix, rescaling it by powers of $\varepsilon$ to align their (local) transcendental weights, 
and finally removing the remaining terms of $\mathcal{O}(\varepsilon^0)$ by a rotation whose entries are fixed by first-order differential equations. This last clean-up step can itself be interpreted in terms of integrand analysis, at the price of having to analyse the integrand to higher orders in $\varepsilon$
to account for the weight drops induced by the existence of differential forms with higher poles~\cite{Gorges:2023zgv,Duhr:2025lbz,Forner:2026vby}.
In this process, new transcendental functions have to be introduced in the canonical differential equation, which are ultimately required to represent the leading singularities of the corresponding integrals.
Besides the periods, in the analysis of the elliptic sectors, we find 16 functions
associated with additional punctures on the curves, which can all be related to complete elliptic integrals of the third kind. 
In addition, we have to further introduce 38 functions
which involve either two different elliptic curves or a pair of an elliptic period and a period of the genus-two curve.
Finally, the differential forms associated with the genus-two curve contain 3 additional functions, one of which is a third-kind form on the genus-two surface, while the remaining two are associated with leading singularities at higher orders in $\varepsilon$.
These functions typically have higher-degree products of the periods in their differential equations. 
We stress here that obtaining an analytic expression for all these transcendental functions in terms of known special functions is not only often impossible but also not necessary: what matters is that they 
are all defined as solutions of first-order inhomogeneous differential equations with rational coefficients.
As an example of this procedure, we discuss the construction of the basis for the genus-two sector in detail in \cref{sec:genustwo}.

\begin{figure}[htbp]
  \centering
  \includegraphics[width=0.80\textwidth]{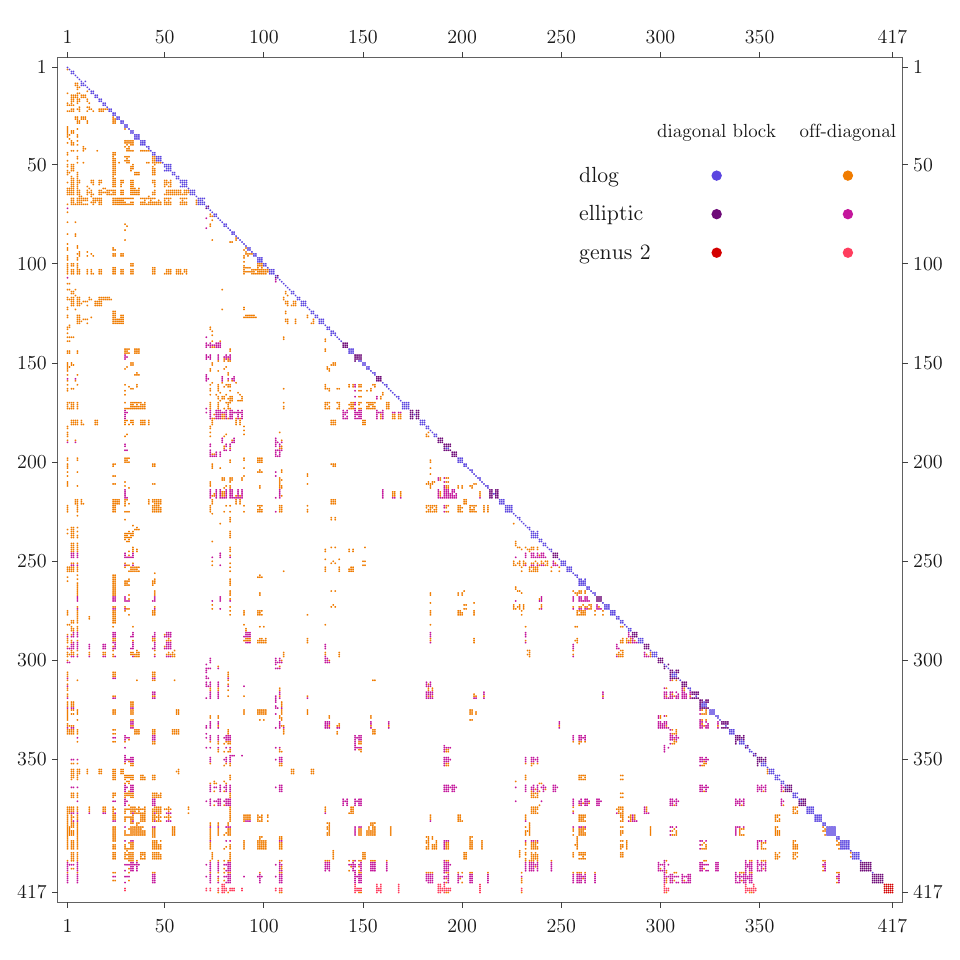}
  \caption{Non-zero entries of the $417\times417$ connection matrix $A$ of the canonical differential
    equation.
    Entries
    within a diagonal block are shown in dark colours, all other entries in light colours.
    Entries built only from dlog forms are shown in blue and orange, entries
    involving elliptic kernels in purple and magenta, and entries involving genus-two kernels
    in red and pink. The genus-two curve enters only in the last block, which corresponds
    to the sector \texttt{nB463}.}
  \label{fig:DEmatrix}
\end{figure}

The resulting connection matrix for the $417$ canonical master integrals, schematically depicted in~\cref{fig:DEmatrix}, is built from the set of linearly independent differential forms summarised in \cref{tab:forms}. The explicit form of the matrices is provided as electronic files in the Zenodo repository~\cite{Zenodo_ttbar} associated to this publication.
We distinguish rational $\mathrm{d}\!\log$-forms, algebraic $\mathrm{d}\!\log$-forms, elliptic forms, and forms associated with the genus-two curve.
The elliptic forms are built from the elliptic periods and the extra transcendental functions described above. Additionally, we call the elliptic forms of \emph{quadratic type} if they are quadratic in one of the periods, and of \emph{simple type} otherwise.
Interestingly, not all of these forms contribute to the finite remainder of the amplitude.
Up to transcendental weight four, the $\varepsilon^0$ coefficient involves only $300$ of the $449$ forms and, following a pattern already observed in previous calculations~\cite{Duhr:2024bzt,Forner:2024ojj,Becchetti:2025rrz,Coro:2025vgn}, none of the 
elliptic forms of quadratic type. Many non-linear forms associated with the genus two geometry also drop, and it will be interesting to investigate how to predict which forms drop from general arguments. We postpone these studies to a future publication.
\begin{table}[htbp]
\centering
\begin{tabular}{lcc}
\toprule
Differential forms & Canonical DE & Finite remainder \\
\midrule
Rational $\mathrm{d}\!\log$-forms & $21$ & $16$ \\
Algebraic $\mathrm{d}\!\log$-forms & $17$ & $17$ \\
Elliptic, simple type & $283$ & $212$ \\
Elliptic, quadratic type & $43$ & $0$ \\
Genus two & $85$ & $55$ \\
\midrule
Total & $449$ & $300$ \\
\bottomrule
\end{tabular}
\caption{Number of linearly independent differential forms in the connection matrix of the canonical differential equation for the $417$ master integrals (second column), and number of those that enter the iterated integrals up to weight four in the $\varepsilon^0$ coefficient of the finite remainder of the amplitude (third column). The iterated integrals in the $\varepsilon$ poles of the bare amplitude contain only rational or algebraic $\mathrm{d}\!\log$-forms but no elliptic or genus-two kernels.}
\label{tab:forms}
\end{table}

Finally, we stress that the $417$ canonical master integrals are not linearly independent at weight four: expressing them in terms of iterated integrals defined over an independent set of differential forms allows us to demonstrate that they satisfy $65$ linear relations, such that the solution can be written in terms of $352$ independent combinations of masters.

\subsection{Genus-two sector}
\label{sec:genustwo}
\begin{figure}[htbp]
\centering
\begin{tikzpicture}[scale=0.6,
  gmass/.style={line width=1.4pt},
  gglue/.style={decorate,decoration={coil,aspect=0.55,segment length=4.5pt,amplitude=2pt},line width=0.6pt},
  lab/.style={fill=white,inner sep=1pt,font=\small}]
\draw[gmass] (0,1.5) to node[pos=0.5,lab] {1} (2.4,1.5);
\draw[gmass] (0,-1.5) to node[pos=0.5,lab] {2} (2.4,-1.5);
\draw[gmass] (0,1.5) to node[pos=0.25,lab] {3} (2.4,-1.5);
\draw[gmass] (2.4,1.5) to node[pos=0.25,lab] {4} (0,-1.5);
\draw[gglue] (2.4,-1.5) to node[pos=0.5,lab] {7} (4.6,-1.5);
\draw[gglue] (2.4,1.5) to node[pos=0.5,lab] {8} (4.6,1.5);
\draw[gmass] (4.6,-1.5) to node[pos=0.5,lab] {9} (4.6,1.5);
\draw[gglue] (0,1.5) -- (-0.85,2.35);
\node at (-1.15,2.65) {$p_1$};
\draw[gglue] (0,-1.5) -- (-0.85,-2.35);
\node at (-1.15,-2.65) {$p_2$};
\draw[gmass,-{Stealth[length=7pt]}] (4.6,-1.5) -- (5.45,-2.35);
\node at (5.75,-2.65) {$p_3$};
\draw[gmass,-{Stealth[length=7pt]}] (4.6,1.5) -- (5.45,2.35);
\node at (5.75,2.65) {$p_4$};
\foreach \v in {(0,1.5),(2.4,1.5),(0,-1.5),(2.4,-1.5),(4.6,-1.5),(4.6,1.5)} \fill \v circle (2.5pt);
\end{tikzpicture}
\caption{The non-planar sector \texttt{nB463}, whose homogeneous differential equation involves the genus-two curve. Thick lines are massive propagators, coiled lines massless ones, and the labels are the propagator indices of \cref{app:families}; $p_1,p_2$ are the incoming gluons, $p_3,p_4$ the outgoing top quarks.}
\label{fig:genus2}
\end{figure}
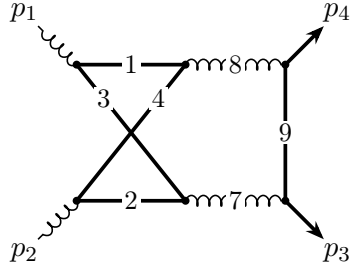

In this section, we discuss in detail the construction of the canonical basis for sector \texttt{nB463}, 
whose graph is shown in \cref{fig:genus2}.
It is the only sector of our problem associated with a curve of genus two, and it illustrates that the method of refs.~\cite{Gorges:2023zgv,Duhr:2024uid,Duhr:2025lbz,Forner:2026vby} applies beyond elliptic curves and beyond the maximal cut, without essential changes.
 
Let us start by considering the corner integral of this sector, which we call  $I_1$.
From a loop-by-loop Baikov representation, it is easy to see that its leading singularities 
are given by the one-fold integral
\begin{equation}
\label{eq:LSgenus2}
  \mathrm{LS}(I_1)\propto\oint\frac{\mathrm{d}x}{\sqrt{P_6(x)}}\,,
\end{equation}
where, in our parametrisation $x=k_1\cdot p_1$, and
\begin{equation}
\label{eq:P6}
  P_6(x)=s\,x(s-2x)\left(2m_t^2s+x(s-2x)\right)\left(m_t^4s-2m_t^2\left(s(t-2x)+8x^2\right)+s(t+2x)^2\right).
\end{equation}
Since $P_6$ is a polynomial of degree six, the equation $y^2=P_6(x)$ defines a hyperelliptic curve of genus two.
The space of holomorphic differentials on this curve is spanned by the two forms of the first kind
\begin{equation}
  \frac{\mathrm{d}x}{\sqrt{P_6(x)}}\qquad\text{and}\qquad\frac{x\,\mathrm{d}x}{\sqrt{P_6(x)}}\,.
\end{equation}
We therefore choose the first two master integrals such that, on the maximal cut, they reduce to these forms, i.e., the corner integral $I_1$ and the integral $I_2$ with the numerator insertion $k_1\cdot p_1$.
The third and fourth master integrals are chosen as their derivatives.
In addition, one can see that the form
\begin{equation}
  \frac{x^2\,\mathrm{d}x}{\sqrt{P_6(x)}}
\end{equation}
has a simple pole with non-vanishing residue at infinity and is thus of the third kind.\footnote{To be precise, there are two points at infinity, and two residues, since the surface is a double cover. The two residues are the same up to an overall constant, which does not matter in our construction.}
Correspondingly, we choose the fifth master integral as the one with the numerator insertion $(k_1\cdot p_1)^2$.
In terms of the integral family in~\cref{tab:nonplanarfamilies1}, our starting basis
reads 
\begin{equation}
\label{eq:basisgenus2}
\begin{split}
  I_1&=\frac{s^2}{t}\,I^{\texttt{nB}}_{1,1,1,1,0,0,1,1,1}\,,\\
  I_2&=\frac{s}{2t}\Big[I^{\texttt{nB}}_{1,1,1,1,-1,0,1,1,1}+I^{\texttt{nB}}_{1,1,1,1,0,-1,1,1,1}+(t-5m_t^2)\,I^{\texttt{nB}}_{1,1,1,1,0,0,1,1,1}\Big]\,,\\
  I_3&=\partial_v I_2\,,\qquad I_4=\partial_v I_1\,,\\
  I_5&=\frac{1}{4}\Big[I^{\texttt{nB}}_{1,1,1,1,-2,0,1,1,1}+2\,I^{\texttt{nB}}_{1,1,1,1,-1,-1,1,1,1}+I^{\texttt{nB}}_{1,1,1,1,0,-2,1,1,1}\\
  &\qquad\ +2(t-5m_t^2)\left(I^{\texttt{nB}}_{1,1,1,1,-1,0,1,1,1}+I^{\texttt{nB}}_{1,1,1,1,0,-1,1,1,1}\right)+(t-5m_t^2)^2\,I^{\texttt{nB}}_{1,1,1,1,0,0,1,1,1}\Big]\,,
\end{split}
\end{equation}
Here and in the following, we use the variables $v$ and $z$ defined by 
\begin{equation}
    v=\frac{s}{t} \quad\text{and}\quad z=\frac{m_t^2}{s} \, .
\end{equation} Additionally we set $t=1$. These are convenient variables when working with the genus-two curve. In particular, they are built as blow-up variables to resolve the singularities at $s=m_t^2=0$
 and allow for series solutions of the genus-two Picard--Fuchs equations close to that point.
 
In this basis, the differential equation has the structure
\begin{equation}
\left(
\begin{array}{ccccc}
 \times & \times & \times & \times & - \\
 \times & \times & \times & \times & - \\
 \times & \times & \times & \times & - \\
 \times & \times & \times & \times & - \\
 \times & \times & \times & \times & \times \\
\end{array}
\right),
\qquad
\left(
\begin{array}{ccccc}
 1 & 1 & 0 & 0 & 1 \\
 1 & 1 & 0 & 0 & 1 \\
 2 & 2 & 1 & 1 & 2 \\
 2 & 2 & 1 & 1 & 2 \\
 1 & 1 & 0 & 0 & 1 \\
\end{array}
\right),
\end{equation}
where, in the first matrix, $\times$ denotes an entry that is non-vanishing at $\varepsilon=0$ and $-$ an entry that vanishes at $\varepsilon=0$, while the second matrix gives the highest power of $\varepsilon$ in each entry.
We observe that the third-kind integral decouples for $\varepsilon=0$ from the $(4\times4)$-block that determines the first- and second-kind periods of the genus two curve, as expected.
 
At $\varepsilon=0$, the differential equations reduce to a system of two coupled second-order differential equations 
 for the maximal cuts of $I_1$ and $I_2$:
\begin{align}
  &D \,(1-vz)\,\partial_v^2\pi^{(1)}_j+N_3\,\partial_v\pi^{(1)}_j-z\,N_1\,\pi^{(1)}_j+2v\,N_4\,\partial_v\pi^{(2)}_j+2N_2\,\pi^{(2)}_j=0\,,
  \label{eq:PFgenus2a}\\
  &2D\,v\,\partial_v^2\pi^{(2)}_j-2v\,M_4\,\partial_v\pi^{(2)}_j-2M_2\,\pi^{(2)}_j-(1-vz)\,M_3\,\partial_v\pi^{(1)}_j+z\,M_1\,\pi^{(1)}_j=0\,,
  \label{eq:PFgenus2b}
\end{align}
where
\begin{align}
    D=&v\,(1-4z)(1+v-vz)(1+v-2vz+v^2z^2)\nonumber\\
  &\times(1+v+6vz-v^2z+9v^2z^2)(1+v-10vz-9v^2z+25v^2z^2)\,,
\end{align}
 and the polynomials $N_i$ and $M_i$ are defined in~\cref{app:polyPic}.
 
The solution space of~\cref{eq:PFgenus2a,eq:PFgenus2b} is four-dimensional, and we denote its basis by $\pi^{(i)}_j$, where $i=1,2$ labels the integral and $j=0,\ldots,3$ the solution.
We choose $\pi^{(i)}_0$ and $\pi^{(i)}_1$ as the solutions free of logarithms, while $\pi^{(i)}_2$ and $\pi^{(i)}_3$ are logarithmic:
\begin{equation}
\begin{aligned}
  \pi^{(1)}_0&=1-4z+36z^2-v\left(\tfrac{3}{2}-3z\right)+\tfrac{3}{2}v^2+\ldots\,,\\[4pt]
  \pi^{(1)}_1&=v-v^2+\ldots\,,\\[4pt]
  \pi^{(1)}_2&=1+z-\tfrac{33}{2}z^2+\tfrac{1}{2}vz-\tfrac{5}{16}v^2
  -\tfrac{5}{16}\Big[2-8z+72z^2-v\left(1-6z\right)+v^2\Big]\log z+\ldots\,,\\[4pt]
  \pi^{(1)}_3&=\tfrac{3}{2}-z+\tfrac{3}{2}z^2+2vz-\tfrac{5}{32}v^2
  -\tfrac{5}{16}\Big[2-8z+72z^2+6vz\Big]\log z
  -\tfrac{5}{16}\Big[v-v^2\Big]\log v+\ldots\,,
\end{aligned}
\end{equation}
and
\begin{equation}
\begin{aligned}
  \pi^{(2)}_0&=\tfrac{3}{4}-z+9z^2-\tfrac{3}{4}v\left(1-z\right)+\tfrac{3}{4}v^2+\ldots\,,\\[4pt]
  \pi^{(2)}_1&=-\tfrac{1}{2}+\tfrac{1}{2}v\left(1-z\right)-\tfrac{1}{2}v^2+\ldots\,,\\[4pt]
  \pi^{(2)}_2&=\tfrac{1}{4}z-\tfrac{33}{8}z^2+\tfrac{5}{32}v\left(1-4z\right)-\tfrac{15}{64}v^2
  -\tfrac{5}{32}\Big[1-4z+36z^2-v\left(1-z\right)+v^2\Big]\log z+\ldots\,,\\[4pt]
  \pi^{(2)}_3&=-\tfrac{5}{32}-\tfrac{1}{4}z+\tfrac{3}{8}z^2+\tfrac{5}{32}v\left(1-4z\right)-\tfrac{5}{32}v^2\nonumber\\
  &\quad+\tfrac{5}{8}\Big[z-9z^2\Big]\log z
  +\tfrac{5}{32}\Big[1-v\left(1-z\right)+v^2\Big]\log v+\ldots\,,
\end{aligned}
\end{equation}
where the dots denote terms of total order three and higher in $v$ and $z$.
 
The solutions $\pi^{(2)}_j$ are chosen such that, together with $\pi^{(1)}_j$, they solve the coupled system at $\varepsilon=0$, and the remaining free coefficients of the logarithmic solutions $\pi^{(1)}_2$ and $\pi^{(1)}_3$ are fixed  by demanding that the intersection matrix takes the form~\cite{Duhr:2024uid}
\begin{equation}
\label{eq:Sigma}
  \Sigma=\left(
\begin{array}{cccc}
 0 & 0 & 1 & 0 \\
 0 & 0 & 0 & 1 \\
 -1 & 0 & 0 & 0 \\
 0 & -1 & 0 & 0 \\
\end{array}
\right).
\end{equation}
With this choice, the Wronskian matrix reads
\begin{equation}
\label{eq:Wgenus2}
  W=\left(
\begin{array}{cccc}
 \pi^{(1)}_0 & \pi^{(1)}_1 & \pi^{(1)}_2 & \pi^{(1)}_3 \\
 \pi^{(2)}_0 & \pi^{(2)}_1 & \pi^{(2)}_2 & \pi^{(2)}_3 \\
 \partial_v \pi^{(2)}_0 & \partial_v\pi^{(2)}_1 & \partial_v\pi^{(2)}_2 & \partial_v\pi^{(2)}_3 \\
 \partial_v \pi^{(1)}_0 & \partial_v \pi^{(1)}_1 & \partial_v \pi^{(1)}_2 &\partial_v \pi^{(1)}_3 \\
\end{array}
\right),
\end{equation}
and the periods satisfy the quadratic (bilinear) relations
\begin{equation}
\label{eq:bilinear}
  W\,\Sigma\,W^{T}=Z\,.
\end{equation}
The entries of the matrix $Z$ can be determined either by combining
Griffiths transversality with the differential equations~\cite{Bonisch:2021yfw, Duhr:2024uid, Duhr:2025lbz},
or directly from series expansions of the periods.
In the latter approach, we expand the bilinear relations in~\cref{eq:bilinear}
to sufficiently high orders in $v$ and $z$.
Since the denominators appearing in these relations are inherited
from those of the differential equations, multiplying by a suitable
common denominator yields polynomial expressions in $v$ and $z$
up to the chosen truncation order.
 
Following~\cite{Duhr:2024uid}, we split the Wronskian into its semi-simple and unipotent parts, $W=W^\text{ss}W^\text{u}$, with
\begin{equation}
\label{eq:Ugenus2}
  W^\text{u} = \left(
\begin{array}{cccc}
 1 & 0 & \tau_1 & \tau_{12} \\
 0 & 1 & \tau_{12} & \tau_2 \\
 0 & 0 & 1 & 0 \\
 0 & 0 & 0 & 1 \\
\end{array}
\right) \, ,
\end{equation}
where the genus-two $\tau$-parameters are given by
\begin{equation}
    \tau_1 = \frac{\pi^{(1)}_2 \pi^{(2)}_1-\pi^{(1)}_1 \pi^{(2)}_2}{\pi^{(1)}_0 \pi^{(2)}_1-\pi^{(1)}_1 \pi^{(2)}_0}\,, \quad 
    \tau_{12} = \frac{\pi^{(1)}_2 \pi^{(2)}_0-\pi^{(1)}_0 \pi^{(2)}_2}{\pi^{(1)}_1 \pi^{(2)}_0-\pi^{(1)}_0 \pi^{(2)}_1} \,, \quad
    \tau_2 = \frac{\pi^{(1)}_3 \pi^{(2)}_0-\pi^{(1)}_0 \pi^{(2)}_3}{\pi^{(1)}_1 \pi^{(2)}_0-\pi^{(1)}_0 \pi^{(2)}_1} \, .
\end{equation}
The unipotent part preserves the intersection matrix, $W^\text{u}\,\Sigma\,(W^\text{u})^{T}=\Sigma$.
As we already stressed, one can understand this splitting as a decomposition of the maximal cuts into their leading singularities, contained in $W^\text{ss}$, and their pure parts, contained in $W^\text{u}$.
 
From here, the rotation into $\varepsilon$-factorised form then proceeds in three steps:
\begin{enumerate}
  \item We rotate the first four master integrals by the inverse of the semi-simple part, $(W^{\rm ss})^{-1}$ and normalise the fifth integral by $v r_z$ with $r_z=\sqrt{1-4 z}$.
  \item We align the local transcendental weights by rescaling our basis as
  \begin{equation}
  \label{eq:Teps}
    T_\varepsilon= \text{diag}(\varepsilon,\varepsilon,1,1,\varepsilon)\, .
  \end{equation}
  \item We remove the remaining terms of $\mathcal{O}(\varepsilon^0)$ by the rotation
  \begin{equation}
  \label{eq:Tgenus2}
  \hspace{-1cm}  T=\left(
  \begin{array}{ccccc}
   1 & 0 & 0 & 0 & 0 \\
   0 & 1 & 0 & 0 & 0 \\
   -\frac{512}{5} G_1^2-\frac{1024}{5}G_1 f_1+f_3 & -G_3-\frac{512}{5} G_1 G_2-\frac{1024}{5}G_1 f_2 +f_5 & 1 & 0 & -\frac{1024}{5} G_1 \\
   G_3-\frac{512}{5} G_1 G_2-\frac{1024}{5}G_2 f_1+f_5 & -\frac{512}{5} G_2^2-\frac{1024}{5}G_2 f_2+f_4 & 0 & 1 & -\frac{1024}{5} G_2 \\
   G_1+f_1 & G_2+f_2 & 0 & 0 & 1 \\
  \end{array}
  \right).
  \end{equation}
\end{enumerate}
The functions entering $T$ are
\begin{equation}
  f_j = -\frac{r_z\left((vz-1)^2\,\pi^{(1)}_{j-1}+4v(vz+1)\,\pi^{(2)}_{j-1}\right)}{4v(4z-1)}\,, \qquad j=1,2\,,
\end{equation}
and
\begin{equation}
  f_3=\Phi(0,0)\,,\qquad f_4=\Phi(1,1)\,,\qquad f_5=\Phi(0,1)\,,
\end{equation}
where $\Phi$ is the symmetric bilinear form
\begin{equation}
  \Phi(k,l) = -\frac{8
  \Big[\,4v\,A\,\pi^{(2)}_k\pi^{(2)}_l
  -2B\left(\pi^{(1)}_k\pi^{(2)}_l+\pi^{(2)}_k\pi^{(1)}_l\right)  +z\,C\,\pi^{(1)}_k\pi^{(1)}_l\,\Big]}{5v(4z-1)\left(v^2z^2-2vz+v+1\right)}\,,
\end{equation}
with the polynomials
\begin{align}
 A &= 2v^5z^3\left(120z^3-122z^2+37z-3\right) \nonumber\\
   &\quad +v^4z\left(-824z^4+1150z^3-602z^2+132z-9\right) +v^3\left(896z^4-928z^3+320z^2-52z+6\right) \nonumber\\
   &\quad -2v^2\left(136z^3-78z^2+33z-9\right)-6v\left(8z^2+2z-3\right)+8z+6\,, \nonumber\\[4pt]
 B &= 2v^6z^4\left(56z^2-27z+3\right)+v^5z^2\left(-370z^3+344z^2-100z+9\right) \nonumber\\
   &\quad +v^4z\left(358z^3-234z^2+48z-3\right)-2v^3\left(18z^3+17z^2-10z+1\right) \nonumber\\
   &\quad -6v^2\left(14z^2-8z+1\right)+v\left(22z-6\right)-2\,,\\[4pt]
 C &= 2v^6z^4\left(30z^2+5z-3\right)+v^5z^2\left(-296z^3+162z^2+14z-9\right) \nonumber\\
   &\quad +2v^4z\left(290z^3-278z^2+111z-15\right)+v^3\left(-560z^3+596z^2-206z+23\right) \nonumber\\
   &\quad +v^2\left(260z^2-222z+40\right)+v\left(10-40z\right)-4\,. \nonumber
\end{align}
 $G_1$, $G_2$ and $G_3$ are new transcendental functions (referred to as $\varepsilon$-functions in~\cite{Duhr:2024uid}), defined by the first-order differential equations
\begin{equation}
\begin{aligned}
  \partial_v G_j &= \frac{(1-vz)\,\pi^{(1)}_{j-1}+2v\,\pi^{(2)}_{j-1}}{4v^2\,r_z}\,,\qquad j=1,2\,,\\
  \partial_v G_3 &= \frac{512}{5}\left(G_1\,\partial_vG_2-G_2\,\partial_vG_1\right)
  +\frac{16\,\left(h_1+h_2+h_3\right)}{5(4z-1)(vz-v-1)}\,,
\end{aligned}
\end{equation}
where
\begin{equation}
\begin{aligned}
  h_1 &= -\frac{(vz-1)\,a_1}{v^2(2vz-v-2)}
        \left(\pi^{(1)}_0\,\partial_v\pi^{(1)}_1-\pi^{(1)}_1\,\partial_v\pi^{(1)}_0\right)\,,\\[4pt]
  h_2 &= \frac{4(vz-1)\,a_2}{v(2vz-v-2)}
        \Big(\pi^{(1)}_0\,\partial_v\pi^{(2)}_1-\pi^{(1)}_1\,\partial_v\pi^{(2)}_0
        +\pi^{(2)}_0\,\partial_v\pi^{(1)}_1-\pi^{(2)}_1\,\partial_v\pi^{(1)}_0\Big)\,,\\[4pt]
  h_3 &= -\frac{64v^2z^2(2z-1)(2vz-v-2)^2}{vz-1}
        \left(\pi^{(2)}_0\,\partial_v\pi^{(2)}_1-\pi^{(2)}_1\,\partial_v\pi^{(2)}_0\right)\,,
\end{aligned}
\end{equation}
and
\begin{equation}
\begin{aligned}
  a_1 =& -2+5v(2z-1)+v^2\left(-3+28z-52z^2\right) +v^3\left(1+23z-134z^2+116z^3\right) \\
      &\quad +v^4\left(1+2z-37z^2+76z^3-106z^4\right) +v^5z\left(-3+29z-79z^2+35z^3+34z^4\right)\,,\\[4pt]
  a_2 =& 1+v(3-4z)+3v^2\left(1-4z+2z^2\right) +v^3\left(1-12z+47z^2-68z^3\right)
            \\
            &+v^4z\left(-4+33z-86z^2+65z^3\right)\,.
\end{aligned}
\end{equation}
Since $\Phi(k,l)=\Phi(l,k)$, the symmetry under the exchange of the two solutions free of logarithms is manifest:
\begin{equation}
 f_1\big|_{\pi^{(i)}_0\leftrightarrow\pi^{(i)}_1}=f_2\,,\qquad
 f_3\big|_{\pi^{(i)}_0\leftrightarrow\pi^{(i)}_1}=f_4\,,\qquad
 f_5\big|_{\pi^{(i)}_0\leftrightarrow\pi^{(i)}_1}=f_5\,,\qquad
 G_1\big|_{\pi^{(i)}_0\leftrightarrow\pi^{(i)}_1}=G_2\,.
\end{equation}
This concludes the $\varepsilon$-factorisation of the genus-two sector.
Finally, we find it convenient to trade $\pi^{(1)}_1$ and $\pi^{(2)}_1$ for the combinations
\begin{equation}
    \alpha_1= \frac{\pi^{(2)}_1}{\pi^{(1)}_0 \pi^{(2)}_1- \pi^{(1)}_1\pi^{(2)}_0}\,,\qquad
    \alpha_2= \frac{\pi^{(2)}_0}{\pi^{(1)}_1 \pi^{(2)}_0- \pi^{(1)}_0\pi^{(2)}_1}\,,
\end{equation}
which avoids spurious denominators in the full differential equation matrix.

Once a full canonical  basis is known, the only missing ingredient to be able to write an analytic
solution for the master integrals is a boundary condition. 
Below, we describe how to determine boundary data at the threshold. These boundary constants allow us to construct, on the one hand, formal solutions for all master integrals in terms of iterated integrals and, on the other hand, high-precision series-expansion representations centred at the threshold point.
For the numerical evaluation of the master integrals and the scattering amplitudes across the whole phase space, we opt then for a mixed approach based on the combination of the  series expansion close to threshold described below, augmented by the choice of a set of Bernoulli-like variables, 
and the solution of the system by numerical transport following~\cite{Czakon:2026tog}, see the discussion in~\cref{sec:numerics}.

\subsection{Solution at threshold}
\label{sec:blowup}

At the production threshold, $s=4m_t^2$ and $t=-m_t^2$, multiple singularities overlap, and one has to perform a blow-up to disentangle them properly. The singularities can be read off from the denominators of the differential equations, including the ones appearing in the equations defining the periods and the new transcendental functions. 
We find that all singularities at the threshold point can  be resolved 
by means of the change of variables
\begin{equation}
\label{eq:blowup}
    s= 4m_t^2 (1 + x_1)\,, \qquad\qquad
    t = -m_t^2 (1 + x_1 x_2)\,,
\end{equation}
such that all vanishing denominators factorise a power of $x_1$.
We use these variables to expand the solution of the differential equations around the threshold.

The complete solutions of the canonical differential equations 
require boundary conditions for the master integrals. We fix them at the production threshold, $s=4m_t^2$ and $t=-m_t^2$, i.e., at $x_1=x_2=0$ in the blow-up variables of \cref{eq:blowup}. Since the master integrals are singular at threshold, where they contain logarithms of $x_1$, we cannot obtain the boundary values by evaluating the integrals exactly at that point. We therefore proceed in a semi-numerical way:
\begin{enumerate}
    \item We compute all the master integrals numerically with \texttt{AMFlow}~\cite{Liu:2017jxz,Liu:2022chg} at a phase-space point close to threshold, $(x_1,x_2)=(10^{-4},2\cdot10^{-4})$ in units of $m_t=1$, with a precision of about $100$ digits, and of up to about $300$ digits for the subset of integrals that enter the constants that are most difficult to fix.
    \item In the vicinity of the threshold, we solve the canonical differential equations in terms of generalised series expansions in $x_1, x_2$, and logarithms thereof. These depend on one constant of integration per master integral and per order in $\varepsilon$. We match these solutions to the numerical values above, and in this way, transport the boundary point exactly back to the threshold point.
    By performing the expansion to suitably high orders in $x_1, x_2$, we can make sure that the truncation error at the matching point is much smaller than the precision of the numerical input.
    \item We fit the resulting numerical constants with integer-relation algorithms (PSLQ) on an ansatz of transcendental numbers up to weight four, built from $\pi$, $\log2$, $\log3$, $\zeta_3$, $\mathrm{Li}_2(-1/2)$, $\mathrm{Li}_3(1/3)$ and $\mathrm{Li}_3(-1/2)$, and their products, as suggested by generalised polylogarithms with letters in $\{0,1,2,3\}$ evaluated at unity~\cite{Duhr:2019tlz}. We also find that two further constants require the 
    letters $1\pm\sqrt{2}$.
\end{enumerate}
Most of the constants are rational numbers, or combinations of zeta values, logarithms, and polylogarithms of the type described above. A remaining set of constants, related to the elliptic curves and to the genus-two curve evaluated at threshold, cannot be easily 
matched to any known basis of numbers. For these, we keep their numerical values, 
known to high precision, and we have verified with PSLQ that they are 
linearly independent of each other and of the constants of the ansatz above.
We provide the results for all 417 master integrals evaluated at threshold in the Zenodo repository associated to this paper~\cite{Zenodo_ttbar}.

The analytic expression for the master integrals in terms of iterated integrals also allows us to test for extra relations among the $\varepsilon$-coefficients of the 417 master integrals.
As already hinted at above, in this way we can prove that, up to the order relevant for the finite remainder of the scattering amplitudes, the coefficients of the masters satisfy $65$ linear relations, leaving only $352$ independent objects. 
This also allows us to write the two-loop results in terms of independent coefficients, which we can use to perform UV renormalisation and IR subtraction analytically, as described in the next section.


%
\section{Ultraviolet renormalisation and infrared structure}
\label{sec:ren}

\subsection{Ultraviolet renormalisation}
\label{sec:uv}

We start from the bare scattering amplitude, expanded in the bare strong coupling constant $\alpha_s^{b}$ with $n_f=n_l+n_h$ active flavours, where $n_l$ is the number of light quarks and $n_h$ the number of heavy quarks, as
\begin{equation}
\label{eq:bareamplitude}
\boldsymbol{\mathcal{A}}^{b}=4\pi\alpha_s^{b}\sum_{\ell=0} \left(\frac{\alpha^{b}_s}{4\pi}\right)^\ell
\boldsymbol{\mathcal{A}}^{b,(\ell)}\,.
\end{equation}
In our case, we have $n_l = 5$ and $n_h=1$.
We renormalise
the top-quark mass, the wave functions of the external gluons, light and heavy quarks in the on-shell scheme.
In particular, we implement the top-quark mass renormalisation through a mass counterterm insertion in the lower-order amplitudes.
The strong coupling constant is instead renormalised in the $\overline{\rm MS}$ scheme with $n_f$ flavours, and the top quark is subsequently decoupled, so that the renormalised amplitude is expressed in terms of $\alpha^{(n_l)}_s(\mu)$ with $n_l$ active flavours at the renormalisation scale $\mu$. The explicit decoupling relation reads $\alpha_s^{(n_f)}=\zeta_{\alpha_s}\,\alpha_s^{(n_l)}$, where the decoupling constant $\zeta_{\alpha_s}$ admits itself a perturbative expansion in $\alpha_s^{(n_l)}$,
\begin{equation}
    \zeta_{\alpha_s} = \sum_{n=0}\left(\frac{\alpha_s^{(n_l)}}{4\pi}\right)^n\zeta^{(n)} 
    \quad\quad\text{with}\quad\quad
    \zeta^{(0)} = 1
    \,. \label{eq:decoupling}
\end{equation}
We provide the explicit expression of the coefficients $\zeta^{(n)}$ in~\cref{app:renorm}.
The relations between the bare parameters and fields, indicated with a superscript $b$, and the renormalised ones 
are given by
\begin{gather}
    \,\alpha_s^{b}
    = S_\varepsilon^{-1}\,\alpha_s^{(n_f)}Z_{\alpha_s}\,,
    \label{eq:alphasren}\\
    m_t^{b} = Z_m\,m_t\,,
    \label{eq:massren}\\
    \psi_t^b = \sqrt{Z_t}\,\psi_t\,,\qquad
    \psi_q^b = \sqrt{Z_q}\,\psi_q\,,\qquad
    G_\mu^b = \sqrt{Z_G}\,G_\mu \,,
    \label{eq:fieldrenorm}
\end{gather}
where $\psi_{t,q}$ refer to the heavy and light quark fields, $G_\mu$ to the gluon field and $S_\varepsilon=(4\pi)^\varepsilon e^{-\gamma_E\varepsilon}$.
Each renormalisation constant $Z_X$ is then expanded in $\alpha_s^{(n_f)}$ as
\begin{equation}
\label{eq:renormparfields}
    Z_{X} = \sum_{n=0}\left(\frac{\alpha_s^{(n_f)}}{4\pi}\right)^n Z^{(n)}_{X} \quad\quad\text{with}\quad\quad
    Z^{(0)}_X = 1\,,
\end{equation}
where $X$ indicates a generic parameter or field. As for the decoupling constant, the perturbative coefficients 
are collected in \cref{app:renorm}.

We implement the top mass renormalisation  
by expanding the bare propagators around the renormalised mass, which is effectively equivalent to inserting appropriate counter-term Feynman rules 
in the lower-loop amplitudes.
In particular, we denote by $\boldsymbol{\mathcal{A}}^{(\ell)}_{[k]}$ the bare $\ell$-loop amplitude with $k$ insertions of the mass counterterm on the internal top-quark propagators. 
With this notation, inserting \cref{eq:decoupling,eq:alphasren,eq:massren,eq:fieldrenorm} into \cref{eq:bareamplitude}, 
we can finally express the renormalised amplitude through an $\alpha^{(n_l)}_s$ expansion as
\begin{equation}
\boldsymbol{\mathcal{A}}=4\pi\alpha_s^{(n_l)}\sum_{\ell=0} \left(\frac{\alpha_s^{(n_l)}}{4\pi}\right)^\ell
\boldsymbol{\mathcal{A}}^{(\ell)}\,.
\label{eq:renamplitude}
\end{equation}
The explicit expressions of the renormalised loop coefficients read
\begin{align}
    \boldsymbol{\mathcal{A}}^{(0)} &= \boldsymbol{\mathcal{A}}^{b,(0)}\,,\label{eq:uv0}\\
    \boldsymbol{\mathcal{A}}^{(1)} &= \boldsymbol{\mathcal{A}}^{b,(1)} + Z_m^{(1)}\,\boldsymbol{\mathcal{A}}^{(0)}_{[1]} + 
    \left(Z_{\alpha_s}^{(1)} +\zeta^{(1)} + Z_{\phi}^{(1)} + Z_{t}^{(1)}\right)\boldsymbol{\mathcal{A}}^{b,(0)}\,,\label{eq:uv1}\\
    \boldsymbol{\mathcal{A}}^{(2)} &= \boldsymbol{\mathcal{A}}^{b,(2)} 
    + \left(Z_m^{(1)}\right)^{2}\boldsymbol{\mathcal{A}}^{(0)}_{[2]}
    + \left[Z_m^{(2)}+Z_m^{(1)}
    \left(Z_{\alpha_s}^{(1)}+ 2\zeta^{(1)} + Z_{\phi}^{(1)} + Z_{t}^{(1)} \right)\right]\boldsymbol{\mathcal{A}}^{(0)}_{[1]}\nn \\
    &+ Z_m^{(1)}\,\boldsymbol{\mathcal{A}}^{(1)}_{[1]}
    + \left(2Z_{\alpha_s}^{(1)} +2\zeta^{(1)}+ Z_{\phi}^{(1)} + Z_{t}^{(1)} \right)\boldsymbol{\mathcal{A}}^{b,(1)} 
    + \bigg[ \left(Z^{(1)}_{\phi}+Z^{(1)}_{t}\right)Z^{(1)}_{\alpha_s} \nn \\
    & + Z^{(1)}_{\phi}Z^{(1)}_{t} + 2 \left(Z_{\alpha_s}^{(1)} + Z_{\phi}^{(1)} + Z_{t}^{(1)}\right)\zeta^{(1)}
    + Z_{\alpha_s}^{(2)} +\zeta^{(2)} + Z_{\phi}^{(2)} + Z_{t}^{(2)}\bigg]\boldsymbol{\mathcal{A}}^{b,(0)}\,,\label{eq:uv2}
\end{align}
where $\phi =q,G$ is a generic field associated with the incoming particles.
These relations hold for each form factor $\mathbf{F}_i$, $\mathbf{G}_i$ and for the polarised amplitudes, since the latter involve simple linear combinations of the scalar form factors. 
Notice that, in the $q\bar q$ channel the tree-level amplitude does not contain top-quark propagators, so that all the amplitudes $\boldsymbol{\mathcal{A}}^{(0)}_{[k]}$ vanish and the mass counterterm enters only at two loops through $\boldsymbol{\mathcal{A}}^{(1)}_{[1]}$.

At one loop, we compared the bare and the renormalised squared amplitudes, summed over colours and polarisations against \texttt{OpenLoops~2}~\cite{Buccioni:2019sur}, and found agreement.

Alongside our results for the one-loop bare form factors through $\mathcal{O}(\varepsilon^4)$ and the two-loop
ones through $\mathcal{O}(\varepsilon^2)$, we also publish the quantities ${\mathcal{A}}^{(\ell)}_{[k]}$
necessary to perform the mass renormalisation in~\cite{Zenodo_ttbar}.

\subsection{Infrared structure}
\label{sec:ir}
The infrared divergences of renormalised amplitudes are universal, and their structure for scattering
amplitudes with arbitrarily many massive external legs at two loops is known~\cite{Catani:1998bh,Becher:2009kw,Becher:2009cu,Ferroglia:2009ii}.\footnote{
Recently, results at three-loops for scattering amplitudes involving a single external massive particle have
been presented~\cite{Liu:2022elt,Gardi:2025lws}.}
In particular, writing the renormalised amplitude as in~\cref{eq:renamplitude},
the finite remainder  can be written as
$$\boldsymbol{\mathcal{A}}^{\rm fin}=\mathbf{Z}^{-1}\boldsymbol{\mathcal{A}}\,,$$ 
where $\mathbf{Z}$ is an operator in colour space. 
By expanding this operator in powers of the strong coupling constant up to two loops
\begin{equation}
\mathbf{Z}=\mathbf{1}+ 
\left(\frac{\alpha_s^{(n_l)}}{4 \pi} \right) \mathbf{Z}^{(1)}+ \left( \frac{\alpha_s^{(n_l)}}{4 \pi} \right)^2\mathbf{Z}^{(2)}\,,
\end{equation}
one obtains
\begin{align}
    \boldsymbol{\mathcal{A}}^{{\rm fin},(1)} &= \boldsymbol{\mathcal{A}}^{(1)} - \mathbf{Z}^{(1)}\boldsymbol{\mathcal{A}}^{(0)}\,, \nn\\
    \boldsymbol{\mathcal{A}}^{{\rm fin},(2)} &= \boldsymbol{\mathcal{A}}^{(2)} - \mathbf{Z}^{(1)}\boldsymbol{\mathcal{A}}^{(1)} - \left(\mathbf{Z}^{(2)}-\mathbf{Z}^{(1)}\mathbf{Z}^{(1)}\right)\boldsymbol{\mathcal{A}}^{(0)}\,,
\end{align}
with $\mathbf{Z}^{(\ell)}$ explicitly given by
\begin{align}
    \mathbf{Z}^{(1)} &= \frac{\boldsymbol{\Gamma}'_1}{4\varepsilon^2}+\frac{\boldsymbol{\Gamma}_1}{2\varepsilon}\,,\\
    \mathbf{Z}^{(2)} &= 
    \frac{\boldsymbol{\Gamma}_1'^{\,2}}{32\varepsilon^4}
    +\frac{\boldsymbol{\Gamma}'_1}{8\varepsilon^3}\left(\boldsymbol{\Gamma}_1-\frac32\,\beta^{(n_l)}_0\right)+\frac{\boldsymbol{\Gamma}_1}{8\varepsilon^2}\left(\boldsymbol{\Gamma}_1-2\beta^{(n_l)}_0\right)+\frac{\boldsymbol{\Gamma}'_2}{16\varepsilon^2}+\frac{\boldsymbol{\Gamma}_2}{4\varepsilon}\,.
    \label{eq:zirtwoloop}
\end{align}
Here, $\beta^{(n_l)}_0$ is the first coefficient of the QCD $\beta$-function
with $n_l$ flavours, and its explicit expression is given in \cref{app:renorm}.

The operators $\boldsymbol{\Gamma}_k$ and $\boldsymbol{\Gamma}'_k$ entering the soft-anomalous dimension~\cite{Ferroglia:2009ii} involve massless and massive coloured partons. 
For the case of $2\to2$ scattering of two massless partons and a $t\bar t$ pair, 
these operators are built from the cusp anomalous dimension $\gamma_{\rm cusp}(\alpha_s)$ and its generalisation to massive partons $\gamma_{\rm cusp}(\beta_{34},\alpha_s)$, 
from the gluon and light-quark collinear anomalous dimensions, $\gamma_{G}$ and $\gamma_{q}$ respectively, the heavy-quark collinear anomalous dimension $\gamma_{Q}$, and from the dipole colour correlators $\mathbf{T}_i\cdot\mathbf{T}_j$. 
Starting from two loops, $\boldsymbol{\Gamma}$ contains also tripole correlations $\mathbf{T}_i\cdot\mathbf{T}_j\cdot\mathbf{T}_k$. 
The explicit expressions for the $\boldsymbol{\Gamma}$ operators, as well as those for the anomalous dimensions 
are collected in \cref{app:renorm}.

We have verified that for both the $gg$ and the $q\bar{q}$ channels the renormalised form factors have the expected pole structure, namely that with the anomalous dimensions above, all the $\varepsilon$ poles of $\boldsymbol{\mathcal{A}}^{\rm fin,(1)}$ and $\boldsymbol{\mathcal{A}}^{\rm fin,(2)}$ cancel throughout, leaving a
result free of infrared singularities.


%
\section{Numerical results}
\label{sec:numerics}

We make our analytic results publicly accessible through the Zenodo repository~\cite{Zenodo_ttbar}, 
as well as through the \texttt{C++} code~\texttt{hqpAmp}~\cite{hqpamp}. 
In the first one, we provide the analytic results, whereas the second, which can be downloaded via
$$
\texttt{git clone https://gitlab.com/hqpamp/hqpamp}\,,
$$
allows  for the numerical evaluation of the polarised $t\bar{t}$ scattering amplitudes up to two loops.
In the same git repository, the interested user will also find \texttt{hqpInt}, 
a separate code for the evaluation of the canonical master integrals up to $\mathcal{O}(\varepsilon^6)$.
\texttt{hqpAmp} has immediate applications in predictions of $t\bar{t}$-production cross sections at hadron colliders at NNLO in QCD. Detailed instructions on the analytic results and on the usage of the code are provided in
the corresponding repositories.

To ensure a numerically reliable point-by-point evaluation,~\texttt{hqpAmp} combines two complementary strategies:
a high-precision series expansion of the scattering amplitude around the production threshold,
and the numerical transport of the integrals from suitably chosen boundary points to the desired
phase-space point by directly solving the canonical differential equations.
We discuss each strategy in turn below.

\begin{figure}[t]
  \centering
  \includegraphics[width=\textwidth]{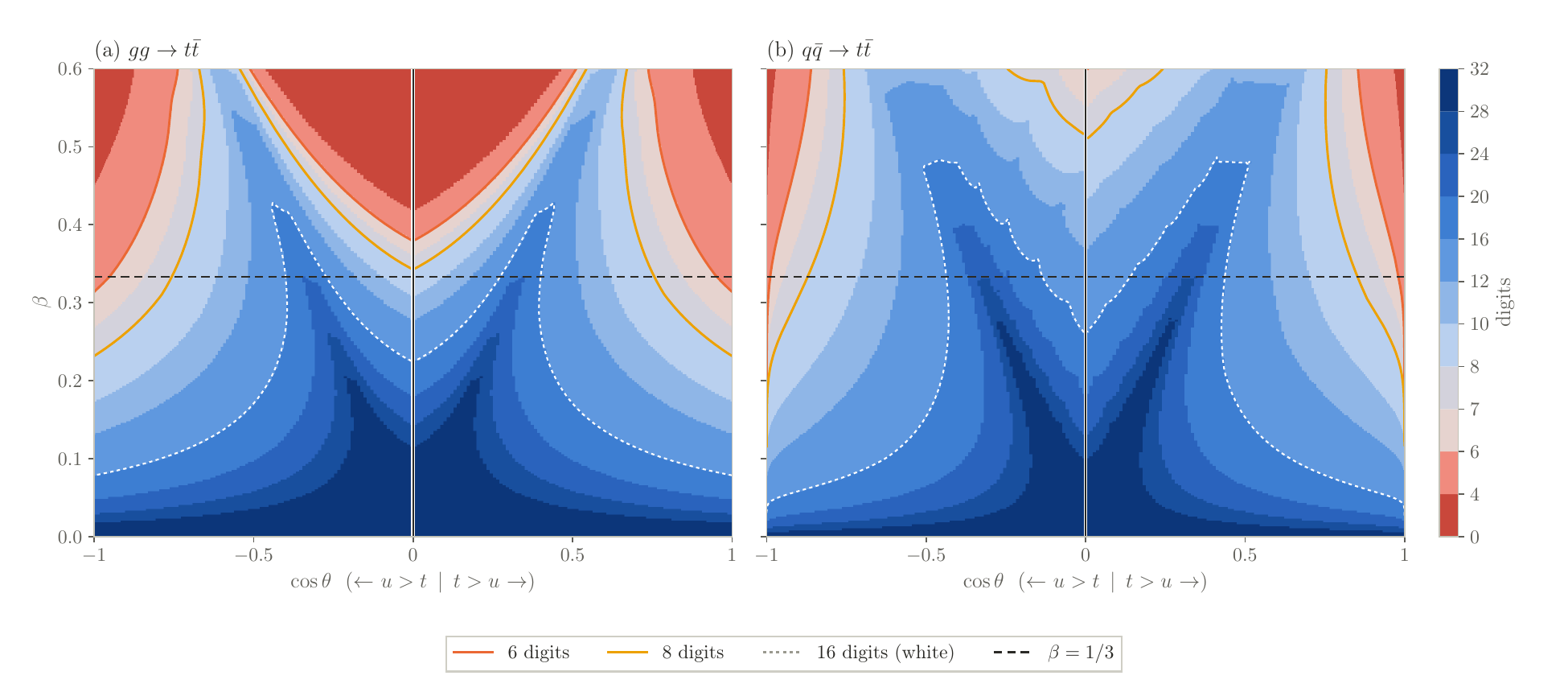}
  \caption{Self-convergence of the threshold expansions for $gg\to t\bar t$ (a) and $q\bar q\to t\bar t$ (b): digits to which the two-loop finite remainder is constant when the last two orders are dropped. Below the dashed line $\beta = 1/3$ the threshold expansions are used in the numerical code.}
  \label{fig:threshold-convergence}
\end{figure}

\paragraph{Series expansions.} We provide a series expansion of both the canonical master integrals and form factors at threshold up to $30$ orders in the blow-up variables $(x_1,x_2)$ of \cref{sec:blowup}. 
In order to improve the convergence of the series, we introduce Bernoulli-type variables. 
Due to the different singularity structures of the two partonic channels, we employ 
two different Bernoulli variables, empirically chosen to maximise the domain of convergence around the production threshold.
In particular, setting $m_t=1$, we use the expansion variables
\begin{equation}
    \begin{aligned}
        gg \to t\bar{t}: \quad
        &(r,x_2),
        & \text{with}\quad r &= \log\left(1+\frac{4x_1}{2-x_1 x_2}\right), \\[1ex]
        q\bar{q} \to t\bar{t}: \quad
        &\begin{cases}
            (z,x_2), & t>u, \\
            (z,x_3), & u>t,
        \end{cases}
        & \quad\text{with}\quad z &= \frac{1}{4}\log(1+4x_1),
    \end{aligned}
\end{equation}
and
\begin{equation}
    x_3 = -\frac{4(m_t^2+u)}{s-4 m_t^2}
        = \left.x_2\right|_{t\leftrightarrow u}.
\end{equation}
As for the $q\bar{q}\to t\bar{t}$ channel, 
we use two different expansion variables since the convergence is not symmetric in $u$ and $t$. 
For the gluon amplitude instead, we can simply use Bose symmetry to evaluate the series only at $t>u$ and map it to
the region $t<u$. 
We find that these expansions yield at least six accurate digits in the amplitudes at all points with $\beta<1/3$, covering a significant portion of the physical phase space.
A graphical representation of the convergence of the series in the Bernoulli variables is provided in~\cref{fig:threshold-convergence}.

\paragraph{Numerical integration of the differential equations.}
For points away from the threshold region, we employ the numerical code published in~\cite{Czakon:2026tog}, which solves the differential equations
by transporting the master integrals from a boundary point to the target point by solving their canonical differential equations together with auxiliary equations for the algebraic and transcendental functions entering the connection matrix.\footnote{Earlier versions of 
this program have also been successfully used in other
collider scattering amplitudes, see e.g.~\cite{Czakon:2020vql,Badger:2025ljy}.}
This approach relies on the fact that all algebraic or transcendental functions appearing in a system of differential equations in canonical
form must, by construction, satisfy rational differential equations. This is because they are defined as solutions of rational coupled blocks of differential equations in the original, non-canonical equations.
Importantly, only boundary conditions for the canonical integrals have to be fixed to physical values, while the boundary values chosen for the evolution of the algebraic and transcendental functions appearing in the canonical basis can be fixed almost entirely arbitrarily, consistently
with the relations adopted to simplify those functions, see the discussion in~\cite{Czakon:2026tog}. 

Provided the boundaries for the canonical integrals 
are available up to the desired $\varepsilon$-order, 
there are no limitations to extend the evaluation of the master integrals to arbitrarily high $\varepsilon$-orders.
As for the two-loop scattering amplitudes required for cross-section predictions, we need 
up to order $\varepsilon^4$, and, as said before, only 352 integrals need to be evaluated.
Furthermore, thanks to the availability of the analytic results expressed in terms of iterated integrals, we can
identify which differential kernels drop at the amplitude level and set them to zero from the get go, thus speeding up the repeated numerical evaluations of the connection matrix. 

\begin{figure}[h!]
  \centering
  \includegraphics[width=\textwidth]{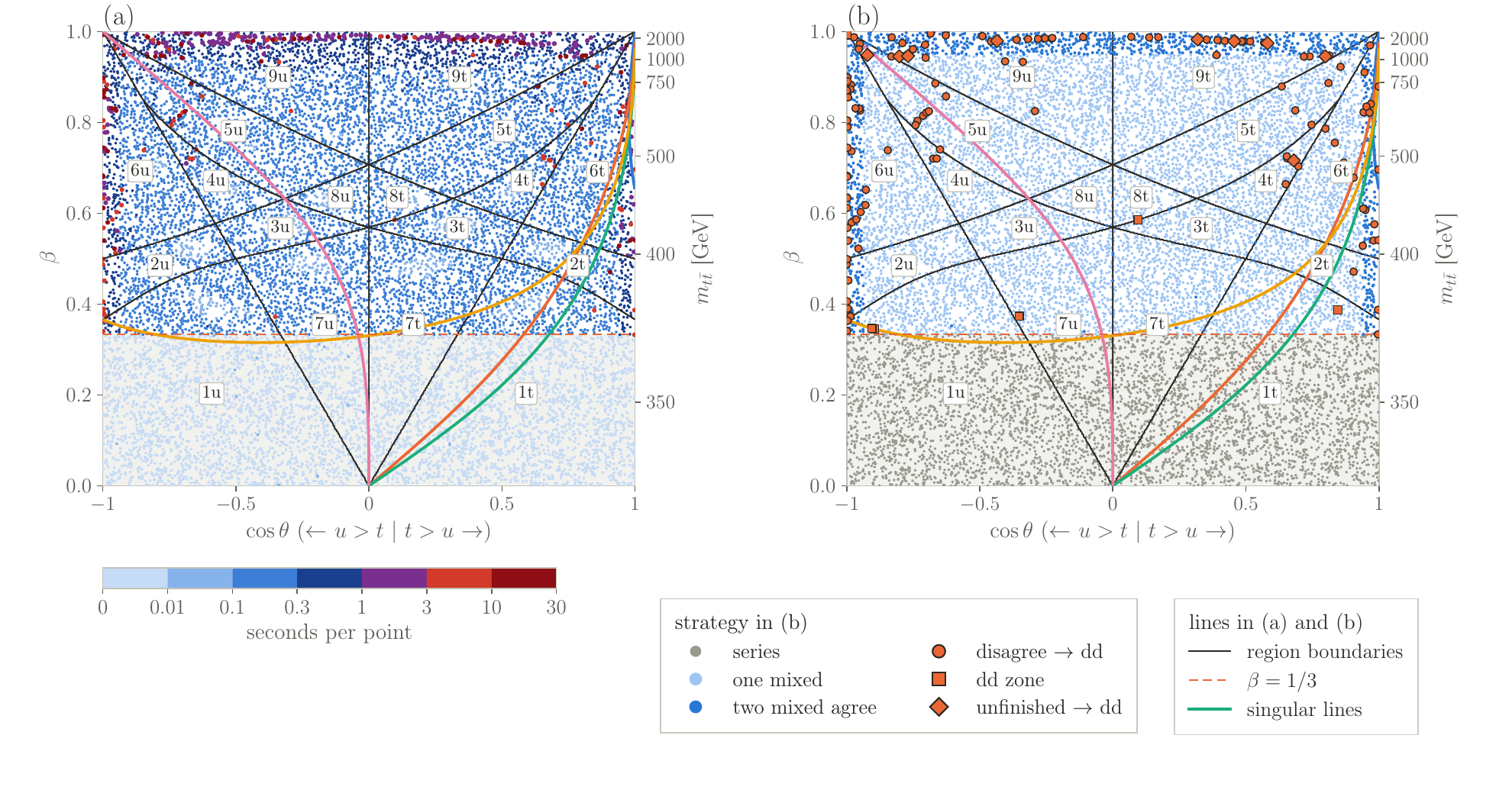}
  \caption{Evaluation time (a) and strategy (b) of the two-loop amplitudes for
  both channels on 15000 points uniformly distributed in $\beta$ and
  $\cos\theta$ (7500 with $t>u$ and 7500 with $u>t$). Black lines delimit the regions $1t$--$9t$ and their
  $t \leftrightarrow u$ images $1u$--$9u$, coloured lines are the singular
  lines of the elliptic and genus-two periods; below the dashed line
  $\beta = 1/3$, the threshold expansions are used.}
  \label{fig:timing-strategy}
\end{figure}

As illustrated in~\cref{fig:timing-strategy}, 
the differential equations contain spurious singular lines which cross into the physical region and can compromise the numerical stability of  the connection matrices.
Through its interface with the \texttt{QD} library~\cite{QD},
\texttt{hqpAmp} supports both double and double-double precision, corresponding to the \texttt{dd\_real} type of the \texttt{QD} library, and allows to 
select the working precision at runtime. 
With this, we present an evaluation strategy
which increases the precision near spurious singular lines and in the forward and backward regimes, in order to meet a target accuracy. 
We aim at 6 digits for all polarised amplitudes \emph{relative to the largest one}. This choice is made to avoid that critically small polarised amplitudes, whose weight on typical observables is expected to be tiny, force the code to switch more often than necessary to double-double precision. In addition, we still require that even the smallest polarised amplitudes are evaluated with at least 2 significant digits.
Since transporting the integrals from a boundary point to the target kinematic point accounts for most of the evaluation time, we perform the transport in double precision and assemble the amplitudes in double-double precision at negligible additional cost. We refer to this setup as \emph{mixed} precision.
At critical points instead, the transport is also upgraded to double-double precision. 

We discuss the details and effectiveness of this strategy below, but 
we stress that all parameters controlling the numerical precision can be adjusted by the user as explained in detail in~\cite{hqpamp}.

\begin{enumerate}
  \item \emph{Refused points.} Points with 
        $\sin\theta < 10^{-4}$ are rejected since close to the forward and backward edges, 
        even double-double precision returns unreliable results.
        We also add a tunable technical cut-off for points with 
        $\beta < 10^{-6}$, since at threshold value $\beta=0$
        loop amplitude are singular.
        
  \item \emph{Threshold expansion.} For $\beta < 1/3$, the scattering amplitudes for both
        partonic channels are obtained from their series expansions around threshold, 
        thus no master integrals need to be transported.
        \label{it:threshold}
  \item \emph{Points on a singular line.} 
      When a kinematic point lies exactly on a singular line, the amplitude is regular, 
      but none of the individual terms entering its assembly are.
      In this case, we displace the point by equal amounts on both sides of the line and average the two mixed
        evaluations, such that the term linear in the displacement cancels out. The displacement relative to the invariants
        is of order $10^{-4}$ and is tunable by the user.
  \item \emph{The dd zone.} Close to a singular line, at a relative distance below
        $10^{-5}$, or close to the edges of the phase space, $\sin\theta < 10^{-2}$, 
        the transport is performed in double-double precision, with a dynamic tolerance which decreases with the distance from the singular line and with the relative size of the smallest polarised amplitude.
  \item \emph{Two mixed evaluations.} 
        For $\sin\theta < 0.35$ or $\beta > 0.95$,
        at a relative distance below $10^{-4}$ from a singular line, or if the
        first transport required more than $100$ steps, we repeat the mixed
        evaluation along a different integration path. We accept the result if
        both evaluations for the worst polarised amplitude relative to the largest one agree to $10^{-6}$ and if each polarised amplitude agrees with itself 
        to at least $1\%$. 
        If one of these two conditions is not met, or if a mixed evaluation
        needs more than $4\times10^{4}$ evaluations of the connection matrix,\footnote{This number loosely corresponds to $5$s on the machine used for these tests.} 
        we upgrade the point to double-double precision.
  \item \emph{One mixed evaluation.} At all other points, a single mixed
        evaluation is used.
        \label{it:onemixed}
\end{enumerate}

As one can see in~\cref{fig:timing-strategy}, the vast majority of points are handled by steps 
\ref{it:threshold} and \ref{it:onemixed}, while double-double precision is only required in narrow strips
near the edges of the phase space and close to singular lines. 

The evaluation times reported in the following are for both channels combined,
on one core of an Apple MacBook Pro with an M1 Pro chip.
For points with $\beta<1/3$, where the threshold expansions apply, 
the evaluation takes $7\,$ms per point on average. 
For all other points, which require the numerical transport, the average evaluation time is
$0.39\,$s (median $0.17\,$s). Averaged over a sample that is uniform in $\beta$
and $\cos\theta$ over the whole phase space, where one third of the points lie below $\beta=1/3$, the mean evaluation time is $0.26\,$s per point (median $0.13\,$s).
We note that on our sample, which is uniformly distributed in phase space and therefore not representative of a realistic Monte Carlo run, 
$0.9\%$ of the points are evaluated in double-double precision 
and they take $36\%$ of the total time. 

\begin{figure}[h]
  \centering
  \includegraphics[width=1\textwidth]{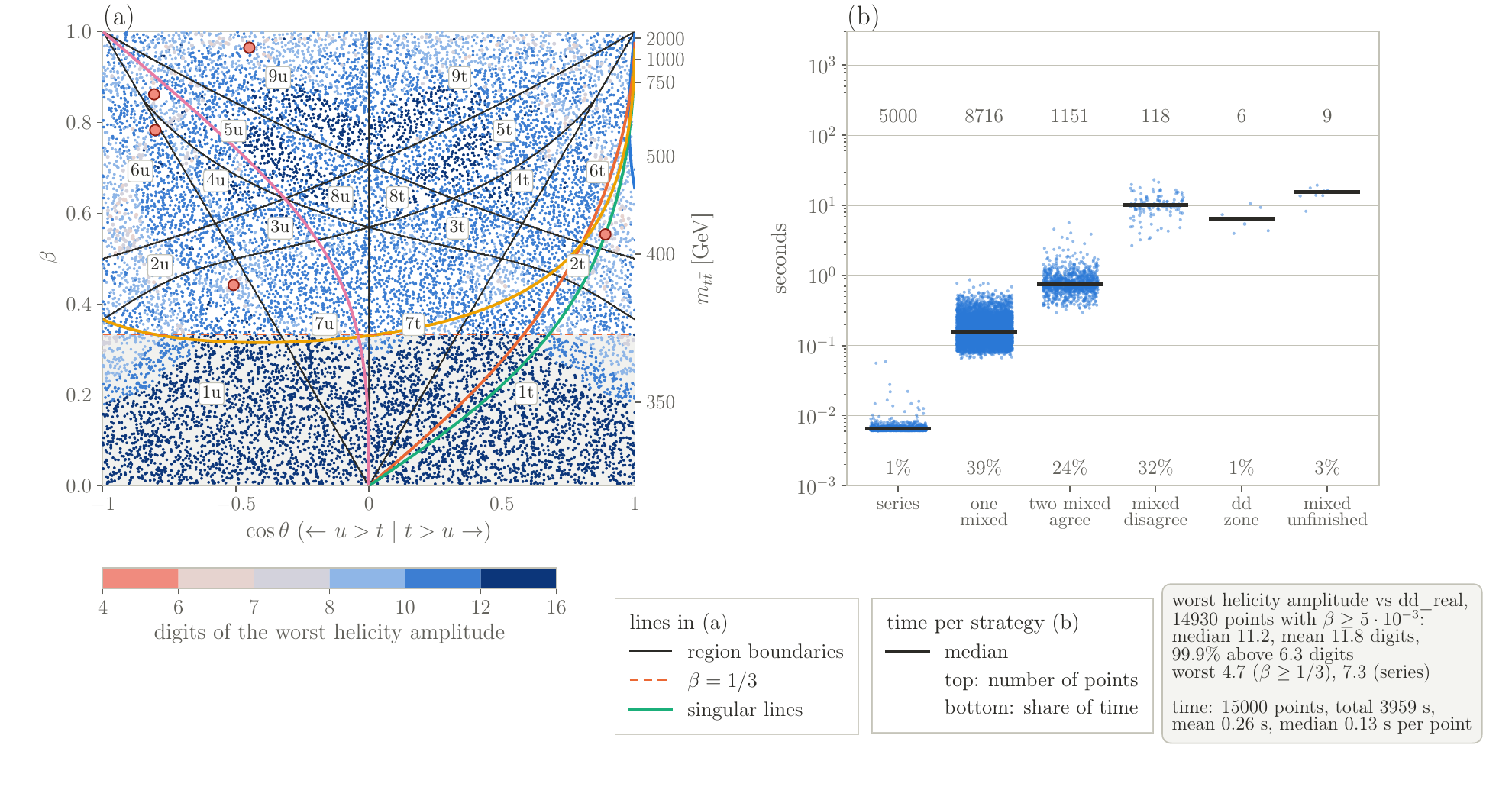}
  \caption{(a) Number of digits of the colour-summed helicity amplitude with the largest discrepancy, normalised to the magnitude of the largest amplitude, obtained with \texttt{hqpAmp} and compared with a double-double evaluation at tolerance $10^{-20}$, for 14930 points of the 15000 in \cref{fig:timing-strategy}. The 70 points missing have $\beta \leq 5 \cdot 10^{-3}$ and are not checked in double-double. The median agreement is $11.2$ digits. (b) Evaluation time per point for each strategy.}
  \label{fig:time-per-strategy}
\end{figure}

In~\cref{fig:time-per-strategy} we can also see the agreement against an independent run in double-double precision with smaller tolerance for the numerical transport.
We can see that 5 out of 15000 points miss the accuracy goal, 
which is only $0.03 \%$ of the total ensemble. 
The worst agreement is $4.7$ digits. The mean agreement is $11.8$ digits (median $11.2$ digits).

To minimise the number of times the transport path crosses a singular line, we provide nine boundary points, with boundary values computed using \texttt{AMFlow}~\cite{Liu:2022chg}, from which the code automatically selects the most suitable one at run time. 
Furthermore, we exploit the fact that our integral basis is closed under a $u\leftrightarrow t$ crossing, 
so that we only need to evaluate our integrals for $t>u$. 
The mirrored points at $u\leftrightarrow t$ are obtained via a constant transformation $C$ of our matrix and by exchanging the special functions $f$ with their crossed counterparts $f^{{\rm x12}}$.
\begin{equation}
  \vec{J}\big(s,u,m_t^2;\varepsilon\big)\Big|_{f\,\to\,f^{{\rm x12}}}
  \;=\; C\,\vec{J}\big(s,t,m_t^2;\varepsilon\big)\,,
  \qquad
  C \in \mathbb{Q}^{417\times 417}\,, \qquad C^2 = 1\,.
  \label{eq:crossing}
\end{equation}
 We observe that these are not necessarily equal to the value of the special functions belonging to the crossed sectors due to a choice of boundary values. However, the amplitudes are insensitive to these choices.
In particular, for the genus-two periods we find that the crossed values are given by:
\begin{equation}
  \begin{pmatrix} \pi^{{\rm x12},(1)}_j \\[2pt] \pi^{{\rm x12},(2)}_j \end{pmatrix}\!(s,t,m_t^2)
  \;=\; \frac{u}{t}
  \begin{pmatrix} 1 & 0 \\[2pt] \tfrac{1}{2} & -1 \end{pmatrix}
  \begin{pmatrix} \pi^{(1)}_j \\[2pt] \pi^{(2)}_j \end{pmatrix}\!(s,t,m_t^2)\,,
  \qquad j = 0,1\,.
\end{equation}

\begin{figure}[h]
  \centering
  \includegraphics[width=\textwidth]{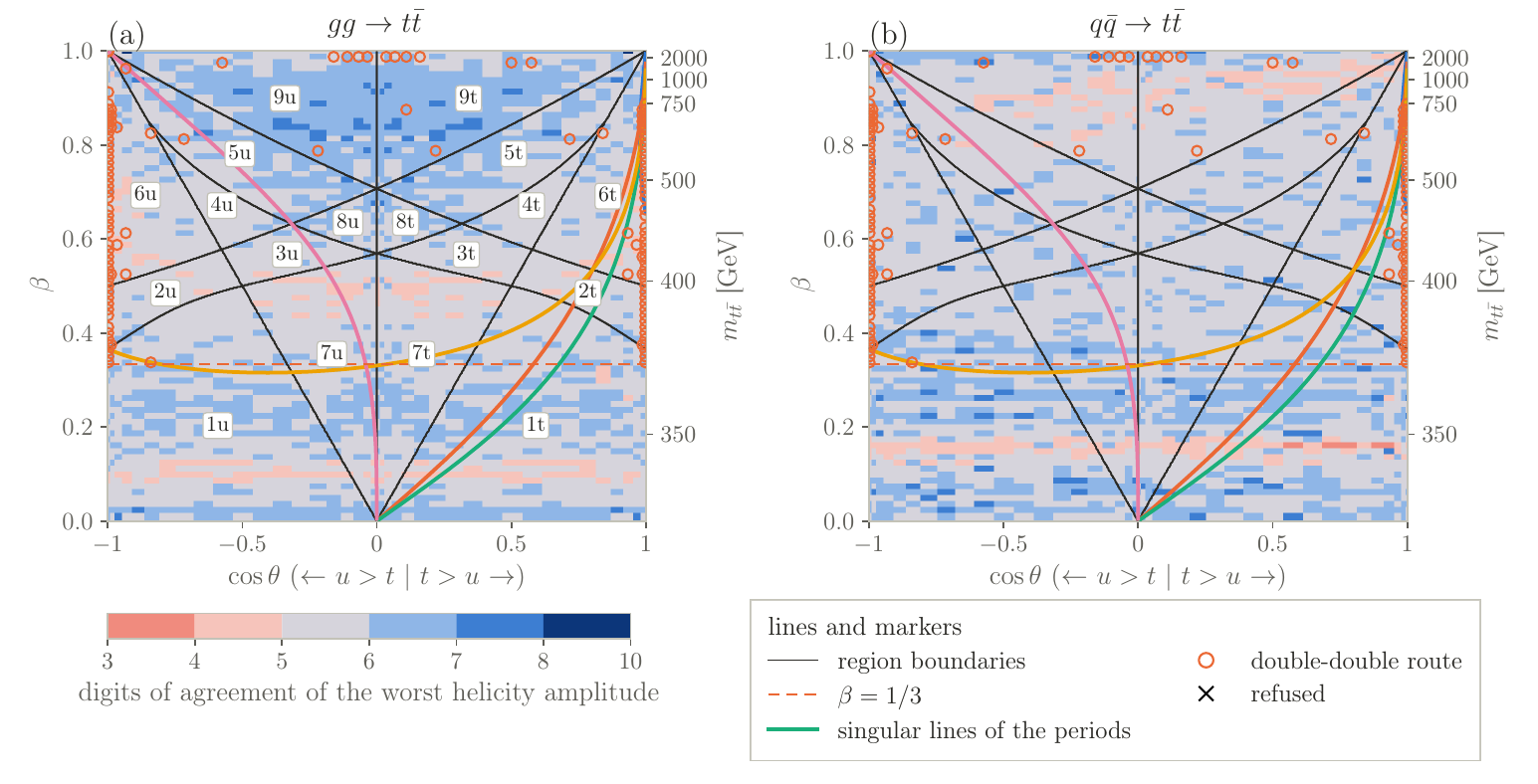}
  \caption{Comparison of \texttt{hqpAmp} against the grid of~\cite{Chen:2017jvi} for $gg\to t\bar t$ (a) and $q\bar q\to t\bar t$ (b) at the $3402$ nodes of each grid: digits of agreement for the colour-summed helicity amplitude with the largest discrepancy, normalised to the magnitude of the largest amplitude.
  Lines and regions as in~\cref{fig:timing-strategy}; nodes evaluated in double-double precision are circled. }
  \label{fig:ccp-agreement}
\end{figure}

Finally, we report on checks that we performed on our results. 
We compared our evaluations against~\cite{Czakon:2007ej, Czakon:2007wk, Barnreuther:2013qvf}, by reproducing the entire grid for the form factors published in~\cite{Chen:2017jvi}.
The comparison is shown in~\cref{fig:ccp-agreement}.
At all $3402$ nodes of both grids, we find agreement within the numerical accuracy of the grid evaluations. 
The median deviation evaluated on the helicity amplitude with largest discrepancy, normalised to the amplitude with the largest magnitude, is $1.5\times10^{-6}$ for $gg\to t\bar t$ and $2.2\times10^{-6}$ for $q\bar q\to t\bar t$, and about $95\%$ of the nodes agree to better than $10^{-5}$. 
Wherever the numerical transport is used, so $\beta\ge1/3$, the largest discrepancies are $3\times10^{-5}$ for $gg\to t\bar t$ and $7\times10^{-5}$ for $q\bar q\to t\bar t$.
The largest deviations detected overall are $9.6\times10^{-5}$ and $1.8\times10^{-4}$ for the two channels respectively, and both occur in the threshold region, where we checked the
results of our series expansion against an evaluation in double-double precision finding
agreement up to the target accuracy.
\paragraph{Results to higher orders in the dimensional regulator.}
In addition to the main program \texttt{hqpAmp}, in the same repository~\cite{hqpamp},
we provide \texttt{hqpInt}, a separate numerical code for the evaluation of
the canonical master integrals up to $\mathcal{O}(\varepsilon^6)$. This $\varepsilon$-order is indeed required to obtain the one-loop amplitudes to $\mathcal{O}(\varepsilon^4)$ and the two-loop ones to $\mathcal{O}(\varepsilon^2)$.
Note that locally, in particular close to a point of maximal unipotent monodromy, where the polylogarithmic transcendental weight counting can be used, this corresponds to transcendental weight 6. 
Combined with the analytic results for the bare
form factors and the mass counterterms expressed in terms of canonical master integrals,
see \cref{eq:uv0,eq:uv1,eq:uv2}, this allows for the evaluation of the bare, renormalised and infrared finite scattering amplitudes for both partonic channels. 
For reference, in \cref{tab:ggeps2,tab:qqeps2} we present the $\varepsilon^2$ coefficients of the independent two-loop bare form factors at the phase-space point $s=5.6\, m_t^2$, $t=-0.9\,m_t^2$. 
\begin{table}[htbp]
\centering
\small
\renewcommand{\arraystretch}{1.2}
\begin{tabular}{cccc}
\toprule
$i$ & $F_i^{(2),[1]}$ & $F_i^{(2),[2]}$ & $F_i^{(2),[3]}$ \\
\midrule
1 & $-4902.8957+1009.3080\,i$ & $-3202.6533-716.20028\,i$ & $-1338.9355-3273.0214\,i$ \\
2 & $\phantom{-}5635.0115-17.608902\,i$ & $-1461.9899-272.01511\,i$ & $-37.209784-240.75995\,i$ \\
3 & $\phantom{-}3223.1511+770.04986\,i$ & $\phantom{-}551.32079+456.74019\,i$ & $\phantom{-}597.77016+1645.2559\,i$ \\
4 & $-4925.4396-1068.3406\,i$ & $\phantom{-}2750.6242+947.32907\,i$ & $\phantom{-}320.18565+1047.7972\,i$ \\
5 & $\phantom{-}1480.0805-254.98876\,i$ & $\phantom{-}8086.0618-1098.2508\,i$ & $\phantom{-}8072.4872-15368.974\,i$ \\
6 & $\phantom{-}736.44539-1021.0054\,i$ & $-5786.4098+1272.0391\,i$ & $-7954.5065+16348.829\,i$ \\
\bottomrule
\end{tabular}
\caption{Coefficients at $\varepsilon^2$ for the two-loop bare form factors $F_i^{(2),[k]}$ of the $gg\to t\bar t$ channel,
see \cref{eq:loopexpansion}, at the phase-space point $s=5.6\, m_t^2$, $t=-0.9\,m_t^2$, for $N_c=3$, $n_l=5$ and $n_h=1$.
The superscript $[k]$ refers to the colour basis element $\big|C_k^{(gg)}\big\rangle$ of \cref{eq:ggcolour}.}
\label{tab:ggeps2}
\end{table}

\begin{table}[htbp]
\centering
\small
\renewcommand{\arraystretch}{1.2}
\begin{tabular}{ccc}
\toprule
$i$ & $G_i^{(2),[1]}$ & $G_i^{(2),[2]}$ \\
\midrule
1 & $\phantom{-}542.42333+43.437917\,i$ & $\phantom{-}77.974877-98.629303\,i$ \\
2 & $-80.976353+54.045931\,i$ & $\phantom{-}66.068143+610.12972\,i$ \\
3 & $\phantom{-}1430.7348-82.824195\,i$ & $-437.53886+106.07042\,i$ \\
\bottomrule
\end{tabular}
\caption{Coefficients at $\varepsilon^2$ for the two-loop bare form factors $G_i^{(2),[k]}$ of the $q\bar q\to t\bar t$ channel
at the same phase-space point and for the same parameters as in \cref{tab:ggeps2}.
The index $[k]$ refers to the colour basis element $\big|C_k^{(q\bar q)}\big\rangle$ of \cref{eq:qqcolour}.}
\label{tab:qqeps2}
\end{table}


%
\section{Conclusions}
\label{sec:conc}

In this paper, we have presented the complete two-loop QCD corrections to top-quark pair production in the gluon-fusion and in the quark--antiquark annihilation channels in analytic form. 
Our results retain full dependence on the top-quark mass and on the number of colours, 
including consistently all contributions from closed loops of light and heavy quarks. 
The amplitudes are expressed in terms of $417$ master integrals, for which we have constructed a basis such that the differential equations take the $\varepsilon$-factorised form. 
The construction of this canonical basis required handling Feynman integrals defined on 13 different elliptic curves and on one hyperelliptic curve of genus two. Our results, therefore, provide a realistic and demanding test of the recently proposed methods to obtain canonical bases beyond polylogarithms.

The $\varepsilon$-factorised form allows us to solve the differential equations in two complementary ways. 
Close to the top-pair production threshold, we have obtained deep series expansions of the master integrals 
and of the polarised scattering amplitudes, 
whose convergence was augmented by the use of suitably chosen Bernoulli-like variables.
This allows us to reliably use the series representation to evaluate the amplitude with at least six accurate digits for all points with $\beta<1/3$, thus well beyond the threshold point.
Elsewhere, we use the recently published code~\cite{Czakon:2026tog} 
to transport the master integrals numerically starting from a set of nine boundary points.
We implement this strategy in the \texttt{C++} code \texttt{hqpAmp} that chooses, point by point, 
between these two evaluation possibilities. Furthermore, we supplement it with a rescue system that guarantees
fast and reliable evaluations across the entire phase space.
Such code delivers bare, ultraviolet-renormalised and infrared-finite polarised scattering amplitudes
up to two-loop orders for all partonic processes relevant to $t\bar{t}$ production at hadron colliders. We cross-checked our results for the finite remainder against the numerical grids that are the current standard input of NNLO top-pair production cross section predictions. 

Moreover, starting from our $\varepsilon$-factorised differential equations, it is easy to extend our results to higher orders in $\varepsilon$, as required for the calculation of 
three-loop $t\bar{t}$-production amplitudes. 
Alongside the analytic expressions for the amplitudes at higher order in $\varepsilon$,
we provide \texttt{hqpInt} for the evaluation of all the necessary master integrals, thus offering the possibility
to obtain bare and UV-renormalised results through $\mathcal{O}(\varepsilon^2)$ at two loops.

Finally, we stress that the methods used here are not specific to this process, 
and we expect them to be applicable to other multi-loop amplitude calculations with several massive internal lines, where elliptic, higher-genus or higher-dimensional geometries appear.


\acknowledgments
We thank Andreas von Manteuffel for collaboration in the initial stages of this project and for sharing definitions of integral families and of some polylogarithmic canonical bases.
We further thank Johann Usovitsch for advice on the use of \texttt{Kira}.

We have used AI agents, including OpenAI Codex and Claude Code, 
for code implementation and help in drafting drawings presented in this article.

This work has been supported in part by the Deutsche Forschungsgemeinschaft (DFG, German Research Foundation) through the Excellence Cluster ORIGINS under Germany’s Excellence Strategy – EXC-2094-390783311 (BS, LT, FJW) and through Projektnummer 417533893/GRK2575 ``Rethinking Quantum Field Theory'' (BS), and in part by the European Research Council (ERC) under the European Union’s research and innovation program grant agreements 949279 (ERC Starting Grant HighPHun) (LT) and 101167287 (ERC Synergy Grant MaScAmp) (CN).
Views and opinions expressed are however those of the author(s) only and do not necessarily reflect those of the European Union or the European Research Council Executive Agency. Neither the European Union nor the granting authority can be held responsible for them.

\appendix

%
\section{Spinor conventions}
\label{app:spinorconventions}
The Dirac-$\gamma$ matrices are expressed in the Weyl representation and they are given by
\begin{equation}
    \gamma^0 = \begin{pmatrix}
    0\;  & \mathbf{1} \\
    \mathbf{1}\;  & 0
\end{pmatrix},\quad\quad
    \gamma^i = \begin{pmatrix}
    0          & -\sigma^i \\
    \sigma^i  & 0
\end{pmatrix} \;\; i=1,\ldots,3,
\end{equation}
where $\sigma^i$ are the Pauli matrices
\begin{equation}
    \sigma^1 =
    \begin{pmatrix}
    0\;  & 1 \\
    1\;  & 0
\end{pmatrix},\quad\quad
    \sigma^2 =
    \begin{pmatrix}
    0\;  & -i \\
    i\;  & 0
\end{pmatrix},\quad\quad
    \sigma^3 =
    \begin{pmatrix}
    1\;  & 0 \\
    0\;  & -1
\end{pmatrix}\,.
\end{equation}
We use the metric $g^{\mu\nu}=\mathrm{diag}(1,-1,-1,-1)$ and write $\slashed{p}=\gamma^\mu p_\mu=p^0\gamma^0-p^i\gamma^i$.
The matrix $\gamma^5$ is defined as
$\gamma^5 = i \gamma^0\gamma^1\gamma^2\gamma^3$ and it satisfies the relations
\begin{equation}
    \gamma^5 = \begin{pmatrix}
    \mathbf{1}\;  &  0 \\
    0\;  & -\mathbf{1}
\end{pmatrix},\quad\quad
    \left\lbrace \gamma^\mu,\gamma^5\right\rbrace = 0,\quad\text{and}\quad
    \mathrm{Tr}\left(\gamma^\mu \gamma^\nu \gamma^\rho \gamma^\sigma \gamma^5 \right) = 4 i \varepsilon^{\mu\nu\rho\sigma},
\end{equation}
where $\varepsilon^{\mu\nu\rho\sigma}$ is the totally anti-symmetric Levi-Civita symbol. In these conventions we have
\begin{equation}
    \varepsilon^{0123} = - 1\,.
\end{equation}
The left- and right-chirality projectors are then defined as
\begin{equation}
    P_R = \frac{1+\gamma^5}{2} =
   \begin{pmatrix}
       \mathbf{1}\;  & 0 \\
       0\;  & 0
   \end{pmatrix}, \quad\quad
   P_L = \frac{1-\gamma^5}{2} =
   \begin{pmatrix}
       0\;  & 0 \\
       0\;  & \mathbf{1}
   \end{pmatrix}\,.
\end{equation}
The Dirac massive spinors are indicated as $u_s(p)$ and $v_s(p)$, where the subscript $s=\pm$ is the helicity and the momentum $p$ is such that $p^2 = m^2$ and $p^0 > 0$: $u_s(p)$ describes a fermion and $v_s(p)$ an antifermion of momentum $p$ and helicity $s$, so that, with $\Sigma^i=\gamma^5\gamma^0\gamma^i$ and $\hat{\mathbf{p}}=\mathbf{p}/|\mathbf{p}|$,
\begin{equation}
    (\boldsymbol{\Sigma}\cdot\hat{\mathbf{p}})\,u_s(p) = s\,u_s(p)\,,\qquad
    (\boldsymbol{\Sigma}\cdot\hat{\mathbf{p}})\,v_s(p) = -s\,v_s(p)\,.
\end{equation}
Explicitly we have
\begin{equation}
  u_{+}(p) =
    \begin{pmatrix}
    r/c \\
    a \,p_\perp/r \\
    m\, b/r \\
    m \,p_\perp/r
    \end{pmatrix},\;
 u_{-}(p) =
    \begin{pmatrix}
     m \,p^*_\perp/r \\
    -m \,b/r \\
     a \,p^*_\perp/r \\
    -r/c
    \end{pmatrix},\;
 v_{+}(p) =
    \begin{pmatrix}
    - m \,p^*_\perp/r \\
      m\, b/r \\
      a \,p^*_\perp/r \\
    - r/c
    \end{pmatrix},\;
  v_{-}(p) =
    \begin{pmatrix}
    r/c \\
    a \,p_\perp/r \\
    -m \,b/r \\
    -m \,p_\perp/r \\
    \end{pmatrix},
\end{equation}
where we have introduced the shorthand notation
\begin{equation*}
a = p^0 + |\mathbf{p}|,\quad
b = p^3 + |\mathbf{p}|,\quad
c = 2 |\mathbf{p}|,\quad
r = \sqrt{abc}\,,\quad \text{and} \quad
p_\perp = p^1 + i \,p^2\,.
\end{equation*}
Barred spinors $\overline{u}_r(p)$ can be simply obtained from the ones above as
$\overline{u}_r(p) = u_r^{\dagger} (p) \gamma^0$.
The same applies to $\overline{v}_r(p)$.
The spinors are normalised as $\overline{u}_r(p)u_s(p)=2m\,\delta_{rs}$ and $\overline{v}_r(p)v_s(p)=-2m\,\delta_{rs}$, and fulfil $\sum_s u_s(p)\overline{u}_s(p)=\slashed{p}+m$ and $\sum_s v_s(p)\overline{v}_s(p)=\slashed{p}-m$.
Spinors for massless fermions can be obtained from the above ones taking the
$m\to 0$ limit. Within our conventions they are given by
\begin{equation}
  u_{+}(p) = v_{-}(p) =
    \begin{pmatrix}
    \sqrt{p^+} \\
    \sqrt{p^-} p_\perp/|p_\perp| \\
    0 \\
    0
    \end{pmatrix}, \quad
  u_{-}(p) = v_{+}(p) =
    \begin{pmatrix}
    0 \\
    0 \\
    \sqrt{p^-} p^*_\perp/|p_\perp| \\
    -\sqrt{p^+}
    \end{pmatrix}, \quad
    p^\pm = p^0 \pm p^3\,.
\end{equation}
When $|p_\perp|=0$, we fix $p_\perp/|p_\perp| = +1$ if $p^-=0$ and
$p_\perp/|p_\perp| = -1$ if $p^+=0$.
We then introduce the bra-ket notation of the spinor-helicity formalism for massless spinors as follows
\begin{align}
    &  |i \rangle \equiv |i+\rangle  \equiv u_{+}(p_i)            = v_{-}(p_i),\;
    && \langle i| \equiv \langle i-| \equiv \overline{u}_{-}(p_i) = \overline{v}_{+}(p_i), \nn \\
    &  |i]        \equiv |i-\rangle  \equiv u_{-}(p_i)            = v_{+}(p_i), \;
    && [i|        \equiv \langle i+| \equiv \overline{u}_{+}(p_i) = \overline{v}_{-}(p_i)\,.
\end{align}
Useful identities that we will use often are the following
\begin{equation}
    \langle i_1\cdots i_n i_1] = \mathrm{Tr}\left( P_L \,\slashed{p}_{i_1}\cdots\,\slashed{p}_{i_n}\right),
    \quad\quad
    [i_1\cdots i_n i_1\rangle  = \mathrm{Tr}\left( P_R \,\slashed{p}_{i_1}\cdots\,\slashed{p}_{i_n}\right)\,.
\end{equation}
Within the spinor-helicity formalism we can express the polarisation vector for an \textit{outgoing} massless gauge boson of momentum $k$ as
\begin{equation}
    \epsilon^{*\mu}_{+}(k,q) =  \frac{1}{\sqrt{2}}\frac{\langle q \gamma^\mu k]}{\langle q k\rangle}, \quad\quad
    \epsilon^{*\mu}_{-}(k,q) = -\frac{1}{\sqrt{2}}\frac{[q\gamma^\mu k\rangle}{[q k]}\,,
\end{equation}
where $q$ is an auxiliary massless momentum.
With the spinors above, $\epsilon^{*\mu}_{-}(k,q)=\big(\epsilon^{*\mu}_{+}(k,q)\big)^*$, so that the polarisation vector of an \textit{incoming} gauge boson of momentum $k$ and helicity $\lambda$ is $\epsilon^{\mu}_{\lambda}(k,q)=\epsilon^{*\mu}_{-\lambda}(k,q)$.
In \cref{sec:ggchannel} we use the vectors $\varepsilon^{*}_{\pm}$ above also for the incoming gluons, \ie, we label the gluon helicities as outgoing.

\section{Helicity coefficients in terms of the form factors}
\label{app:helcoeff}
In this appendix, we give the explicit expression of the polarised amplitudes in terms of form factors, within the spinors conventions of \cref{app:spinorconventions}. 
The incoming quarks and gluons have definite helicities, and the top quarks definite spin states. 
We label the latter by $h,\bar h=\pm$ and use the spinors $\bar{u}_{h}(p_3)=u_{h}^\dagger(p_3)\gamma^0$ and $v_{-\bar h}(p_4)$ of \cref{app:spinorconventions}, 
where the subscripts are helicities, \ie $\bar{u}(p_3)$ and $v(p_4)$ in the amplitudes are replaced by $\bar{u}_{h}(p_3)$ and $v_{-\bar h}(p_4)$. In the $t\bar t$ centre-of-mass frame $h$ and $\bar h$ are then the spin projections of the top and of the antitop along the direction of motion of the \emph{top} quark: $h$ is the helicity of the top, $\bar h$ \emph{minus} the helicity of the antitop, and $h=\bar h$ is a pair of opposite helicities. 
In the following, we
assume the partonic centre-of-mass frame of \cref{eq:kinexample}, with $\hat n_1$ along $+\hat{z}$, the top quark at polar angle $\theta$ and azimuth $\phi$ measured from the $x$-axis, explicitly
\begin{equation}
\label{eq:partoniccentreofmassframe}
p_{1,2} = \frac{\sqrt{s}}{2}\left(1,0,0,\pm 1\right),
\quad\quad
p_{3,4}=\frac{\sqrt{s}}{2}\left(1,\pm\beta\cos\phi\sin\theta,\pm\beta\sin\phi\sin\theta,\pm\beta\cos\theta\right)\,.
\end{equation}
In this frame, the Dirac structures appearing in \cref{eq:Appgg,eq:Apmgg}, as well as in \cref{eq:Aqqred}, for the two polarisations $(h,\bar h) = (+,+)$ and
$(h,\bar h) = (+,-)$ read 
\begin{center}
\small\renewcommand{\arraystretch}{1.5}
    \begin{tabular}{c|cccc}
$(h,\bar h)$ & $\bar u_h(p_3)\,v_{-\bar h}(p_4)$ & $\bar u_h(p_3)\,\slashed{p}_1\,v_{-\bar h}(p_4)$ & $\bar u_h(p_3)\,\slashed{a}\,v_{-\bar h}(p_4)$ & $\bar u_h(p_3)\,\slashed{p}_1\slashed{a}\,v_{-\bar h}(p_4)$\\ \hline
$(+,+)$ & $0$ & $-\,s\sin\theta/2$ & $-\sqrt{s}$ & $-m_t\sqrt{s}$\\
$(+,-)$ & $-\sqrt{s}\,\beta\,e^{-i\phi}$ & $-m_t\sqrt{s}\,\cos\theta\,e^{-i\phi}$ & $0$ & $-\,s\beta\sin\theta\,e^{-i\phi}/2$
\end{tabular}\;.
\end{center}
The other two top-quark polarisations follow from
charge conjugation as shown explicitly below for
the two partonic channels.

\subsection*{Gluon channel}
\label{app:helcoeffgg}

The polarised amplitudes of \cref{eq:Appgg,eq:Apmgg} can be evaluated using the spinor-helicity conventions 
of \cref{app:spinorconventions} and the
explicit spinor structures for the top-quarks derived above. 
Working in the partonic centre-of-mass frame, one has $\langle12\rangle^2=[21]^2=s$, and $\langle1|\slashed{p}_3|2]= s\beta\sin\theta\,e^{i\phi}/2$, thus for the phases
\begin{equation}
\Phi_{++}=\Phi_{--}=1,\quad\quad
\Phi_{+-}=-e^{-2i\phi},\quad\quad
\Phi_{-+}=-e^{2i\phi}\,.
\end{equation}
Starting from \cref{eq:Appgg,eq:Apmgg}, 
and using \cref{eq:FFlindep},
the polarised amplitudes
for the four possible gluon-helicity configurations, and equal top quarks polarisations read
\begin{align}
    \boldsymbol{\mathcal{A}}_{++}(+,+) &= \boldsymbol{\mathcal{A}}_{--}(+,+) = -g_s^2\frac{s\sin\theta}{4m_t^2}\,\big(\mathbf{F}_4-\mathbf{F}_2\big)\,, \nn\\
    \boldsymbol{\mathcal{A}}_{+-}(+,+) &= g_s^2e^{-2i\phi}\left[\frac{s\sin\theta}{4m_t^2}\,\big(\mathbf{F}_2+\mathbf{F}_4\big)-\frac{2m_t^2}{s\beta\sin\theta}\,\big(\mathbf{F}_5+\mathbf{F}_6\big)\right]\,, \nn\\
    \boldsymbol{\mathcal{A}}_{-+}(+,+) &= g_s^2e^{2i\phi}\left[\frac{s\sin\theta}{4m_t^2}\,\big(\mathbf{F}_2+\mathbf{F}_4\big)+\frac{2m_t^2}{s\beta\sin\theta}\,\big(\mathbf{F}_5+\mathbf{F}_6\big)\right]\,,
    \label{eq:Aggexp1}
\end{align}
while for opposite top-quark polarisations
\begin{align}
    \boldsymbol{\mathcal{A}}_{++}(+,-) &= -g_s^2\frac{\sqrt{s}\,e^{-i\phi}}{2m_t}\left[\beta\,\big(\mathbf{F}_3-\mathbf{F}_1\big)+\cos\theta\,\big(\mathbf{F}_4-\mathbf{F}_2\big)-\frac{2m_t^2}{s}\,\mathbf{F}_6\right]\,, \nn\\
    \boldsymbol{\mathcal{A}}_{--}(+,-) &= -g_s^2\frac{\sqrt{s}\,e^{-i\phi}}{2m_t}\left[\beta\,\big(\mathbf{F}_3-\mathbf{F}_1\big)+\cos\theta\,\big(\mathbf{F}_4-\mathbf{F}_2\big)+\frac{2m_t^2}{s}\,\mathbf{F}_6\right]\,, \nn\\
    \boldsymbol{\mathcal{A}}_{+-}(+,-) &= g_s^2\frac{\sqrt{s}\,e^{-3i\phi}}{2m_t}\left[\beta\,\big(\mathbf{F}_1+\mathbf{F}_3\big)+\cos\theta\,\big(\mathbf{F}_2+\mathbf{F}_4\big)\right]\,, \nn\\
    \boldsymbol{\mathcal{A}}_{-+}(+,-) &= g_s^2\frac{\sqrt{s}\,e^{i\phi}}{2m_t}\left[\beta\,\big(\mathbf{F}_1+\mathbf{F}_3\big)+\cos\theta\,\big(\mathbf{F}_2+\mathbf{F}_4\big)\right]\,,
    \label{eq:Aggexp2}
\end{align}
where the arguments are $(h,\bar h)$, $g_s^2 = 4\pi\alpha_s$, and each identity clearly holds for every colour component of the form factors. 
For the remaining two top-quark polarisations one has
\begin{align}
    \boldsymbol{\mathcal{A}}_{\lambda_1\lambda_2}(-,-) &= \boldsymbol{\mathcal{A}}_{\lambda_1\lambda_2}(+,+)\Big|_{\mathbf{F}_5+\mathbf{F}_6\to-(\mathbf{F}_5+\mathbf{F}_6)}\,, \nn\\
    \boldsymbol{\mathcal{A}}_{\lambda_1\lambda_2}(-,+) &= -\,\boldsymbol{\mathcal{A}}_{\lambda_2\lambda_1}(+,-)\Big|_{\phi\to-\phi,\;\mathbf{F}_6\to-\mathbf{F}_6}\,.
\end{align}
For equal gluon helicities and $h=\bar h$ only $\mathbf{F}_4-\mathbf{F}_2$ appears, so that at tree level, where $\mathbf{F}_4^{(0)}=\mathbf{F}_2^{(0)}$, the amplitudes $\boldsymbol{\mathcal{A}}_{++}^{(0)}(h,h)$ and $\boldsymbol{\mathcal{A}}_{--}^{(0)}(h,h)$ vanish, as for massless quarks.

\subsection*{Quark channel}
\label{app:helcoeffqq}

In the explicit frame of~\cref{eq:partoniccentreofmassframe}, we find that 
$J_{\mp\pm}^\mu=\sqrt{s}\,(0,-1,\pm i,0)$.
With the explicit spinor structures of the top-quark,
the amplitudes of \cref{eq:Aqqred} are given by
\begin{align}
    \boldsymbol{\mathcal{A}}_{-+}(-,-) &= g_s^2\frac{e^{-i\phi}}{\beta^2\sin^2\theta}\Big[-\beta\sin^2\theta\,\mathbf{G}_2-\frac{4m_t^2}{s}\,(1+\cos\theta)\,\mathbf{G}_3\Big]\,, \nn\\
    \boldsymbol{\mathcal{A}}_{-+}(+,+) &= g_s^2\frac{e^{-i\phi}}{\beta^2\sin^2\theta}\Big[-\beta\sin^2\theta\,\mathbf{G}_2+\frac{4m_t^2}{s}\,(1-\cos\theta)\,\mathbf{G}_3\Big]\,, \nn\\
    \boldsymbol{\mathcal{A}}_{-+}(+,-) &= g_s^2\frac{2m_t\,e^{-2i\phi}}{\sqrt{s}\,\beta^2\sin\theta}\Big[-\beta^2\mathbf{G}_1-\beta\cos\theta\,\mathbf{G}_2+\frac{4m_t^2}{s}\,\mathbf{G}_3\Big]\,, \nn\\
    \boldsymbol{\mathcal{A}}_{-+}(-,+) &= g_s^2\frac{2m_t}{\sqrt{s}\,\beta^2\sin\theta}\Big[\beta^2\mathbf{G}_1+\beta\cos\theta\,\mathbf{G}_2-\frac{4m_t^2}{s}\,\mathbf{G}_3\Big]\,,
    \label{eq:Aqqexp}
\end{align}
where the arguments are $(h,\bar h)$. 
Under a parity transformation,
since $\cos\theta$ is invariant,
we then obtain
\begin{align}
    \boldsymbol{\mathcal{A}}_{+-}(+,+) &= \boldsymbol{\mathcal{A}}_{-+}(-,-)\big|_{\phi\to-\phi}\,,&
    \boldsymbol{\mathcal{A}}_{+-}(-,-) &= \boldsymbol{\mathcal{A}}_{-+}(+,+)\big|_{\phi\to-\phi}\,,\nn\\
    \boldsymbol{\mathcal{A}}_{+-}(-,+) &= -\,\boldsymbol{\mathcal{A}}_{-+}(+,-)\big|_{\phi\to-\phi}\,,&
    \boldsymbol{\mathcal{A}}_{+-}(+,-) &= -\,\boldsymbol{\mathcal{A}}_{-+}(-,+)\big|_{\phi\to-\phi}\,,
    \label{eq:Aqqpar}
\end{align}
\ie $\phi\to-\phi$ in the phases only, 
with the form-factor coefficients unchanged. 
At tree level only $\mathbf{G}_3^{(0)}$ is non-vanishing
and it is simply given by
\begin{equation}
\mathbf{G}_3^{(0)} = 
\frac{s\beta^2\sin^2\theta}{8m_t^2}
\left(|C_1^{(q\bar q)}\rangle-\frac{1}{N_c}|C_2^{(q\bar q)}\rangle\right)\,.
\end{equation}

\section{Integral family definitions}
\label{app:families}
We define the integral families by the nine propagators $D_i=q_i^2-m_i^2$ listed in \cref{tab:planarfamilies,tab:nonplanarfamilies1,tab:nonplanarfamilies2}, where $k_1$ and $k_2$ are the loop momenta and $p_4=p_1+p_2-p_3$. The external kinematics is given by
\begin{equation}
    p_1^2=p_2^2=0\,,\quad p_3^2=p_4^2=m_t^2\,,\quad (p_1+p_2)^2=s\,,\quad (p_1-p_3)^2=t\,.
\end{equation}

\begin{table}[h!]
\centering
\footnotesize
\begin{tabular}{clllll}
\toprule
$i$ & \texttt{pA} & \texttt{pB} & \texttt{pC} & \texttt{pD} & \texttt{pE} \\
\midrule
1 & $k_1$ & $k_1$ & $\mathbf{k_1}$ & $k_1$ & $k_1$ \\
2 & $k_2$ & $\mathbf{k_2}$ & $\mathbf{k_2}$ & $k_2$ & $\mathbf{k_2}$ \\
3 & $k_1 - k_2$ & $\mathbf{k_1 - k_2}$ & $k_1 - k_2$ & $k_1 - k_2$ & $\mathbf{k_1 - k_2}$ \\
4 & $k_1 - p_1$ & $k_1 - p_1$ & $\mathbf{k_1 - p_1}$ & $k_1 - p_1$ & $k_1 - p_1$ \\
5 & $k_2 - p_1$ & $\mathbf{k_2 - p_1}$ & $\mathbf{k_2 - p_1}$ & $k_2 - p_1$ & $\mathbf{k_2 - p_1}$ \\
6 & $k_1 - p_1 - p_2$ & $k_1 - p_1 - p_2$ & $\mathbf{k_1 - p_1 - p_2}$ & $\mathbf{k_1 + p_2 - p_3}$ & $\mathbf{k_1 + p_2 - p_3}$ \\
7 & $k_2 - p_1 - p_2$ & $\mathbf{k_2 - p_1 - p_2}$ & $\mathbf{k_2 - p_1 - p_2}$ & $\mathbf{k_2 + p_2 - p_3}$ & $k_2 + p_2 - p_3$ \\
8 & $\mathbf{k_1 - p_3}$ & $\mathbf{k_1 - p_3}$ & $k_1 - p_3$ & $\mathbf{k_1 - p_3}$ & $\mathbf{k_1 - p_3}$ \\
9 & $\mathbf{k_2 - p_3}$ & $k_2 - p_3$ & $k_2 - p_3$ & $\mathbf{k_2 - p_3}$ & $k_2 - p_3$ \\
\bottomrule
\end{tabular}
\caption{Propagators of the planar families. Each entry is the momentum $q_i$ flowing through the propagator $D_i=q_i^2-m_i^2$; entries in bold are massive ($m_i^2=m_t^2$), the others are massless.}
\label{tab:planarfamilies}
\end{table}

\begin{table}[h!]
\centering
\footnotesize
\begin{tabular}{clllll}
\toprule
$i$ & \texttt{nA} & \texttt{nB} & \texttt{nC} & \texttt{nD} & \texttt{nE} \\
\midrule
1 & $k_1$ & $\mathbf{k_1}$ & $k_1$ & $k_1$ & $k_1$ \\
2 & $k_2$ & $\mathbf{k_2}$ & $k_2$ & $k_2$ & $\mathbf{k_2}$ \\
3 & $k_1 + p_1$ & $\mathbf{k_1 + p_1}$ & $\mathbf{k_1 - p_3}$ & $\mathbf{k_1 - p_3}$ & $\mathbf{k_1 - p_3}$ \\
4 & $k_2 + p_2$ & $\mathbf{k_2 + p_2}$ & $\mathbf{k_2 - p_4}$ & $k_2 + p_1$ & $k_2 - p_4$ \\
5 & $k_1 - p_3$ & $k_1 - p_3$ & $k_1 + p_1$ & $k_1 - p_4$ & $k_1 + p_1$ \\
6 & $k_2 - p_4$ & $k_2 - p_4$ & $k_2 + p_2$ & $k_2 + p_2$ & $k_2 + p_2$ \\
7 & $k_1 - k_2 + p_1$ & $k_1 - k_2 + p_1$ & $\mathbf{k_1 - k_2 - p_3}$ & $\mathbf{k_1 - k_2 - p_3}$ & $k_1 - k_2 - p_3$ \\
8 & $k_1 - k_2 - p_2$ & $k_1 - k_2 - p_2$ & $\mathbf{k_1 - k_2 + p_4}$ & $k_1 - k_2 - p_1$ & $k_1 - k_2 + p_4$ \\
9 & $\mathbf{k_1 - k_2 + p_1 - p_3}$ & $\mathbf{k_1 - k_2 + p_1 - p_3}$ & $\mathbf{k_1 - k_2 - p_3 + p_1}$ & $k_1 - k_2 - p_1 - p_2$ & $k_1 - k_2 - p_3 + p_1$ \\
\bottomrule
\end{tabular}
\caption{Propagators of the non-planar families \texttt{nA}--\texttt{nE}, with the same conventions as in \cref{tab:planarfamilies}.}
\label{tab:nonplanarfamilies1}
\end{table}

\begin{table}[h!]
\centering
\footnotesize
\begin{tabular}{cllll}
\toprule
$i$ & \texttt{nF} & \texttt{nG} & \texttt{nH} & \texttt{nI} \\
\midrule
1 & $\mathbf{k_1}$ & $\mathbf{k_1}$ & $k_1$ & $k_1$ \\
2 & $k_2$ & $\mathbf{k_2}$ & $\mathbf{k_2}$ & $\mathbf{k_2}$ \\
3 & $k_1 - p_3$ & $k_1 - p_3$ & $k_1 + p_1$ & $\mathbf{k_1 - p_3}$ \\
4 & $k_2 + p_1$ & $\mathbf{k_2 + p_1}$ & $\mathbf{k_2 + p_2}$ & $\mathbf{k_2 + p_1}$ \\
5 & $k_1 - p_4$ & $k_1 - p_4$ & $k_1 - p_3$ & $k_1 - p_4$ \\
6 & $k_2 + p_2$ & $k_2 + p_2$ & $k_2 - p_4$ & $k_2 + p_2$ \\
7 & $k_1 - k_2 - p_3$ & $\mathbf{k_1 - k_2 - p_3}$ & $\mathbf{k_1 - k_2 + p_1}$ & $k_1 - k_2 - p_3$ \\
8 & $\mathbf{k_1 - k_2 - p_1}$ & $k_1 - k_2 - p_1$ & $\mathbf{k_1 - k_2 - p_2}$ & $\mathbf{k_1 - k_2 - p_1}$ \\
9 & $\mathbf{k_1 - k_2 - p_1 - p_2}$ & $k_1 - k_2 - p_1 - p_2$ & $k_1 - k_2 + p_1 - p_3$ & $\mathbf{k_1 - k_2 - p_1 - p_2}$ \\
\bottomrule
\end{tabular}
\caption{Propagators of the non-planar families \texttt{nF}--\texttt{nI}, with the same conventions as in \cref{tab:planarfamilies}.}
\label{tab:nonplanarfamilies2}
\end{table}
\newpage
With these definitions, we find a total of $417$ master integrals, including all the crossings needed for the full physical amplitude. The masters are distributed among the integral families of \cref{app:families} and their crossings as in \cref{tab:mastercounts}.
\begin{table}[h!]
\centering
\footnotesize
\begin{tabular}{lccc}
\toprule
family & uncrossed & crossed (\texttt{x12}) & total \\
\midrule
\texttt{pA} & 35 & 22 & 57 \\
\texttt{pB} & 84 & 57 & 141 \\
\texttt{pC} & 30 & 22 & 52 \\
\texttt{pD} & 18 & 11 & 29 \\
\texttt{pE} & 23 & 18 & 41 \\
\midrule
\texttt{nA} & 4 & -- & 4 \\
\texttt{nB} & 16 & 5 & 21 \\
\texttt{nC} & 9 & 7 & 16 \\
\texttt{nD} & 8 & 4 & 12 \\
\texttt{nE} & 5 & -- & 5 \\
\texttt{nF} & 4 & 4 & 8 \\
\texttt{nG} & 7 & 7 & 14 \\
\texttt{nH} & 5 & -- & 5 \\
\texttt{nI} & 6 & 6 & 12 \\
\midrule
total & 254 & 163 & 417 \\
\bottomrule
\end{tabular}
\caption{Number of canonical master integrals per family. The second column counts the masters of the family itself, the third those of its crossing $p_1\leftrightarrow p_2$ (\texttt{x12}), and the last column is the sum of the two.}
\label{tab:mastercounts}
\end{table}

\section{Polynomials in genus-two Picard--Fuchs equations}
\label{app:polyPic}
The polynomials in \cref{eq:PFgenus2a,eq:PFgenus2b} read
\begin{align}
  N_1&=1+4z+4v\,(1-z-4z^{2})+v^{2}\,(8-33z+3z^{2}+12z^{3})\nonumber\\
  &\quad +2v^{3}\,(4-25z+11z^{2}+119z^{3}-108z^{4})+v^{4}\,(3-37z+116z^{2}-66z^{3}+23z^{4}-100z^{5})\nonumber\\
  &\quad -4zv^{5}\,(3-31z+164z^{2}-552z^{3}+964z^{4}-664z^{5})\nonumber\\
  &\quad +z^{2}v^{6}\,(9-209z+1460z^{2}-4485z^{3}+6693z^{4}-4140z^{5})\nonumber\\
  &\quad +2z^{4}v^{7}\,(1-z)(1-4z)(1-9z)(9-25z)\,,\nonumber\\[2pt]
  N_2&=1+4z+4v\,(1-2z)(1+3z)+v^{2}\,(5-18z+57z^{2}-132z^{3})\nonumber\\
  &\quad +2v^{3}\,(1-2z)(1-14z+96z^{2}-204z^{3})-zv^{4}\,(14-200z+1032z^{2}-2171z^{3}+1476z^{4})\nonumber\\
  &\quad +4z^{2}v^{5}\,(1-2z)(8-76z+207z^{2}-141z^{3})\nonumber\\
  &\quad +z^{3}v^{6}\,(12-125z+386z^{2}-245z^{3}-316z^{4})\,,\nonumber\\[2pt]
  N_3&=1+4z+v\,(4-7z-12z^{2})+v^{2}\,(6-61z+149z^{2}-28z^{3})\nonumber\\
  &\quad +v^{3}\,(4-106z+453z^{2}-311z^{3}-316z^{4})\nonumber\\
  &\quad +v^{4}\,(1-77z+484z^{2}-686z^{3}-601z^{4}+1116z^{5})\nonumber\\
  &\quad -zv^{5}\,(21-239z+1016z^{2}-2286z^{3}+3421z^{4}-2524z^{5})\nonumber\\
  &\quad +z^{2}v^{6}\,(47-587z+3306z^{2}-10281z^{3}+16399z^{4}-10548z^{5})\nonumber\\
  &\quad -z^{3}v^{7}\,(27-585z+4046z^{2}-12473z^{3}+18405z^{4}-10860z^{5})\nonumber\\
  &\quad -4z^{5}v^{8}\,(1-z)(1-4z)(1-9z)(9-25z)\,,\nonumber\\[2pt]
  N_4&=3+4z+3v\,(1-2z)(3+4z)+v^{2}\,(9-42z+133z^{2}-260z^{3})\nonumber\\
  &\quad +v^{3}\,(1-2z)(3-44z+286z^{2}-600z^{3})-3zv^{4}\,(6-84z+432z^{2}-911z^{3}+620z^{4})\nonumber\\
  &\quad +z^{2}v^{5}\,(1-2z)(35-336z+921z^{2}-628z^{3})\nonumber\\
  &\quad +z^{3}v^{6}\,(12-125z+386z^{2}-245z^{3}-316z^{4})\,,\nonumber\\[2pt]
  M_1&=1+v\,(3-10z)+v^{2}\,(5-34z+51z^{2})+v^{3}\,(5-49z+122z^{2}-28z^{3})\nonumber\\
  &\quad +v^{4}\,(2-36z+193z^{2}-416z^{3}+375z^{4})-zv^{5}\,(11-133z+643z^{2}-1483z^{3}+1338z^{4})\nonumber\\
  &\quad +z^{2}v^{6}\,(9-134z+686z^{2}-1414z^{3}+949z^{4})\,,\nonumber\\[2pt]
  M_2&=1+v\,(3-6z-4z^{2})+v^{2}\,(2-15z+41z^{2}-36z^{3})-v^{3}\,(1+z-62z^{2}+270z^{3}-368z^{4})\nonumber\\
  &\quad -v^{4}\,(1-19z+69z^{2}+44z^{3}-327z^{4}+72z^{5})\nonumber\\
  &\quad +zv^{5}\,(11-105z+523z^{2}-1843z^{3}+3562z^{4}-2596z^{5})\nonumber\\
  &\quad -z^{2}v^{6}\,(18-337z+2117z^{2}-5915z^{3}+7777z^{4}-4140z^{5})\nonumber\\
  &\quad -2z^{4}v^{7}\,(1-z)(1-4z)(1-9z)(9-25z)\,,\nonumber\\[2pt]
  M_3&=1+v\,(3-10z)+v^{2}\,(3-34z+83z^{2})+v^{3}\,(1-43z+186z^{2}-124z^{3})\nonumber\\
  &\quad -zv^{4}\,(24-139z+160z^{2}+41z^{3})-zv^{5}\,(5-45z+209z^{2}-651z^{3}+858z^{4})\nonumber\\
  &\quad +z^{2}v^{6}\,(9-134z+686z^{2}-1414z^{3}+949z^{4})\,,\nonumber\\[2pt]
  M_4&=1+4z+2zv\,(9-20z)-v^{2}\,(6-57z+105z^{2}+100z^{3})\nonumber\\
  &\quad -v^{3}\,(8-107z+264z^{2}+464z^{3}-1216z^{4})\nonumber\\
  &\quad -v^{4}\,(3-95z+429z^{2}-206z^{3}-751z^{4}+132z^{5})\nonumber\\
  &\quad +zv^{5}\,(31-266z+1245z^{2}-4548z^{3}+8970z^{4}-6408z^{5})\nonumber\\
  &\quad -z^{2}v^{6}\,(36-695z+4401z^{2}-12321z^{3}+16391z^{4}-9060z^{5})\nonumber\\
  &\quad -4z^{4}v^{7}\,(1-z)(1-4z)(1-9z)(9-25z)\,.
\label{eq:PFgenus2polys}
\end{align}

\section{Renormalisation constants and details on the infrared subtraction}
\label{app:renorm}
First, we collect the renormalisation constants used in \cref{sec:uv}  through two loops, expanded in $\eps$ to the order required to obtain the finite remainder of the two-loop amplitude.
Throughout this section, we use $L_\mu=\ln(\mu^2/m_t^2)$, and for the quadratic Casimir coefficients
we have
$$
C_A = N_c,\quad\quad C_F = \frac{N_c^2-1}{2 N_c}\,.
$$
We start with the decoupling constant of \cref{eq:decoupling} given by
\begin{align}
    \zeta^{(1)} &= \frac{4}{3}T_F n_h
\left[
\left(\frac{\mu^2}{m_t^2}\right)^\varepsilon
\frac{\Gamma(1+\varepsilon)e^{\varepsilon\gamma_E}}{\varepsilon}-\frac{1}{\varepsilon}
\right], \\
    \zeta^{(2)} &= T_F n_h \left[
    \frac{16}{9} T_F n_h L_\mu^2
    + C_F \left(15 + 4 L_\mu\right)
    + C_A \left(-\frac{32}{9} + \frac{20}{3} L_\mu\right) + \mathcal{O}(\varepsilon)
\right]\,.
\end{align}
Results for the decoupling relations have been presented up to third loop order in~\cite{Chetyrkin:1997un} and 
through four loops in~\cite{Chetyrkin:2005ia}.
Renormalisation constants for parameters and fields are expanded in $\alpha_s^{(n_f)}$, see \cref{eq:renormparfields}.
As for the strong coupling in the $\overline{\rm MS}$ scheme we have
\begin{equation}
    Z^{(1)}_{\alpha_s} = -\frac{\beta_0}{\varepsilon},\quad\quad\quad
    Z^{(2)}_{\alpha_s} = \frac{\beta_0^2}{\varepsilon^2} - \frac{\beta_1}{2\varepsilon},
\end{equation}
where the coefficients of the QCD $\beta$-function are 
\begin{equation}
    \beta_0 = \frac{11}{3} C_A - \frac{4}{3} T_F \,n_f\,, \quad\quad\quad
    \beta_1 = \frac{34}{3} C_A^2 - \frac{20}{3} C_A n_f T_F - 4 C_F n_f T_F\,.
\end{equation}
The coefficient $\beta^{(n_l)}_0$ appearing in \cref{eq:zirtwoloop} is equivalent to $\beta_0$ in the equation
above with $n_f$ replaced by $n_l$.
The renormalisation constants for the top-quark mass read explicitly
\begin{align}
    Z_m^{(1)} &= -\left(\frac{\mu^2}{m_t^2}\right)^\varepsilon C_F \frac{3-2\varepsilon}{1-2\varepsilon}\,
\frac{\Gamma(1+\varepsilon)\,e^{\varepsilon\gamma_E}}{\varepsilon}\,, \\
Z_m^{(2)} &= \left(\frac{\mu^2}{m_t^2}\right)^{2\varepsilon} C_F \bigg[
T_F n_h \left(
    -\frac{2}{\varepsilon^2}
    + \frac{5 + 12L_\mu}{3\varepsilon}
    + \frac{143}{6}
    + \frac{16}{3}L_\mu
    - 2L_\mu^2
    - \frac{8\pi^2}{3}
\right) \nn \\
&\quad\quad\quad\quad\quad\quad\quad+ T_F n_l \left(
    -\frac{2}{\varepsilon^2}
    + \frac{5 + 12L_\mu}{3\varepsilon}
    + \frac{71}{6}
    + \frac{16}{3}L_\mu
    - 2L_\mu^2
    + \frac{4\pi^2}{3}
\right)\nn\\
&\quad\quad\quad\quad\quad\quad\quad+ C_F \left(
    \frac{9}{2\varepsilon^2}
    + \frac{45}{4\varepsilon}
    + \frac{199}{8}
    - \frac{17\pi^2}{4}
    + 8\pi^2\log 2
    - 12\zeta_3
\right)\nn \\
&\quad\quad\quad\quad\quad\quad\quad
+ C_A \bigg(
    \frac{11}{2\varepsilon^2}
    - \frac{97 + 132L_\mu}{12\varepsilon}
    - \frac{1111}{24}
    - \frac{44}{3}L_\mu
    + \frac{11}{2}L_\mu^2 \nn \\
&\quad\quad\quad\quad\quad\quad\quad
+ \frac{4\pi^2}{3} - 4\pi^2\log 2 + 6\zeta_3 \bigg) + \mathcal{O}(\varepsilon)
\bigg]\,.
\end{align}
Next, we report the expression for the field renormalisation constants, starting with the
top quark
\begin{align}
    Z^{(1)}_{t} &= Z_m^{(1)}\,, \\
    Z^{(2)}_{t} &= C_F \left(\frac{\mu^2}{m_t^2}\right)^{2\varepsilon} \bigg[
T_F n_h \left(
    \frac{1 + 4L_\mu}{\varepsilon}
    + \frac{947}{18}
    + \frac{16}{3}L_\mu
    - 2L_\mu^2
    - 5\pi^2
\right) \nn \\
&\quad\quad\quad\quad\quad\quad\quad+ T_F n_l \left(
    -\frac{2}{\varepsilon^2}
    + \frac{11 + 12L_\mu}{3\varepsilon}
    + \frac{113}{6}
    + \frac{16}{3}L_\mu
    - 2L_\mu^2
    + \frac{4\pi^2}{3}
\right) \nn \\
&\quad\quad\quad\quad\quad\quad\quad+ C_F \left(
    \frac{9}{2\varepsilon^2}
    + \frac{51}{4\varepsilon}
    + \frac{433}{8}
    - \frac{49\pi^2}{4}
    + 16\pi^2\log 2
    - 24\zeta_3
\right) \nn \\
&\quad\quad\quad\quad\quad\quad\quad+ C_A \bigg(
    \frac{11}{2\varepsilon^2}
    - \frac{127 + 132L_\mu}{12\varepsilon}
    - \frac{1705}{24}
    - \frac{44}{3}L_\mu
    + \frac{11}{2}L_\mu^2 \nn \\
&\quad\quad\quad\quad\quad\quad\quad
+ 5\pi^2
    - 8\pi^2\log 2
    + 12\zeta_3
\bigg) + \mathcal{O}(\varepsilon)
\bigg]\,,
\end{align}
followed by the gluon
\begin{align}
    Z^{(1)}_{G} &= -\left(\frac{\mu^2}{m_t^2}\right)^\varepsilon \frac{4}{3} T_F n_h\,
\frac{\Gamma(1+\varepsilon)e^{\varepsilon\gamma_E}}{\varepsilon}\,,\\
    Z^{(2)}_{G} &= \left(\frac{\mu^2}{m_t^2}\right)^{2\varepsilon} T_F n_h \bigg[
C_F \left(- \frac{2}{\varepsilon}-15\right)
+ C_A \left(
    \frac{35}{9\varepsilon^2}
    - \frac{45 + 88L_\mu}{18\varepsilon}
    + \frac{13}{12}
    + \frac{22}{9}L_\mu^2
    + \frac{13\pi^2}{54}
\right)\nn \\
&\quad\quad\quad\quad\quad\quad\quad\quad+ T_F n_l \left(
    -\frac{16}{9\varepsilon^2}
    + \frac{16L_\mu}{9\varepsilon}
    - \frac{8}{9}L_\mu^2
    - \frac{4\pi^2}{27}
\right)\nn \\
&\quad\quad\quad\quad\quad\quad\quad\quad+ T_F n_h \left(
    \frac{16L_\mu}{9\varepsilon}
    - \frac{8}{9}L_\mu^2
    + \frac{4\pi^2}{27}
\right) + \mathcal{O}(\varepsilon)
    \bigg]\,
\end{align}
and finally the light-quark
\begin{align}
Z_{q}^{(1)} &= 0\,, \\
Z_{q}^{(2)} &= \left(\frac{\mu^2}{m_t^2}\right)^{2\varepsilon} T_F \,n_h C_F\left( \frac{1}{\varepsilon} - \frac{5}{6} + \mathcal{O}(\varepsilon) \right)\,.
\end{align}

We now provide the expressions for the operators and quantities relevant to the infrared subtraction discussed in
\cref{sec:ir}.
The operator $\boldsymbol{\Gamma}$ for a $2\to2$ scattering with two massive particles, labelled with $(3,4)$, 
up to two-loop order reads~\cite{Ferroglia:2009ii}
\begin{align}
\boldsymbol{\Gamma} &= \mathbf{T}_1\!\cdot\!\mathbf{T}_2\;\gamma_{K}(\alpha_s)\left[\ln\frac{\mu^2}{s}+i\pi\right] - \mathbf{T}_3\!\cdot\!\mathbf{T}_4\;\gamma_{K}(\beta_{34},\alpha_s)
+ \sum_{i=1,2}\gamma_{i}(\alpha_s) + 2\gamma_{Q}(\alpha_s) \nn\\
& +\sum\limits_{\substack{i=1,2 \\ J=3,4}}^N \mathbf{T}_i\!\cdot\!\mathbf{T}_J\;\gamma_{K}(\alpha_s)\,\ln\frac{m_t\,\mu}{2\,p_i\!\cdot\!p_J}
+ \left(\frac{\alpha_s^{(n_l)}}{4\pi}\right)^2\!\!\sum_{\substack{I,J=3,4\\ I\neq J}}\ \sum_{k=1,2}\mathbf{T}_{IJk}\;f_2\!\left(\beta_{IJ},\ln\frac{v_J\!\cdot\!p_k}{v_I\!\cdot\!p_k}\right)\,,
\label{eq:softanomdim}
\end{align}
with the tripole operator defined as 
$\mathbf{T}_{ijk} = i f^{abc}\,\mathbf{T}^a_i\cdot\mathbf{T}^b_j\cdot\mathbf{T}^c_k$, and
the velocities as $v^\mu_J = p^\mu_{J}/m_J$.
The operator $\boldsymbol{\Gamma}'$ is obtained from the derivative of $\boldsymbol{\Gamma}$ with respect to
the scale $\mu$, thus one has
\begin{equation}
    \boldsymbol{\Gamma}'\equiv\frac{\partial\boldsymbol{\Gamma}}{\partial\ln\mu}=-\gamma_{K}(\alpha_s)\left(C_1+C_2\right)\,\mathbf{1}\,,
\end{equation}
with $C_{1,2}$ the Casimir constant associated with the incoming particles.
The cusp angle between the top and the antitop is
\begin{equation}
    \beta_{34} = \ln\frac{1+\beta}{1-\beta}-i\pi\,, 
    \qquad \text{with} \qquad \beta=\sqrt{1-\frac{4m_t^2}{s}}\,.
\end{equation}
The cusp and collinear anomalous dimensions are expanded perturbatively as
\begin{gather}
    \gamma_K(\alpha_s) = \sum_{n=1}\left(\frac{\alpha_s^{(n_l)}}{4\pi}\right)^n \gamma^{(n)}_K\,,\quad\quad
    \gamma_K(\beta,\alpha_s) = \sum_{n=1}\left(\frac{\alpha_s^{(n_l)}}{4\pi}\right)^n \gamma^{(n)}_{K}(\beta)\,, \nn \\
    \gamma_{q,G,Q}(\alpha_s) = \sum_{n=1}\left(\frac{\alpha_s^{(n_l)}}{4\pi}\right)^n \gamma^{(n)}_{q,G,Q}\,,
\end{gather}
with explicit coefficients 
\begin{align}
  \gamma_{K}^{(1)} & = 4\,, \nn \\
  \gamma_{K}^{(2)} &= 8\left[\left(\frac{67}{18}-\zeta_2\right)C_A-\frac{10}{9}\,T_Fn_l\right]\,,\nn\\
  \gamma^{(1)}_{K}(\beta) &= \gamma_{K}^{(1)} \beta \coth\beta,\nn\\
  \gamma^{(2)}_{K}(\beta) &= \gamma_{K}^{(2)} \beta \coth\beta + 
  8\,C_A\,\Big\{\zeta_2+\zeta_3+\beta^2 \nn\\
&\quad+\coth^2\!\beta\left[\mathrm{Li}_3(e^{-2\beta})+\beta\,\mathrm{Li}_2(e^{-2\beta})-\zeta_3+\zeta_2\beta+\frac{\beta^3}{3}\right] \nn\\
    &\quad+\coth\beta\left[\mathrm{Li}_2(e^{-2\beta})-2\beta\ln(1-e^{-2\beta})-\zeta_2(1+\beta)-\beta^2-\frac{\beta^3}{3}\right]\Big\}\,,\nn\\
    \gamma_{q}^{(1)} &= -3C_F\,, \nn\\
    \gamma_{q}^{(2)} &= C_F^2\left(-\frac32+2\pi^2-24\zeta_3\right)+C_FC_A\left(-\frac{961}{54}-\frac{11\pi^2}{6}+26\zeta_3\right) \nn\\
    &\quad+C_FT_Fn_l\left(\frac{130}{27}+\frac{2\pi^2}{3}\right),\nn\\[0.3em]
    \gamma_{G}^{(1)} &= -\beta^{(n_l)}_0\,, \nn\\
    \gamma_{G}^{(2)} &= C_A^2\left(-\frac{692}{27}+\frac{11\pi^2}{18}+2\zeta_3\right)+C_AT_Fn_l\left(\frac{256}{27}-\frac{2\pi^2}{9}\right)+4C_FT_Fn_l\,,\nn\\[0.3em]
    \gamma_{Q}^{(1)} &= -2C_F \, ,\nn\\
    \gamma_{Q}^{(2)} &= 4C_F\left[C_A\left(-\frac{49}{18}+\zeta_2-\zeta_3\right)+\frac{10}{9}\,T_Fn_l\right]\,.
\end{align}
The function $f_2$ appearing in \cref{eq:softanomdim} is given by
\begin{equation}
f_2\!\left(\beta_{IJ},\ln\frac{v_J\!\cdot\!p_k}{v_I\!\cdot\!p_k}\right) = 
-\gamma_K^{(1)} g\left(\beta_{IJ}\right)\ln\frac{v_J\!\cdot\!p_k}{v_I\!\cdot\!p_k}\,,
\end{equation}
with~\cite{Ferroglia:2009ii}
\begin{equation}
    g(\beta)=\coth\beta\left[\beta^2+2\beta\ln(1-e^{-2\beta})-\mathrm{Li}_2(e^{-2\beta})+\zeta_2\right]-\beta^2-\frac{\pi^2}{6}\,.
\end{equation}


\bibliographystyle{JHEP}
\bibliography{ttb2loop}

\end{document}